\documentclass[prd, reprint,
 amsmath,amssymb,
 aps,
]{revtex4-2}

\usepackage{graphicx}
\usepackage{dcolumn}
\usepackage{bm}
\usepackage{color}
\usepackage{braket}
\usepackage{booktabs}
\usepackage{comment}
\usepackage{url}
\usepackage{aas_macros}
\usepackage{xcolor}
\usepackage[colorlinks=true, linkcolor=blue, citecolor=blue, urlcolor=blue]{hyperref}

\begin{document}


\title{Simulation-based inference from the Lyman-alpha\\forest 1D power spectrum with \texttt{CAMELS}}

\author{Francesco Sinigaglia$^{1,2,3,4}$}
\email{IFPU fellow: fsinigag@sissa.it}
\author{Patricia Iglesias-Navarro$^{5,6}$}
\author{Matteo Viel$^{1,2,3,4,7}$}
\vspace{0.5cm}
\affiliation{
\vspace{0.2cm}$^{1}$Institute for Fundamental Physics of the Universe (IFPU), Via Beirut 2, I-34151 Trieste, Italy}
\vspace{0.2cm}
\affiliation{
$^{2}$SISSA - International School for Advanced Studies, Via Bonomea 265, 34136 Trieste, Italy}
\vspace{0.2cm}
\affiliation{
$^{3}$INAF - Osservatorio Astronomico di Trieste, Via G. B. Tiepolo 11, I-34131 Trieste, Italy}
\vspace{0.2cm}
\affiliation{
$^{4}$INFN – National Institute for Nuclear Physics, Via Valerio 2, I-34127 Trieste, Italy}
\vspace{0.2cm}
\affiliation{
$^{5}$Instituto de Astrof\'isica de Canarias, Calle via L\'actea s/n, E-38205, La  Laguna, Tenerife, Spain
}
\vspace{0.2cm}
\affiliation{
$^{6}$Departamento  de  Astrof\'isica, Universidad de La Laguna,  E-38206, La Laguna, Tenerife, Spain}
\vspace{0.2cm}
\affiliation{
$^{7}$ICSC - Centro Nazionale di Ricerca in High Performance Computing, Big Data e Quantum Computing, Via Magnanelli 2, Bologna, Italy\vspace{0.2cm}}

\date{\today}

\begin{abstract}
We perform for the first time full simulation-based inference on the Lyman-$\alpha$ forest 1D power spectrum. In particular, we consider the prediction of the Lyman-$\alpha$ forest $P_{\rm 1D}(k)$ at $2.0<z<3.5$ from the \texttt{CAMELS} cosmological hydrodynamic simulations run with the \texttt{IllustrisTNG} and \texttt{SIMBA} galaxy formation models. We train a normalizing flow to perform neural posterior estimation of two cosmological parameters ($\Omega_m$ and $\sigma_8$) and four astrophysical parameters parametrizing supernova and AGN feedback.
When training and testing the neural network on the same baryon physics model, the posterior distributions of the cosmological parameters are found to be in excellent agreement with the true parameters values (within $10\%$ deviations in $\gtrsim 75\%$ and $\gtrsim 90\%$ of the cases for $\Omega_m$ and $\sigma_8$, and a precision better than $10\%$ in both), while the astrophysical parameters converge to the prior mean due to the limited probed volume. When training on one model and testing on the other (e.g., training on \texttt{IllustrisTNG} and testing on \texttt{SIMBA}, or viceversa), the performance is significantly worse, both in accuracy and in precision, resulting in a $\sim 10\%$ positive bias on the predicted values for $\sigma_8$. We show that a multi-domain training based on the combination of simulations from both models recovers unbiased constraints, offering an effective solution to cope with the complex problem of the lack of convergence in the predictions from different galaxy formation models. This study represents a promising way forward to constrain cosmology and fundamental physics with the Lyman-$\alpha$ forest with artificial intelligence.

\end{abstract}

\maketitle

\section{Introduction}
\label{sec:intro}

Over the past two decades, the one-dimensional (1D) flux power spectrum of the Lyman-$\alpha$ (hereafter Ly$\alpha$) forest has emerged as one of the most sensitive probes of small-scale structure in the high-redshift Universe. The forest—produced by intervening neutral hydrogen absorption in the spectra of distant quasars—encodes information about the matter power spectrum on comoving scales of $\sim 0.1-10~h~{\rm Mpc}^{-1}$ at redshifts $2\lesssim z \lesssim 5$ \citep[see e.g.,][for a review]{McQuinn2016}. Because these scales and redshifts are difficult to access through galaxy surveys or cosmic microwave background measurements, the Ly$\alpha$ forest provides a uniquely powerful window into the nature of dark matter \citep[e.g.,][]{Viel2004,Viel2005,Wang2013,Yeche2017,Irsic2017,Murgia2018,Villasenor2023,GarciaGallego2025}, the thermal and ionization history of the intergalactic medium (IGM) \citep{Lidz2010,Garzilli2012,Nasir2016,Boera2019,Walther2019,Garzilli2020,Gaikwad2021,Villasenor2022,Nasir2024}, the mass of neutrinos \citep[e.g.,][]{Palanque2015,Palanque2015b,Rossi2017,Pedersen2020,Palanque2020,Ivanov2025}, primordial magnetic fields \citep{Chongchitnan2014,Pavicevic2025}, modified theories of gravity \cite{Rossi2017}, and primordial black holes \cite{Inman2019}, among others.
Early measurements of the 1D flux power spectrum were presented by the Sloan Digital Sky Survey (SDSS) \citep{McDonald2006}, the Baryon Oscillation Spectroscopic Survey (BOSS) \citep{Palanque2013}, and extended Baryon Oscillation Spectroscopic Survey (eBOSS) \citep{Chabanier2019}. Currently, the most recent state-of-the-art $P_{\rm 1D}(k)$ measurement has been provided by the Dark Energy Spectroscopic Instrument (DESI) collaboration \citep{Ravoux2025,Karacayli2025}, and has achieved exquisite precision and high degree of control over a variety of systematics, including continuum subtraction, high column density systems (HCDs), and metal contamination. Alternatively to large scale structure redshift surveys, observational campaigns of hundreds of high-redshift quasars conducted with very high-resolution spectrographs---such as XQ-100 \citep{Lopez2016}, the Keck Observatory Database of Ionized Absorption toward Quasars \citep[KODIAQ,][]{Omeara2017}, and the UVES Spectral Quasar Absorption Database \citep[SQUAD,][]{Murphy2019} ---have delivered additional and complementary measurements, probing the power spectrum to highly nonlinear smaller scales ($k\sim 10~h~{\rm Mpc}^{-1}$), in a regime which is very sensitive to the fundamental physics phenomena listed above. Interpreting these measurements, however, requires detailed modeling of nonlinear structure formation, gas dynamics, thermal broadening, peculiar velocities, and astrophysical feedback processes—phenomena that are inherently complex and strongly degenerate.

As cosmological datasets continue to grow in size and precision, the need for an accurate and precise inference framework able to efficiently marginalize over astrophysical uncertainties has emerged as a priority in the field. Traditionally, inference from the Ly$\alpha$ forest has relied on grids of hydrodynamical simulations combined with interpolation schemes to construct approximate likelihoods \citep[e.g.][]{McDonald2006,Becker2011,Viel2013,Palanque2013,Bolton2014,Palanque2015,Palanque2015b,Yeche2017,Irsic2017,Palanque2020,Walther2019,Pavicevic2025}. While successful, this approach can incur into interpolation errors, which would potentially bias posterior estimates. More advanced numerical interpolation techniques such as Gaussian processes \citep{Bird2019,Pederson2021,Cabayol2023,Bird2023} or neural networks \citep{ChavesMontero2024,Walther2025}---often referred to as `emulators'---have been shown to reproduce the Ly$\alpha$  forest $P_{\rm 1D}(k)$ from high-fidelity cosmological hydrodynamic simulation down into the nonlinear regime with percent accuracy and are being used in state-of-the-art inference analysis \citep[see e.g.,][]{ChavesMontero2026}. These approaches have delivered competitive results, but still present important limitations. First, they rely on Gaussian likelihoods, and thereby fail to capture potential deviations beyond this assumption, particularly when incorporating systematics or nontrivial covariance structures. In particular, even robust sampling methods are bottlenecked by this analytical approximation, which can artificially constrain complex degeneracies or obscure true multimodal solutions in the resulting parameter space \citep[see e.g.,][]{Hahn2019}. Second, they typically rely on sets of simulations limited to just one code and to a relatively restricted cosmological and astrophysical parameter space. It was shown that this assumption does not seem to induce significant biases in the face values of the cosmological parameters when applied to simulations run with codes and models not used during training \citep{ChavesMontero2026}. However, the situation may change when considering different baryon physics models, and/or when extending the analysis to higher $k$s. 

In this context, simulation-based inference \citep[hereafter SBI, see e.g.,][]{Akeret2015,Papamakarios2016,Brehmer2018,Alsing2018,Charnock2018,Alsing2019} offers a promising alternative framework. Rather than explicitly constructing a likelihood function, SBI methods learn the mapping between parameters and observables directly from simulations. By amortizing inference across the parameter space, SBI can efficiently handle high-dimensional models, complex summary statistics, and non-Gaussian features of the data, and naturally accommodates nonlinearities and degeneracies while avoiding restrictive assumptions about the analytic form of the likelihood.

In this work, we perform for the first time SBI of the 1D Ly$\alpha$ forest flux power spectrum using the large \texttt{CAMELS} suite of cosmological hydrodynamical simulations \citep{VillaescusaNavarro2021}, exploiting its unprecedented number of simulations spanning different cosmological and astrophysical parameters, as well as the availability of predictions from different simulation codes and baryon feedback prescriptions. Specifically, we train a neural network to approximate the posterior distributions of the model parameters given the Ly$\alpha$ forest $P_{\rm 1D}(k)$, and test the resulting inference framework on two different state-of-the-art galaxy formation models (\texttt{IllustrisTNG} \cite{Weinberger2017,Pillepich2018,Nelson2019} and \texttt{SIMBA} \cite{Dave2019}). We quantify the robustness of our results by measuring the accuracy and precision of the predicted posteriors compared to the true parameters values, as well as diagnose the reliability of the posteriors by means of a variety of statistical tests. By combining high-fidelity simulations with modern amortized inference techniques, this work aims to establish a flexible and scalable inference framework in the view of current and forthcoming  next-generation Ly$\alpha$ forest data, as an alternative to traditional explicit likelihood-based analysis.

The paper is structured as follows. In Section~\ref{sec:camels}, we introduce the cosmological hydrodynamic simulations used to train the networks. In Section~\ref{sec:sbi}, we describe the SBI framework that we use in this work. Section~\ref{sec:results} presents the results of our analysis and their discussion. We conclude in \S\ref{sec:conclusions}.    

\section{The CAMELS simulations}
\label{sec:camels}

The {\it Cosmology and Astrophysics with MachinE Learning Simulations} project) \citep[hereafter \texttt{CAMELS},][]{VillaescusaNavarro2021,CAMELS_DR1} is a large suite of state-of-the-art cosmological N-body and magneto-hydrodynamic simulations. Originally comprising $4,233$ simulations ($2,184$ hydrodynamic simulations and $2,049$ N-body), they were obtained using two different codes and baryon physics models: (i) one set run with \texttt{AREPO} and the \texttt{IllustrisTNG} model, and (ii) one set run with the \texttt{GIZMO} code and the \texttt{SIMBA} model. 
The simulations were run in cosmological boxes of volume $V=(25~h^{-1}~{\rm Mpc})^3$ with periodic boundary conditions, following the evolution of $256^3$ dark matter particles of mass $m_{\rm dm}=6.49\times 10^7(\Omega_m-\Omega_b)/0.251~h^{-1}~{\rm M_\odot}$ and of $256^3$ gas resolution elements with an initial mass $m_{\rm gas}=1.27 \times 10^7~h^{-1}~{\rm M_\odot}$. The initial conditions were generated at $z=127$ using second-order Lagrangian Perturbation Theory and the cosmological parameters (except for $\Omega_m$ and $\sigma_8$) were fixed to the following values: $\Omega_b=0.049$, $h=6711$, $n_s=0.9624$, $M_\nu=0.0~{\rm eV}$, $w=-1$, and $\Omega_k=0$. Both the \texttt{IllustrisTNG} and \texttt{SIMBA} sets used herein comprise $1,000$ simulations, run varying $6$ different parameters---the two cosmological parameters $\Omega_m$ and $\sigma_8$ and four astrophysical parameters $A_{\rm SN1}$, $A_{\rm AGN1}$, $A_{\rm SN2}$, $A_{\rm AGN2}$---sampled following a latin hypercube strategy with the following priors: $\Omega_m\in[0.1,0.5]$, $\sigma_8\in[0.6,1.0]$, $A_{\rm SN1}\in[0.25,4.0]$, $A_{\rm AGN1}\in[0.25,4.0]$, $A_{\rm SN2}\in[0.5,2.0]$, $A_{\rm AGN2}\in[0.5,2.0]$. We hereafter refer to this dataset as the `latin hypercube' (or `LH') suite. The specific meaning of the astrophysical parameter will be discussed in detail in Section~\ref{sec:camels_tng} and \ref{sec:camels_simba}, in the context of each model. Since the first data release, the project has expanded significantly and a number of different codes have been employed to run additional \texttt{CAMELS} simulation suites. However, since the Ly$\alpha$ forest spectra have been publicly released only for the initial dataset described above---namely the \texttt{IllustrisTNG} and \texttt{SIMBA} LH suites---we stick to it in this paper and describe it in the following sections. Specifically, we retrieve Ly$\alpha$ forest skewers at $2\le z <3.5$, and in particular at $z=\{2.00,~2.15,~2.30,~2.46,~2.63,~2.80,~3.01,~3.49\}$. Importantly, the \texttt{CAMELS} LH simulations do not explore different IGM thermal and reionization history models. This implies that they share the same $T-\rho$ relation of the intergalactic gas, except for marginal variations induced by baryon feedback. We refer to the \texttt{CAMELS} documentation ({\url{https://camels.readthedocs.io/en/latest/}}) for a thorough description of all the simulations available. 

In what follows, we briefly describe the galaxy formation models underlying the simulations used in this work and the method used to compute the Ly$\alpha$ forest skewers, as well as discuss the strengths and weaknesses of these simulations for this application of SBI. We refer to original \texttt{CAMELS} papers \citep[][and references therein]{VillaescusaNavarro2021,CAMELS_DR1} for a detailed summary of the main physical prescriptions implemented in the \texttt{IllustrisTNG} and \texttt{SIMBA} models.

\subsection{The IllustrisTNG model}
\label{sec:camels_tng}

The \texttt{IllustrisTNG} model \cite{Weinberger2017,Pillepich2018,Nelson2019}---which consists of an improved version of the original \texttt{Illustris} \citep{Vogelsberger2013,Torrey2014}---is a state-of-the-art cosmological galaxy formation model employed to run the \texttt{IllustrisTNG} simulation suite. This model makes use of the \texttt{AREPO} code \citep{Springel2010,Weinberger2020}, which adopts an N-body tree-particle-mesh algorithm to solve the equations of gravity and a moving-mesh approach to solve for magneto-hydrodynamics. In addition, \texttt{IllustrisTNG} includes advanced subgrid models to account for radiative cooling and heating from primordial elements and metals \cite{Katz1996,Wiersma2009} assuming a spatially-uniform UV ionization background \cite{FaucherGiguere2009} and neutral hydrogen self-shielding \citep{Rahmati2013}, star formation from a multi-phase interstellar medium model and chemical enrichment from different stellar evolution processes and tracking the abundance of nine different elements (H, He, C, N, O, Ne, Mg, Si, and Fe) \cite{SpringelHernquist2003}, the formation and growth of supermassive black holes (hereafter SMBH) and the associated low-mode kinetic feedback effect \cite{Weinberger2017}, and feedback from galactic winds \cite{Pillepich2018}. In the \texttt{CAMELS} LH simulations run with this model, all the subgrid physics parameters are fixed to the fiducial values, except for the $4$ astrophysical parameters mentioned above. Specifically, in the \texttt{IllustrisTNG} model, the wind mass loading factor is defined as $\eta_v=\dot{M}_{\rm wind}/{\rm SFR}=1.8~v_w^{-2}~e_w$, where $M_{\rm wind}$ is the mass ejected by the galactic wind, SFR is the star formation rate, $e_w\propto A_{\rm SN1}$ is the total energy injection rate (power) per unit star-formation and $v_w\propto A_{\rm SN2}$ is the wind speed. Therefore, $A_{\rm SN1}$ and $A_{\rm SN2}$ regulate the galactic wind effect. On the other hand, the parameters $A_{\rm AGN1}$ and $A_{\rm AGN2}$ modulate the low accretion rate kinetic SMBH feedback mode. Specifically, the power injected in the kinetic mode $\dot{E}_{\rm low}\propto A_{\rm AGN1}$ and the energy accumulated from the last feedback event  $E_{\rm inj}\propto A_{\rm AGN2}$ are released into the surrounding medium in discrete events in a random direction. In this sense, $A_{\rm AGN1}$ and $A_{\rm AGN2}$ regulate the energy accumulation by the SMBH and the burstiness and speed of its re-injection into the interstellar medium, respectively. 

\subsection{SIMBA}
\label{sec:camels_simba}

\texttt{SIMBA} \cite{Dave2019} is a galaxy formation model  which builds upon \texttt{MUFASA} \cite{Dave2016}. It relies on the \texttt{GIZMO} code \cite{Hopkins2015}, which adopts the tree-particle-mesh algorithm from \texttt{GADGET-III} \cite{Springel2005} with adaptative softening lengths and the Meshless Finite Mass hydrodynamics mode. In \texttt{SIMBA}, radiative cooling and photoionization heating are solved by employing the \texttt{Grackle-3.1} library \cite{Smith2017}, including non-equilibrium evolution of primordial metals and metal cooling, a spatially-uniform ionizing background \cite{HaardtMadau2012}, and neutral hydrogen self-shielding \cite{Rahmati2013}. Star formation is regulated by the ratio between the $H_2$ gas density and the dynamical time, and the $H_2$ abundance is computed following a prescription based on the metallicity and the local column density \cite{Krumholz2011}. The chemical enrichment by stellar evolution is tracked by following the evolution of 11 elements (H, He, C, N, O, Ne, Mg, Si, S, Ca, and Fe), and the formation, growth and destruction of dust grains is modeled on the fly during the simulations. In \texttt{SIMBA}, the $A_{\rm SN1}$ parameter acts as normalization factor of the mass loading factor of galactic winds $\eta \propto A_{\rm SN1}$, while $A_{\rm SN2}$ represents the normalization of the wind speed $v_w\propto A_{\rm SN2}$ as in \texttt{IllustrisTNG}. The black hole growth follows a two-phase model: a cold phase accretion given by the transport of angular momentum by gravitational torques from the stars \citep{Hopkins2011}, whereas for the hot phase the gas follows the spherical Bondi accretion. The feedback from SMBH is modeled in a two-mode fashion \citep{Heckman2014}: a `QSO mode' with high mass loading outflows, and a `jet mode' with lower mass loading but faster outflows at low Eddington ratios. For both feedback modes, the gas ejection from the black hole is bipolar, along the parallel/antiparallel direction to the angular momentum vector defined by the black hole particles.

\subsection{Cosmological parameterization to describe the $P_{\rm 1D}(k)$}
\label{sec:cosmo:pars}

In recent works, \citep[e.g.,][]{Pederson2021,Pedersen2023,ChavesMontero2026} the cosmological $P_{\rm 1D}(k)$ emulation and the resulting inference was performed in terms of the two following `compressed' parameters \citep{McDonald2006}:
\begin{equation}
    \Delta_\star^2=\frac{k_\star^3~P_{\rm lin}(k_\star,z_\star)}{2\pi}
\end{equation}
\begin{equation}
    n_\star = \frac{ d\log P_{\rm lin}(k.z)}{d\log k}\bigg\rvert_{k=k_\star,z=z_\star}
\end{equation}
representing the amplitude and the slope of the linear power spectrum $P_{\rm lin}$ at some pivot scale $k_\star$ and redshift $z_\star$, in contrast to the usage of a `standard' $\Lambda$CDM cosmological parameters basis in other analyses \citep[e.g.,][]{Viel2006,Borde2014,Palanque2015,Palanque2015b,Palanque2020}. This parameter compression is motivated by the fact that at the redshift relevant for the Lya forest ($z\gtrsim 2$), the Universe is practically Einstein-de Sitter and therefore it is reasonable to assume there is no need to marginalize over the growth function and take constraints at a single pivot redshift. Since the aforementioned studies sought to perform cosmological inference by emulating the $P_{\rm 1D}(k)$ starting from a number of reference simulations and applying a variety of numerical techniques to reproduce them, a parametrization that minimizes the number of free parameters is desirable. 

Here, however, we are in a different situation for two reasons: (i) we do not aim at emulating the $P_{\rm 1D}(k)$ to be able to perform an explicit likelihood analysis, but rather to directly approximate the posteriors using neural networks, and (ii) the simulations that we use in this work were run varying explicitly only the parameters $\Omega_m$ and $\sigma_8$, so there is no specific gain in passing from the $(\Omega_m,\sigma_8)$ to the $(\Delta_\star^2,n_\star)$ parametrization and we prefer to stick here to the standard cosmological parameters to ease the interpretability of the results when addressing the performance of the models. In future studies, in which we will make use of \texttt{CAMELS} simulations varying $5$ cosmological parameters (see Section \ref{sec:camels_pros_cons}), we will reassess this assumption and explore whether the aforementioned parameters compression is still advantageous over using the standard $\Lambda$CDM parametrization. In addition, the argument presented above regarding the validity of the parameter compression in an Einstein-de Sitter Universe may no longer hold true in a dynamical dark energy scenario, as proposed by DESI \citep[see e.g.,][]{DESI_DR2}.

\subsection{Ly$\alpha$ forest $P_{\rm 1D}(k)$ and mean flux}
\label{sec:camels_lya}

\begin{figure}
    \centering
    \includegraphics[width=\columnwidth]{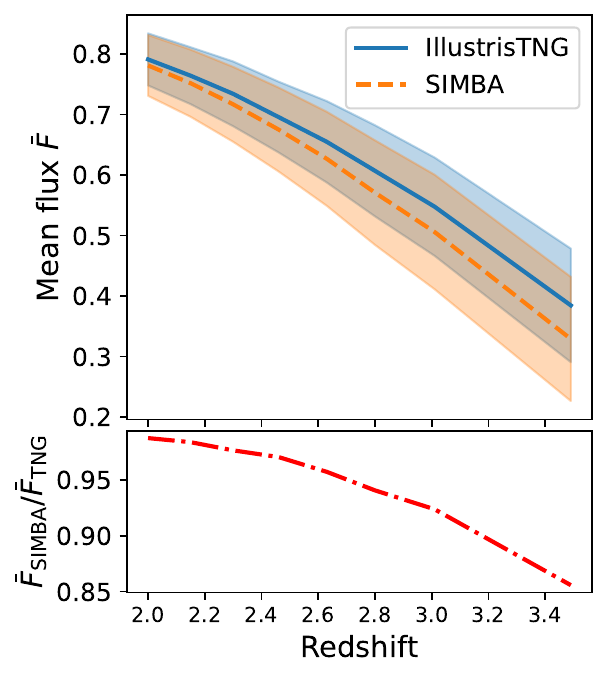}
    \caption{Top: Mean flux $\bar{F}$ as a function of redshift, for \texttt{IllustrisTNG} (blue solid) and \texttt{SIMBA} (orange dashed), averaged over all the $1,000$ available simulations and with shaded regions indicating the standard deviation. Bottom: Mean flux ratios between \texttt{SIMBA} and \texttt{IllustrisTNG}, as a function of redshift.}
    \label{fig:meanflux}
\end{figure}

\begin{figure*}
    \centering
    \includegraphics[width=0.9\textwidth]{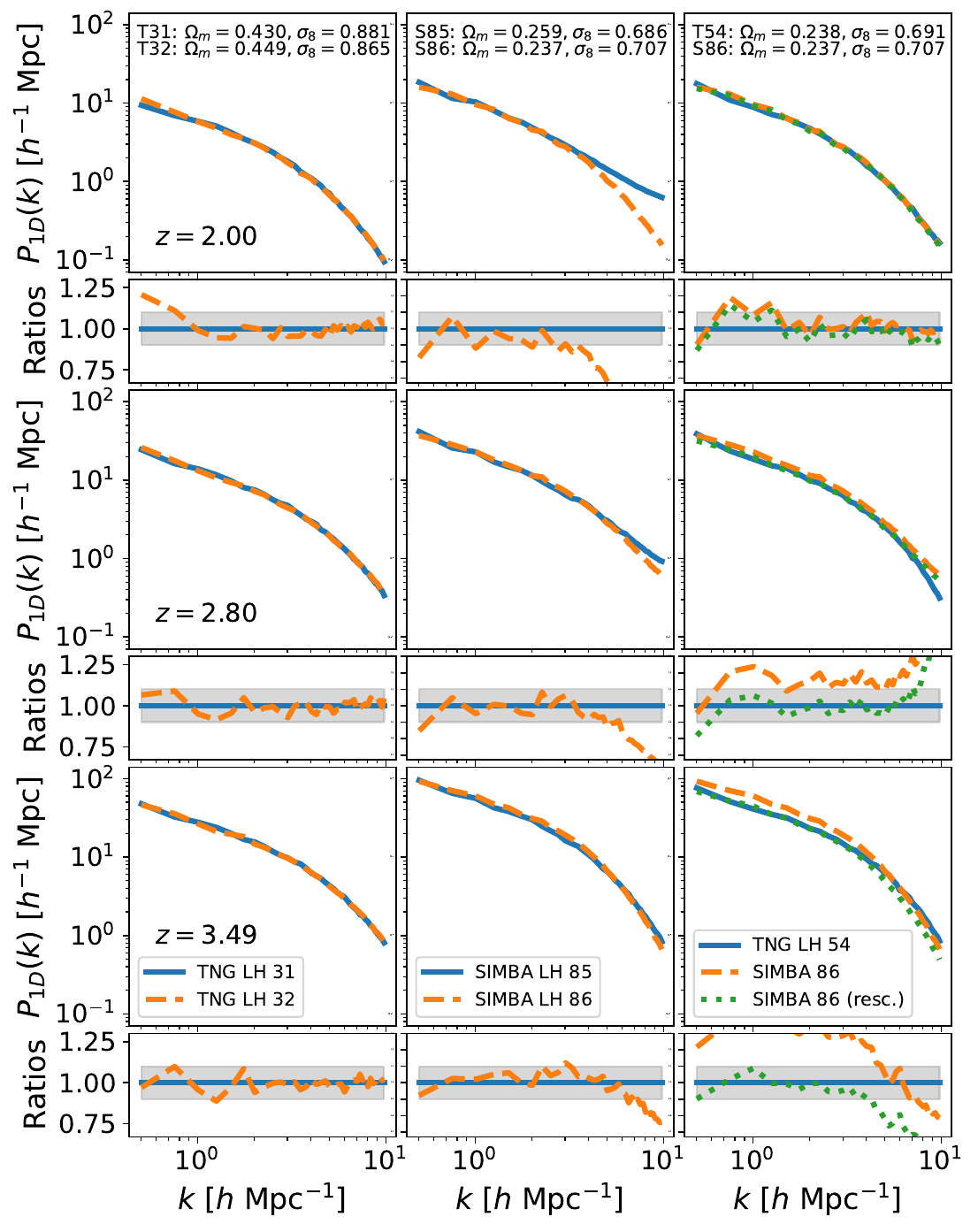}
    \caption{Comparison of $P_{\rm 1D}(k)$ from pairs of realizations with approximately same cosmology (within the same latin hypercube voxel), but different astrophysical parameters, at $z=2.00$ (top), $z=2.80$ (mid), and $z=3.49$ (bottom). Per each redshift, we show the power spectra in the top sub-row, and the ratios with respect to one of the $P_{\rm 1D}(k)$ in the bottom sub-row, where the gray shaded region indicate the $10\%$ deviation region. Left: $P_{\rm 1D}(k)$ for realizations $31$ (blue solid) and $32$ (orange dashed), both from \texttt{IllustrisTNG}. Center: $P_{\rm 1D}(k)$ from realizations $85$ (blue solid) and $86$ (orange dashed), both from \texttt{SIMBA}. Right: $P_{\rm 1D}(k)$ from \texttt{IllustrisTNG} realization $54$ (blue solid), from \texttt{SIMBA} realization $86$ (orange dashed), and from \texttt{SIMBA} realization $86$ but including mean flux rescaling (green dotted).}
    \label{fig:pk_comparison}
\end{figure*}

The Ly$\alpha$ forest skewers from the \texttt{CAMELS} simulations were generated by relying on the \texttt{FSFE} (available at {\url{https://github.com/sbird/fake_spectra}}) C++/Python package \cite{Bird2015,Bird2017}. Briefly, the code operates directly on simulation outputs (e.g., particle-based or moving-mesh formats) and preserves the full phase-space information of individual gas elements.
For each selected line of sight, gas elements within a smoothing length of the skewer are identified. The neutral hydrogen number density is computed from the gas density and ionization fraction, and the contribution of each element to the optical depth is evaluated in redshift space. The line-of-sight velocity includes both Hubble expansion and peculiar motion. The total optical depth $\tau$ at velocity coordinate $v$ is obtained by summing the Voigt-broadened contributions from all intersecting gas elements:
\begin{equation}
\tau(v)=\sum_i \tau_i(v)
\end{equation}

where each term depends on the element’s column density, Doppler parameter (set by its temperature), and the atomic transition parameters (oscillator strength and rest wavelength). The transmitted flux fraction is then computed as 
\begin{equation}
  F(v)=\exp[-\tau(v)] \quad .  
\end{equation}

By directly summing particle-level Voigt profiles rather than projecting the density field onto a one-dimensional grid prior to spectral construction, \texttt{FSFE} minimizes artificial smoothing and retains small-scale velocity and thermal structure. This approach is particularly important for accurate modeling of narrow absorption features and the high-wavenumber behavior of the Lya forest $P_{\rm 1D}(k)$. Per each simulation box, $5,000$ skewers were extracted at random locations along the $z$ coordinate direction. 

Afterwards, we compute the Lya forest $P_{\rm 1D}(k)=\braket{|\delta_F(k)|^2}$, where $\delta_F(k)$ is obtained via Fast Fourier Transforms from the configuration-space flux contrast field $\delta_F=F/\bar{F}-1$, where $\bar{F}$ is the mean transmitted flux. We notice that in this formulation the $P_{\rm 1D}(k)$ is sensitive to potential variations in $\bar{F}$, which in real observational analysis can be due to  e.g. an imperfect continuum subtraction, or other systematics. In state-of-the-art emulators \citep[e.g.,][]{ChavesMontero2026}, this effect is taken into account by marginalizing over the mean flux and has been shown to re-absorb the $P_{\rm 1D}(k)$ mismatch between the eBOSS and DESI measurements, without impacting the cosmological inference \citep{ChavesMontero2026}. In this paper, where the skewers are extracted from simulations and hence have no noise nor systematics, in principle we do not have to deal with such an issue, and any flux mismatch will therefore stem from the underlying astrophysics. This aspect has important implications. On the one hand, it encodes important astrophysical information, which one would ideally like to exploit to obtain more stringent constraints on the underlying baryon physics model. On the other hand, if different galaxy formation models used in combination in the inference framework fail to converge in the mean flux prediction, this may affect our analysis adding a source of uncertainty and potentially bias the cosmological inference. In this case, a marginalization over $\bar{F}$ is desired. In the present case, all the simulations run with the same code share the same astrophysical prescription, and hence, the predicted $\bar{F}$ is mainly determined by variations in the astrophysical parameters, with a negligible contribution arising from cosmic variance. When we consider, however, simulations run with different codes and astrophysical subgrid prescriptions, the situation changes. Figure~\ref{fig:meanflux} shows $\bar{F}$ as a function of redshift, averaged over all the $1,000$ LH simulations available per each model. Clearly, \texttt{SIMBA} underpredicts $\bar{F}$  by a factor $\sim 5-15\%$ at $z\gtrsim 3$ with respect to \texttt{IllustrisTNG}, potentially owing to the slightly different choice for the UV background prescription. This induces a systematic overprediction in the $P_{\rm 1D}(k)$ at fixed cosmological and astrophysical parameters.  This poses the problem of how to deal with this issue. As will be shown later in Section \ref{sec:results} and in Appendix \ref{app:joint}, the astrophysical parameters are very poorly constrained, due to the small volume spanned by each realization and the resulting cosmic variance effect. Figure~\ref{fig:pk_comparison} shows a comparison between the prediction of the power spectrum between pairs of realization having similar cosmological parameters (within the same latin hypercube voxel), but having different astrophysical parameters. One can clearly see that the dependence on astrophysics is weak on large scales, having practically no impact in the two \texttt{IllustrisTNG} simulations, and being consistent with noise at $k\lesssim 3.0~h~{\rm Mpc}^{-1}$ at all redshifts for the two \texttt{SIMBA} simulations (mid) and for the cross-comparison between \texttt{IllustrisTNG} and \texttt{SIMBA} (after the flux rescaling described below). For this reason, we decide here to renormalize the mean flux of \texttt{SIMBA} to match the mean flux of \texttt{IllustrisTNG} (the inverse would have been equivalent). Specifically, we compute the mean flux $\bar{F}_{\rm TNG}$ and $\bar{F}_{\rm SIMBA}$ from the average of the $1,000$ \texttt{IllustrisTNG} and \texttt{SIMBA} LH simulations, respectively, and rescale all the \texttt{SIMBA} flux skewers by a factor $\bar{F}_{\rm TNG}/\bar{F}_{\rm SIMBA}$. We illustrate the effect of the rescaling in the right column of Figure~\ref{fig:pk_comparison}: the original \texttt{SIMBA} $P_{\rm 1D}(k)$ (orange dashed), which has a $\sim 16\%$ offset on large scales with respect to \texttt{IllustrisTNG}, becomes fully unbiased after the rescaling (green dotted).

Due to the limited box size, the minimum probed wavenumber is $k_{\rm min}\sim 0.5~h~{\rm Mpc}^{-1}$, with a Fourier-space sampling $\Delta k = 0.25~h~{\rm Mpc}^{-1}$. To avoid accuracy and convergence issues, as will be explained more in details in Section \ref{sec:camels_pros_cons}, we perform our baseline analysis by conservatively restricting the power spectrum to the following three scale cuts---$k_{\rm max}=1.5~h~{\rm Mpc}^{-1}$, $k_{\rm max}=2.0~h~{\rm Mpc}^{-1}$, and $k_{\rm max}=3.0~h~{\rm Mpc}^{-1}$---and study the robustness of the results against this choice. We notice that the available skewers resolutions enables measurements of the $P_{\rm 1D}(k)$ up to $k\sim 10~h~{\rm Mpc}^{-1}$. The chosen scale cuts encode a relatively small amount of information, as they comprise only $4$, $6$, and $10$ data points per each redshift for the three different $k_{\rm max}$. As will be shown in Section \ref{sec:results}, this suboptimal situation guarantees already stable and converged results, so we expect the results to be even more favorable in a case in which the simulations in the training dataset have larger volumes (as will be the case of future \texttt{CAMELS} simulations, see Section \ref{sec:camels_pros_cons}) and/or higher resolution, which allow probe larger and smaller scales, respectively.

\subsection{The effect of feedback on $P_{\rm 1D}(k)$}
\label{sec:camels_feedback}

\begin{figure*}
    \centering
    \includegraphics[width=0.9\linewidth]{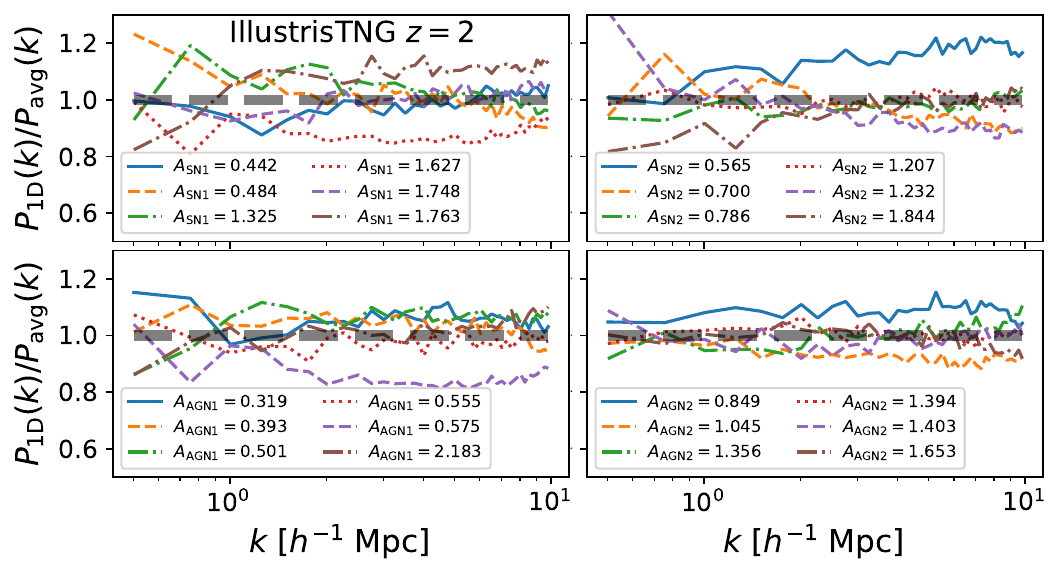}
    \includegraphics[width=0.9\linewidth]{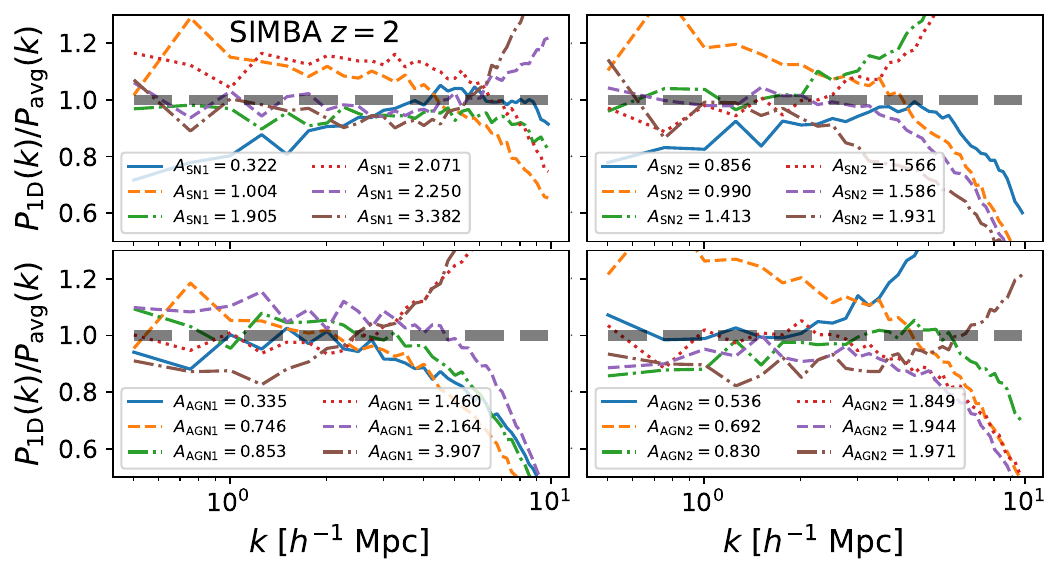}
    \caption{Ratios between the $P_{\rm 1D}(k)$ at $z=2$ from sets of six simulations varying only one astrophysical parameter, and their average, for \texttt{IllustrisTNG} (top set of plots) and \texttt{SIMBA} (bottom set of plots). Per each set of plots, we vary $A_{\rm SN1}$ in the top left panel,  $A_{\rm AGN1}$ in the bottom left panel,  $A_{\rm SN2}$ in the top right panel, and  $A_{\rm AGN2}$ in the bottom right panel.}
    \label{fig:pk_feedback}
\end{figure*}

In this section, we study the effect of supernova and AGN feedback on the power spectrum. Previous works \citep[see e.g.,][]{Viel2013b,Bolton2017,Khaire2024,Dong2024,Tillman2025} showed that the Ly$\alpha$ forest $P_{\rm 1D}(k)$ is sensitive to both the galactic winds driven by supernova feedback and feedback from AGNs on scales relevant for cosmological analysis. To develop an intuitive understanding of the effect of baryon feedback on the $P_{\rm 1D}(k)$ predictions in \texttt{CAMELS}, we consider here subsets of simulations varying just one astrophysical parameter at a time and keeping the other parameters fixed (within a narrow interval). We remind the reader that, because the \texttt{CAMELS} simulations were run with different random seeds, at this stage the effect of feedback is degenerate with cosmic variance, whose impact will be addressed in the next section. We consider six simulations per each case, and plot the ratio against their average in Figure~\ref{fig:pk_feedback} for \texttt{IllustrisTNG} (top) and \texttt{SIMBA} (bottom), at $z=2$, i.e. the redshift at which the feedback effect should be maximized. In particular, per each subsets of plots, we display the variations with $A_{\rm SN1}$ in the top left subpanel, with $A_{\rm AGN1}$ in the bottom left subpanel, with $A_{\rm SN2}$ in the top right subpanel, and with $A_{\rm AGN2}$ in the bottom right subpanel. A detailed analysis of these results reveals that:
\begin{itemize}
    \item the effect on the power spectrum are stronger for \texttt{SIMBA} than for \texttt{IllustrisTNG};
    \item the variations between the simulations have non-negligible effects on the $P_{\rm 1D}(k)$, in many cases well beyond the $10\%$ level;
    \item most importantly, we do not identify any obvious correlation between the variations in the astrophysical parameters and in the power spectrum.
\end{itemize}

From these results, it is therefore not surprising that the inference framework that we have built in this work is not able to constrain the astrophysical parameters (see Appendix~\ref{app:joint}). This result may be either due to an intrinsic lack of sensitivity of the Ly$\alpha$ forest to variations of the feedback parameters, or to the fact that cosmic variance is the dominant effect here. We further diagnose the latter point in the next section.  

\subsection{Cosmic variance}
\label{sec:camels_cv}

\begin{figure*}
    \centering
    \includegraphics[width=0.9\linewidth]{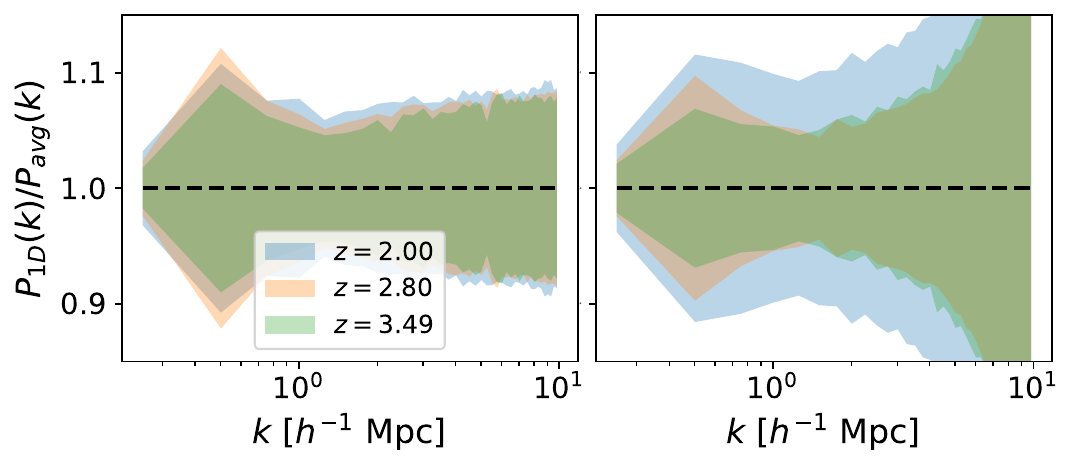}
    \caption{Standard deviation of the $P_{\rm 1D}(k)$ from the $27$ CV realizations at $z=2.00$ (blue), $z=2.80$ (orange), and $z=3.49$ (green), for \texttt{IllustrisTNG} (left) and \texttt{SIMBA} (right).}
    \label{fig:cosmic_variance}
\end{figure*}

One typical issue associated with large scale structure analysis is cosmic variance, stemming from the intrinsic variance in the initial conditions. The limited volume of the \texttt{CAMELS} simulations represents a potential point of concern in this sense. In other works \citep{Pedersen2023,ChavesMontero2026}, which aim at emulating the $P_{\rm 1D}(k)$ and require predictions which are as most free from cosmic variance as possible, this problem has been addressed by using as reference training set cosmological simulations with fixed-and-paired variance-suppressed initial conditions \citep{Angulo2016}. In this way, the prediction from the emulator is free from variance effects, and the error budget corresponding to cosmic variance is then reincorporated into the covariance matrices used to define the likelihood in the Markov Chain Monte Carlo inference. 

To study the effect of cosmic variance on the $P_{\rm 1D}(k)$ predictions, we employ the `CV' \texttt{CAMELS}, consisting of $27$ simulations run with fixed cosmological and astrophysical parameters and varying only the initial random seed. Figure~\ref{fig:cosmic_variance} display the standard deviations from the CV simulations for \texttt{IllustrisTNG} (left) and \texttt{SIMBA} (right), at $z=2.00$ (blue), $z=2.80$ (orange), and $z=3.49$ (green). We notice that in this case we are working with 1D spectra, so the Fourier space modes are counted along the line of sight and not in spherical shells (as would be the case of a 3D power spectrum). Therefore, one does not expect a scale-dependent reduction of cosmic variance due to a larger number of available Fourier modes towards small scales. In contrast, the variance towards high $k$s increases significantly in \texttt{SIMBA}, where the feedback effects dominate. In light of these results, we ascribe the insensitivity of the inference framework on feedback (see Appendix~\ref{app:joint}) to cosmic variance effects, which dominate over the specific impact of any variations in the astrophysical parameters. In future work, that we will carry out with larger-volume simulations, we will re-assess the impact of cosmic variance and whether it still dominates over feedback.

In this work, we perform an implicit likelihood analysis (see Section~\ref{sec:sbi}); therefore, the effect of cosmic variance needs to be modeled directly at the level of individual $P_{\rm 1D}(k)$ predictions. In this sense, the \texttt{CAMELS} simulations from the LH sample were designed to satisfy this requirement, as they were run with different random seeds. This means that training on the \texttt{CAMELS} sample effectively incorporates the additional cosmic variance error contribution, and therefore the inference framework is robust against this effect.

\subsection{Strengths and limitations of the training dataset}
\label{sec:camels_pros_cons}

The \texttt{CAMELS} dataset offers several advantages over other simulation suites, listed here below:
\begin{itemize}
    \item it comprises an unprecedentedly large number of simulations, representing an ideal training dataset for cosmological SBI;
    \item it jointly varies cosmological and astrophysical parameters with wide priors around the fiducial values, thereby enabling a thorough exploration of the parameter space;
    \item it employs different codes and methods to solve the equations of gravity and hydrodynamics, as well as a variety of subgrid physics models and feedback prescriptions. This allows the practitioners to study the convergence of the predictions depending on the specific setup used in each simulation and the impact of this aspect on SBI. 
\end{itemize}

On the other hand, the \texttt{CAMELS} LH suite has the following limitations, which restricts some of the science goals that can be pursued:
\begin{itemize}
    \item it explores only $2$ cosmological parameters ($\Omega_m$ and $\sigma_8$), although future \texttt{CAMELS} datasets will vary $5$ of them ($\Omega_m$, $\Omega_b$, $\sigma_8$, $h$, $n_s$);
    \item the simulations have fairly low resolution and small volume, due to the need of striking a balance between number of realizations and the computational cost. Future \texttt{CAMELS} simulation suites will produce $V=(50~h^{-1}~{\rm Mpc})^3$ ($2nd$ generation) and $V=(100~h^{-1}~{\rm Mpc})^3$ ($3rd$ generation) volumes, which will allow to include larger cosmological scales, to better probe astrophysical phenomena related to massive objects, such as AGN feedback, and to significantly reduce cosmic variance. The limited resolution---which will be kept fixed in future \texttt{CAMELS} generations---has important implications for our analysis, as simulating the physics of the intergalactic medium at $z>2$ requires increasingly high resolution at increasing redshift to obtain converged results \citep[see e.g.,][]{Bolton2017,Bird2023,Walther2025}. Specifically, the resolution of \texttt{CAMELS} is lower than the one used in the (low-resolution) simulations used to train state-of-the-art emulators \citep{Pederson2021,ChavesMontero2026}. Since in this study we only have available one resolution, we cannot perform convergence tests. According to the findings by \cite{Bolton2017}, resolution correction are still of order $\sim 10\%$ up to $k=3.0~h~{\rm Mpc}^{-1}$ in this $z$ range. These effects are similar in magnitude to those induced by cosmic variance. We heuristically limit our study to $z<3.5$ and to scales $k_{\rm max}<3.0~h~{\rm Mpc}^{-1}$ for the main analysis. By showing that the results are stable across these (conservative) conditions, we make sure that the analysis is robust and the inference could be in principle applied to real data within the same scales tested herein. Nonetheless, to study the constraining power stemming from a larger data vector and having access to wider range of scales, in Appendix~\ref{app:large_kmax} we illustrate the potential gain in constraining power that we get by pushing the analysis to larger $k_{\rm max}$. 
\end{itemize}

With these caveats in mind, we proceed in what follows to describe the methodology underlying the inference framework used herein and the main results of this work.

\section{Simulation-based inference with normalizing flows}
\label{sec:sbi}



The aim of this work is to perform SBI jointly of the cosmological and astrophysical parameters sampled by the \texttt{CAMELS} LH simulations. In particular, given the underlying set of parameters $\vec{x}$ and some data $\vec{\theta}$, we aim to estimate the posterior distributions $p(\vec{x}|\vec{\theta})$. As outlined previously in the paper, the parameters that we try to constrain are $\{\Omega_m, \sigma_8, A_{\rm SN1}, A_{\rm AGN1}, A_{\rm SN2}, A_{\rm AGN2}\}$, while the data $\vec{\theta}$ consist of the Ly$\alpha$ forest 1D power spectrum at different redshifts from \texttt{CAMELS}.     

To estimate the posterior distributions, we follow a SBI approach. Briefly, SBI consists of an alternative to traditional likelihood-based analysis, which uses the neural density estimation technique \citep[e.g.,][]{Paramakarios2019} to approximate complex conditional probability distributions and make the inference tractable without the need of an explicit analytical form for the likelihood. Specifically, we adopt a neural posterior estimation (hereafter NPE) approach to directly emulate the posterior distributions, which has the advantage of allowing for a quick sampling directly from the posteriors, at the cost of requiring an analytical knowledge of the priors  of the parameters $\vec{x}$. In our case, this is given directly from the \texttt{CAMELS} setup, which adopts uniform priors for all the parameters, as presented in Section~\ref{sec:camels}. 

\begin{figure*}
    \centering
    \includegraphics[width=\linewidth]{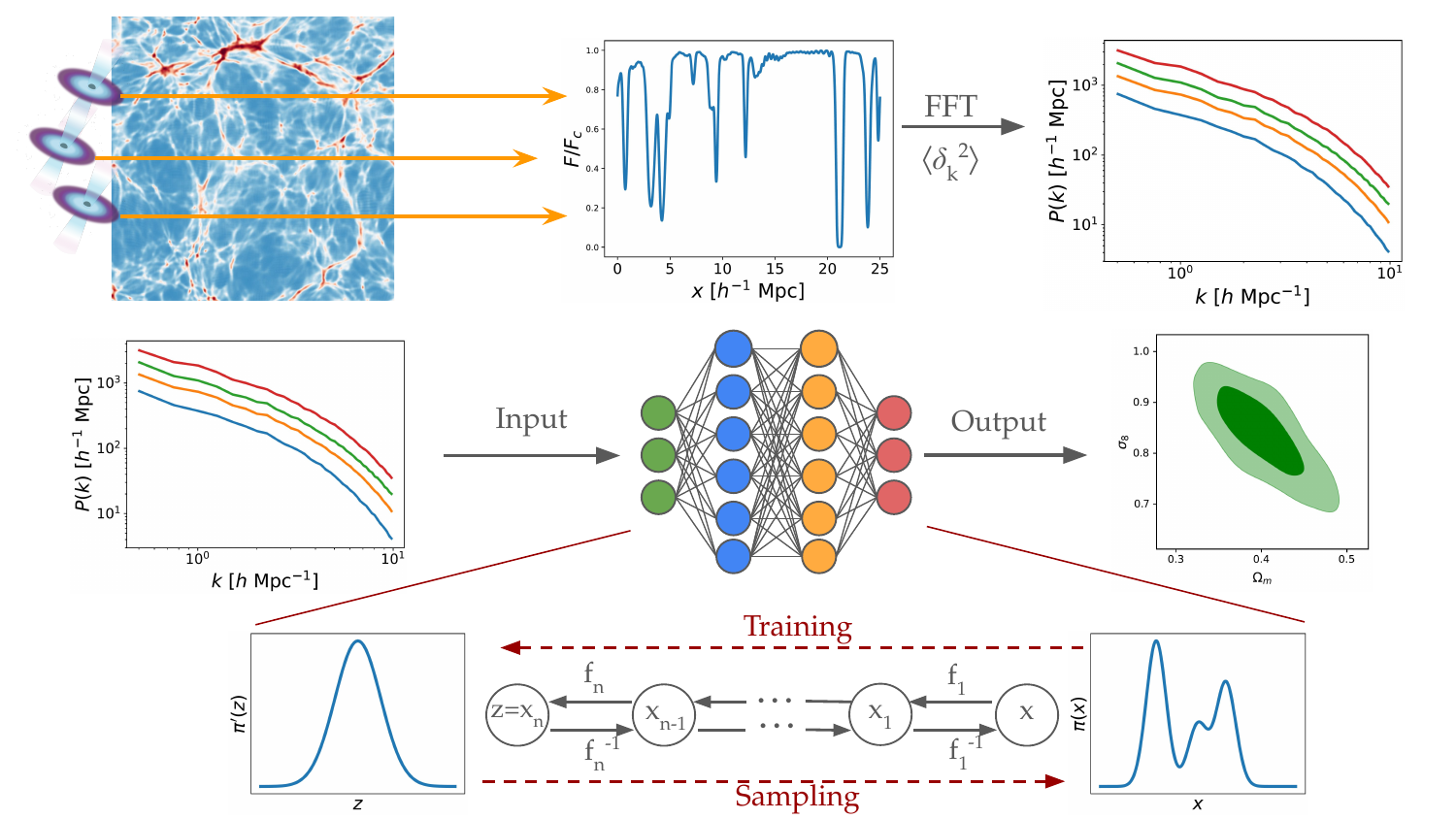}
    \caption{Graphical representation of the full workflow underlying this work. Top: extraction of the Ly$\alpha$ forest and computation of the $P_{\rm 1D}(k)$. Mid: input/output setup of the neural network. Bottom: schematic representation of the normalizing flow.}
    \label{fig:workflow}
\end{figure*}

To perform NPE, we adopt a normalizing flow model \cite{Tabak2010,Tabak2013,JimenesRezende2015}. Normalizing flows employ neural networks to learn a series of bijective transformations $f: x \rightarrow z$ defined to have a tractable Jacobian, to pass from complex distributions $\pi(x)$ to some distributions $\pi^\prime(z)$ which are easy to evaluate. Once the mapping has been learnt, one samples from the distribution $\pi^\prime(z)$ and obtains the target distribution $\pi(x)$ by inverting the mapping $f$ defined above. Specifically, we adopt Masked Autoregressive Flows (hereafter MAF) \cite{Papamakarios2017}, consisting of a stack of autoregressive models \cite{Uria2016}. In practice, we rely on the MAF implementation from the \texttt{sbi} package \cite{TejeroCantero2020,BoeltsDeistler_2025} through the \texttt{LtU-ILI} interface \cite{Ho2024}. In all the cases, we use as fiducial setup MAF with $10$ transforms, $100$ hidden units, and a batch size $n_b=64$. We adopt the \texttt{Adam} optimizer \citep{Kingma2015} with a learning rate $\eta=5\times 10^{-4}$. Out of the full set of $1,000$ simulations, we always use $950$ as training set, and the remaining $50$ for testing, unless stated otherwise. The training set is split into $90\%$/$10\%$ training/cross-validation. To prevent overfitting, the training is stopped if no improvement is found after 20 consecutive epochs. We have explicitly tested other network configurations with more transforms and/or hidden units, and found no improvement in the results. Per each simulation, we feed as input a one-dimensional data vector obtained by stacking the $P_{\rm 1D}(k)$ measurements at the $8$ different available redshifts, and as output the true model parameters. In this sense, we exploit the information from the full redshift range considered in this work.  While \texttt{LtU-ILI} is in principle able to cope with missing or corrupted data (such as NaNs) during the training, extra care must be paid not to incur in such a situation during testing. To prevent it from happening in either case, when one $P_{\rm 1D}(k)$ contains NaNs (which happens in $\ll 1\%$ of the cases, due to the presence of NaNs in the spectra from the corresponding \texttt{CAMELS} realization), we replace the corresponding measurements of the power spectrum via spline interpolation from the $P_{\rm 1D}(k)$ of the same realization but at different redshifts. All the runs were performed using CPUs on an ordinary MacBook Pro with a 10-core 16GB M2 processor. The training took $<5$ minutes. The testing was performed by drawing $20,000$ posterior samples per each parameter and discarding the first $1,000$ to avoid the burn-in phase and took $<1$ minute per simulation, amounting to $\sim 0.5$ hours for the full sample. Figure~\ref{fig:workflow} provides a graphical representation of the full workflow: the extraction of the Ly$\alpha$ forest and the computation of the $P_{\rm 1D}(k)$ in the top row, input/output setup of the neural network in the mid row, and a schematic representation of the normalizing flow in the bottom row.

\begin{figure*}
    \centering
    \includegraphics[width=0.45\linewidth]{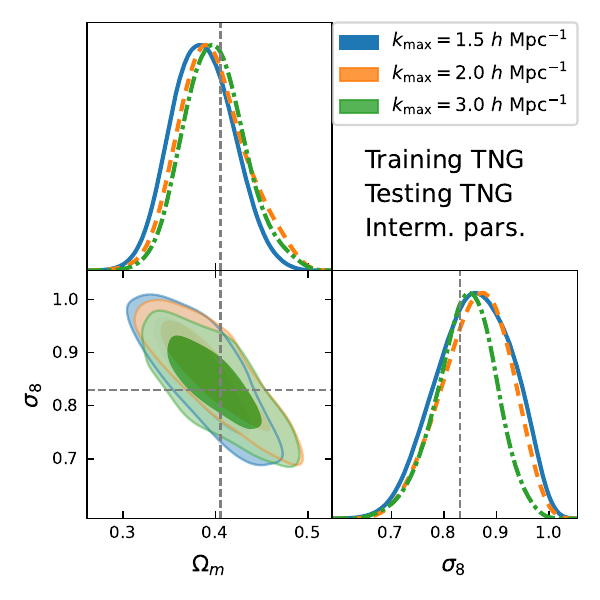}
    \includegraphics[width=0.45\linewidth]{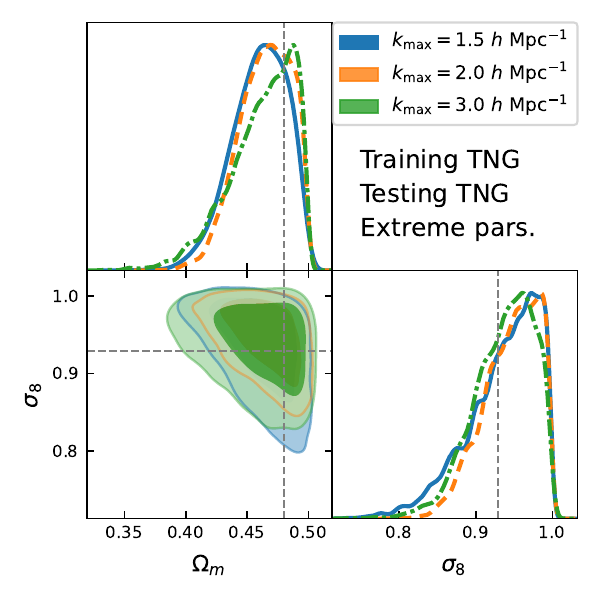}

    \caption{Posterior distributions of the $\Omega_m$ and $\sigma_8$ parameters for different test realizations. Training and testing were performed using simulations run with the \texttt{IllustrisTNG} model. The posteriors shown on the left correspond to simulations having underlying parameters with intermediate values, while the ones shown in the right column correspond to extreme cases, at the edge of the priors. The neural networks were trained setting $k_{\rm max}=1.5~h~{\rm Mpc}^{-1}$ (blue), $k_{\rm max}=2.0~h~{\rm Mpc}^{-1}$ (orange), and  $k_{\rm max}=3.0~h~{\rm Mpc}^{-1}$ (green). The gray dashed lines indicate the true parameters underlying the considered realization.}
    \label{fig:posteriors_om_s8_tngtng}
\end{figure*}

\section{Results and discussion}
\label{sec:results}

In this section, we report the main results of this work and present a discussion of them. The baseline analysis that we perform is to perform training and testing in three different scale cuts and evaluate the performance by looking at the accuracy and precision in the reproduction of the test set, as well as diagnose the resulting posterior distribution through a direct visual inspection, the univariate posterior coverage per each parameter \citep[e.g.,][]{Miller2021,Deistler2022,Hermans2022}, and the combined multivariate `Test of Accuracy with Random Points' \citep[hereafter TARP,][]{Lemos2023}.

\begin{figure*}
    \centering
    \includegraphics[width=0.75\linewidth]{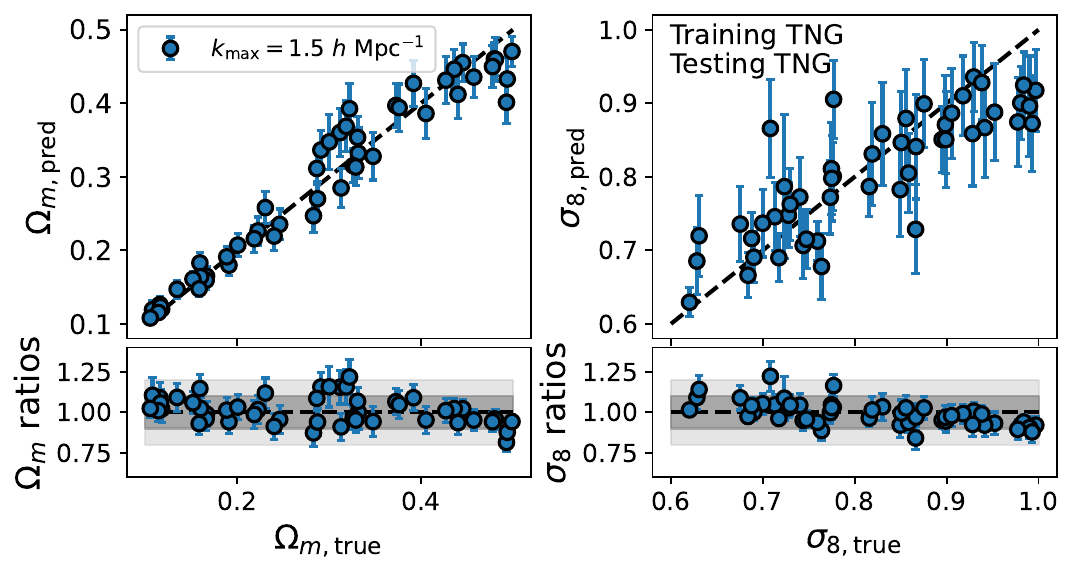}
    \includegraphics[width=0.75\linewidth]{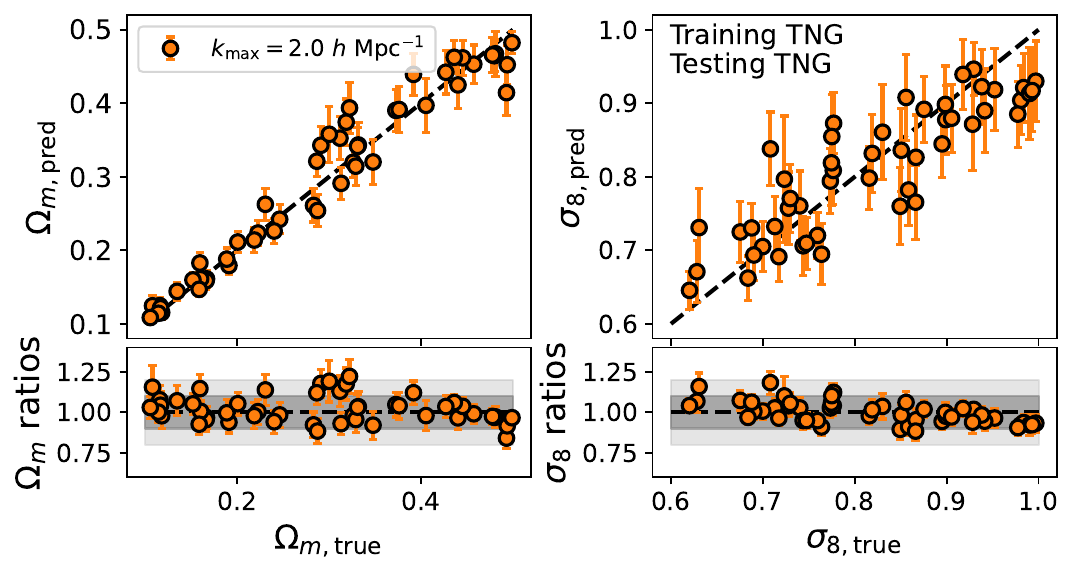}
    \includegraphics[width=0.75\linewidth]{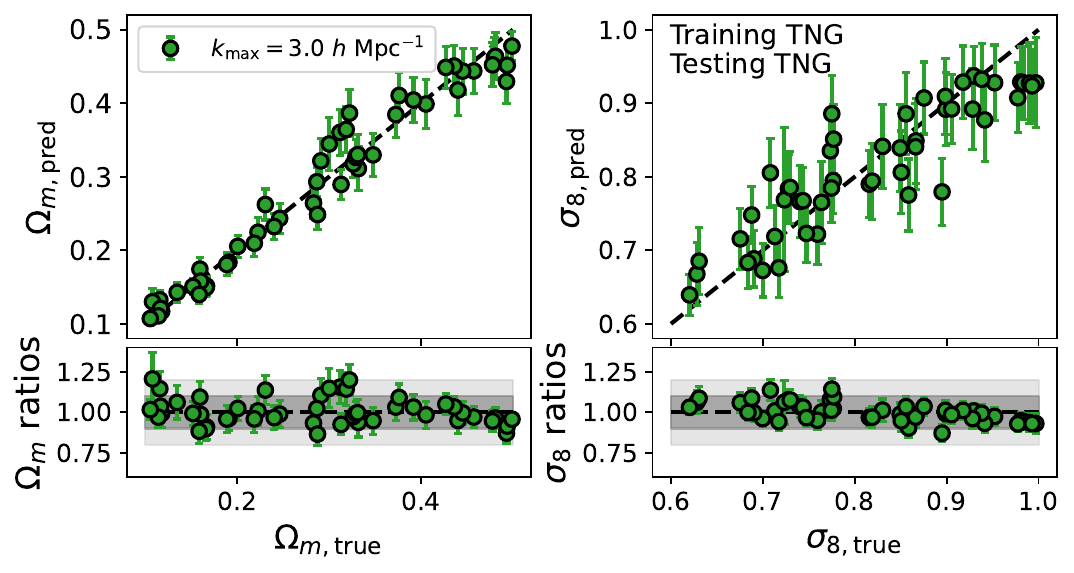}
    \caption{Predicted values for $\Omega_m$ (left column) and $\sigma_8$ (right column) against the true parameters values, for $k_{\rm max}=1.5~h~{\rm Mpc}^{-1}$ (blue), $k_{\rm max}=2.0~h~{\rm Mpc}^{-1}$ (orange), and $k_{\rm max}=3.0~h~{\rm Mpc}^{-1}$ (green). Both training and testing were performed on simulations from the \texttt{IllustrisTNG} suite. The top panels show the 1:1 comparison. The bottom panels display the ratios between the predicted and the true values, and the gray shaded regions stand for $10\%$ (darker) and $20\%$ (lighter) deviations.}
    \label{fig:residuals_allk_tngtng}
\end{figure*}

\begin{figure*}
    \centering
    \includegraphics[width=0.75\linewidth]{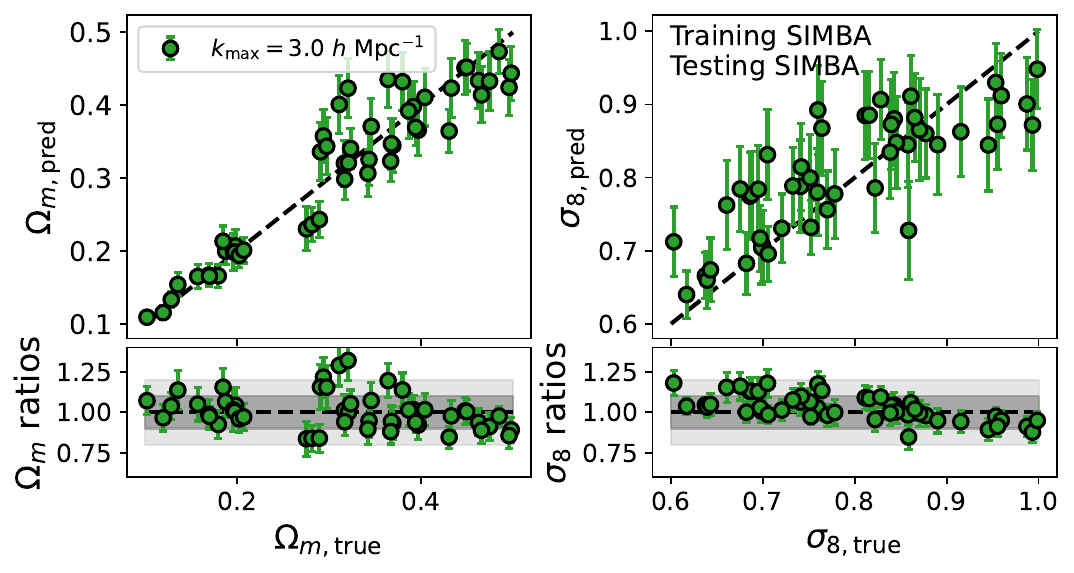}
    \includegraphics[width=0.75\linewidth]{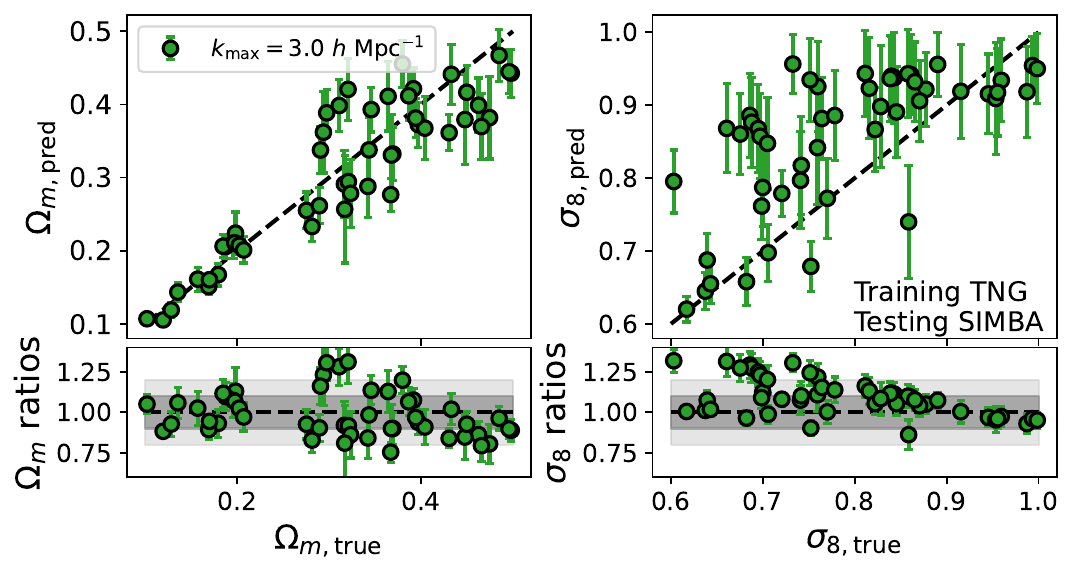}
    \includegraphics[width=0.75\linewidth]{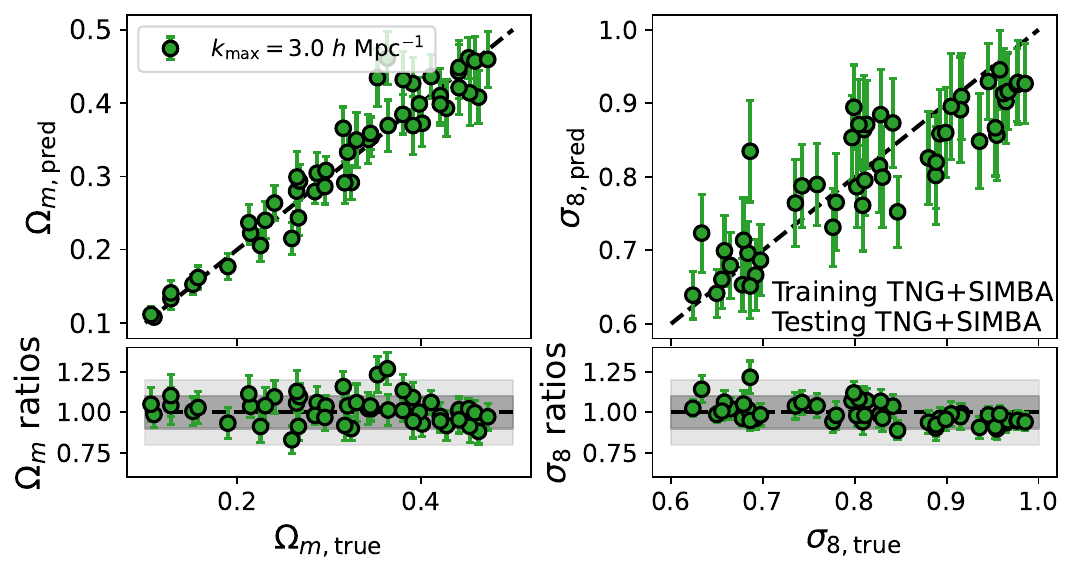}
    \caption{Predicted values for $\Omega_m$ (left column) and $\sigma_8$ (right column) against the true parameters values for $k_{\rm max}=3.0~h~{\rm Mpc}^{-1}$ and for the following case: training and testing performed on \texttt{SIMBA} (top), training performed on \texttt{IllustrisTNG} and testing on \texttt{SIMBA} (mid), training and testing performed on the combination of the two models (bottom). The top panels show the 1:1 comparison. The bottom panels display the ratios between the predicted and the true values, and the gray shaded regions stand for $10\%$ (darker) and $20\%$ (lighter) deviations.}
    \label{fig:residuals_allmodels}
\end{figure*}

\begin{figure*}
    \centering
    \includegraphics[width=0.85\linewidth]{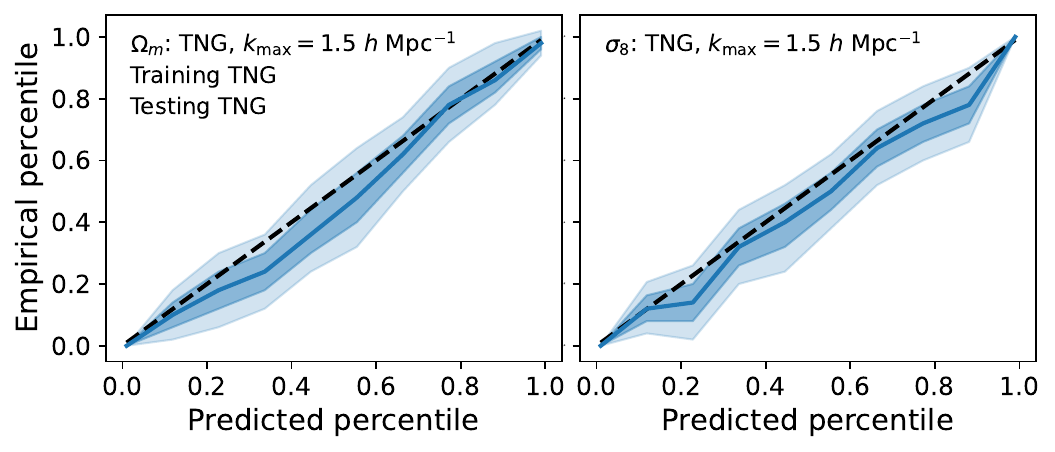}
    \includegraphics[width=0.85\linewidth]{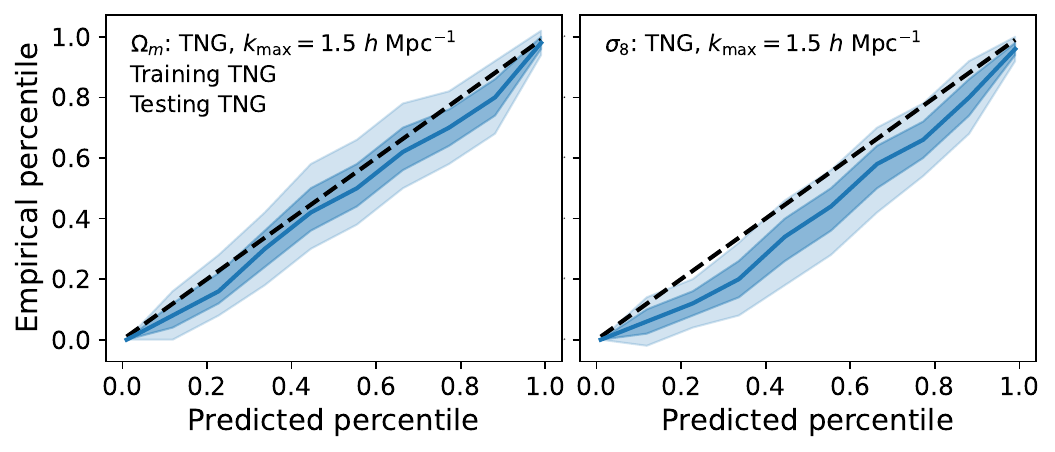}
    \includegraphics[width=0.85\linewidth]{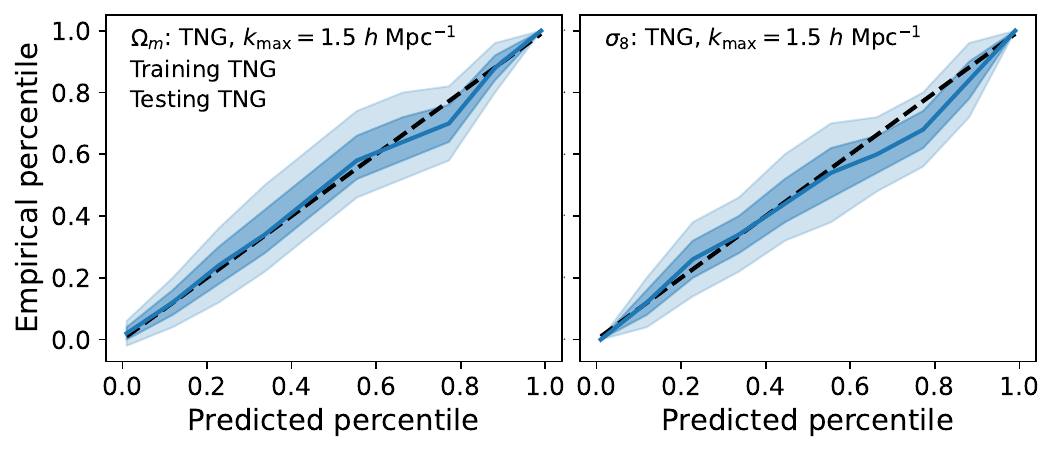}
    \caption{Univariate posterior coverage tests, showing the empirical versus predicted percentiles for the $\Omega_m$ (left) and $\sigma_8$ (right) posteriors, in the case in which both training and testing were performed on \texttt{IllustrisTNG} and with the following scale cuts: $k_{\rm max}=1.5~h~{\rm Mpc}^{-1}$ (top), $k_{\rm max}=2.0~h~{\rm Mpc}^{-1}$ (mid), and $k_{\rm max}=3.0~h~{\rm Mpc}^{-1}$ (bottom). The blue shaded regions represent the $68\%$ (darker) and $95\%$ (lighter) confidence levels.}
    \label{fig:posterior_coverage_tngtng}
\end{figure*}

\begin{figure*}
    \centering
    \includegraphics[width=0.85\linewidth]{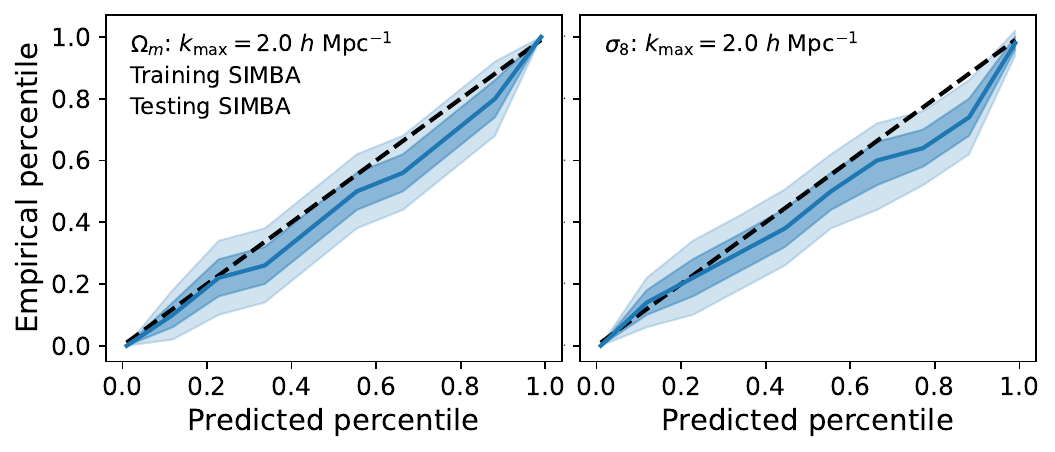}
    \includegraphics[width=0.85\linewidth]{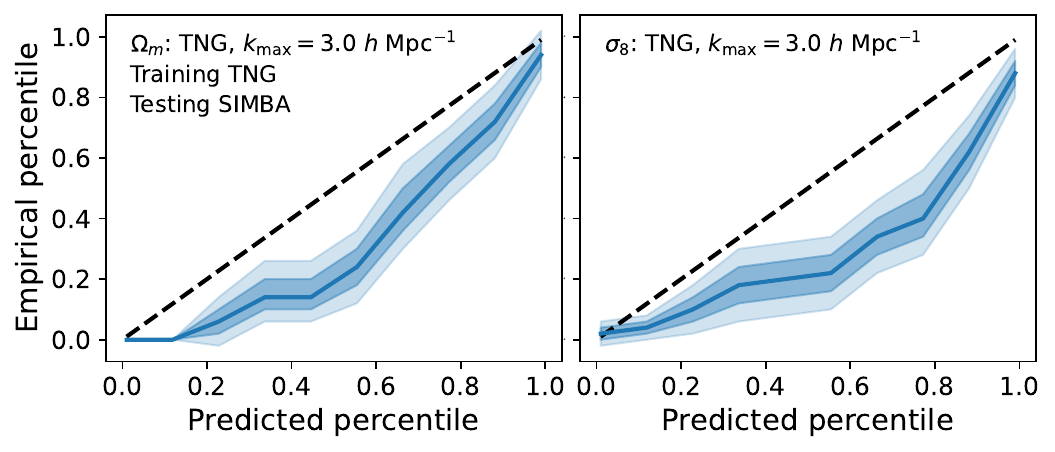}
    \includegraphics[width=0.85\linewidth]{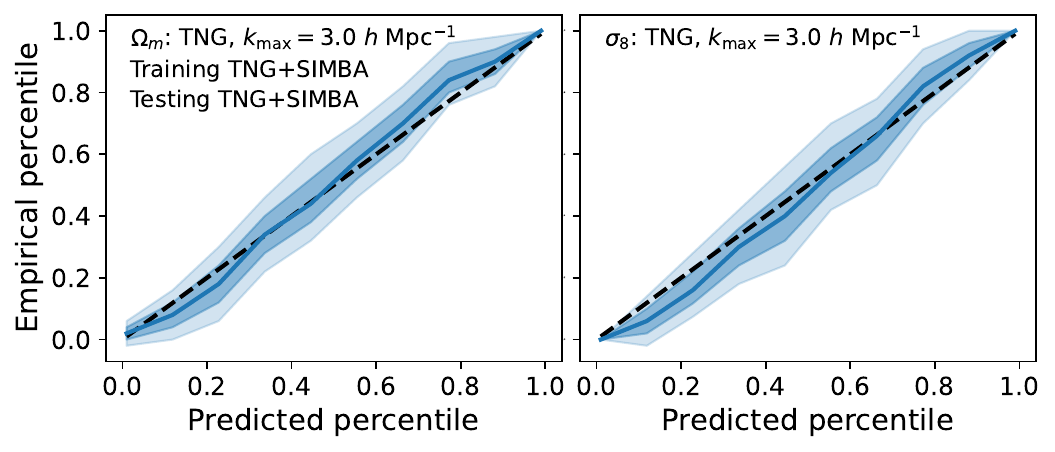}
    \caption{Univariate posterior coverage tests, showing the empirical versus predicted percentiles for the $\Omega_m$ (left) and $\sigma_8$ (right) posteriors, for $k_{\rm max}=3.0~h~{\rm Mpc}^{-1}$ and for the following case: training and testing performed on \texttt{SIMBA} (top), training performed on \texttt{IllustrisTNG} and testing on \texttt{SIMBA} (mid), training and testing performed on the combination of the two models (bottom). The blue shaded regions represent the $68\%$ (darker) and $95\%$ (lighter) confidence levels.}
    \label{fig:posterior_coverage_allmodels}
\end{figure*}

\begin{figure*}
    \centering
    \vspace{-0.5cm}
    \includegraphics[width=0.76\linewidth]{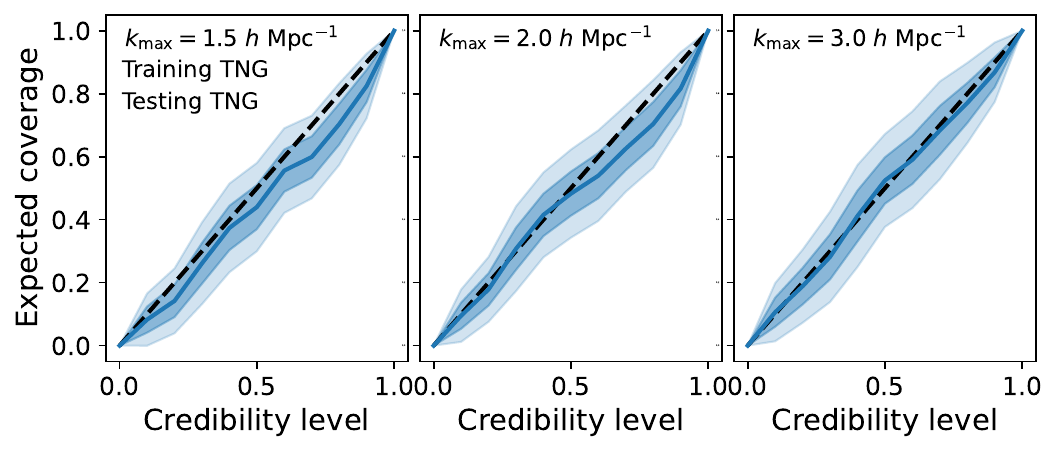}
    \includegraphics[width=0.76\linewidth]{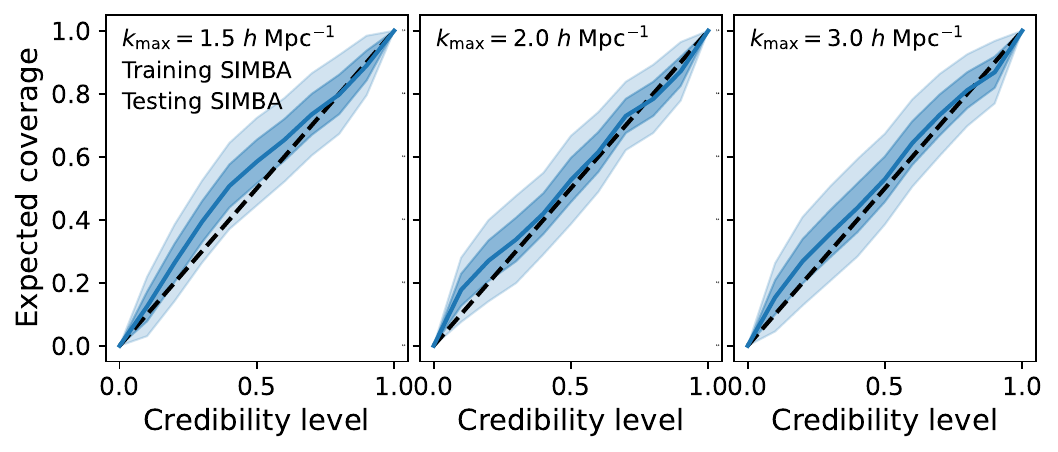}
    \includegraphics[width=0.76\linewidth]{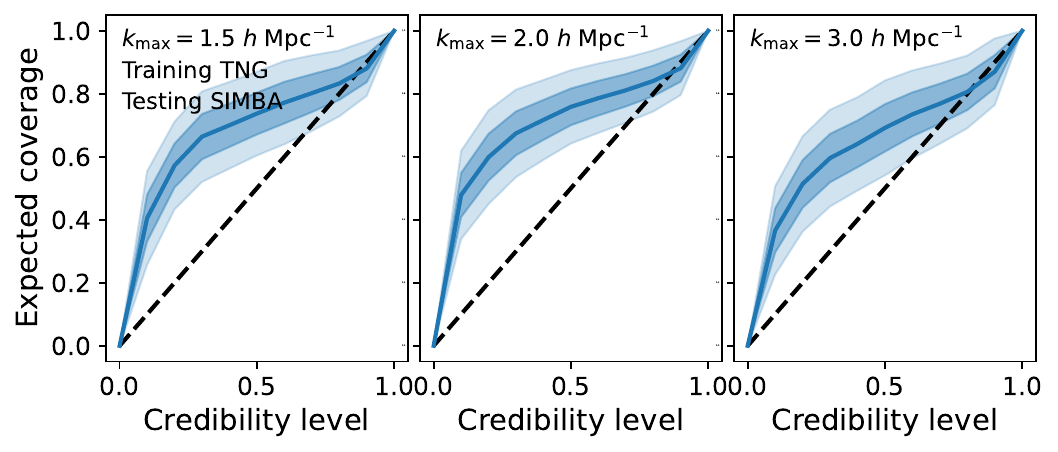}
    \includegraphics[width=0.76\linewidth]{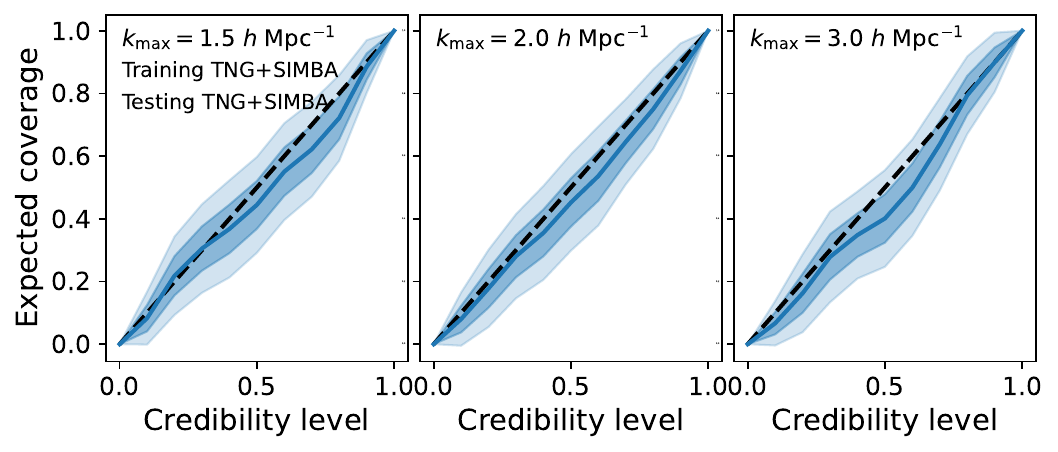}
    \caption{Expected coverage versus credibility level results from TARP, for $k_{\rm max}=1.5~h~{\rm Mpc}^{-1}$ (left), $k_{\rm max}=2.0~h~{\rm Mpc}^{-1}$ (center), and $k_{\rm max}=3.0~h~{\rm Mpc}^{-1}$ (right) and the following cases: training and testing performed on \texttt{IllustrisTNG} (first row), training and testing performed on \texttt{SIMBA} (second row), training performed on \texttt{IllustrisTNG} and testing on \texttt{SIMBA} (third row), training and testing performed on the combination of the two models (fourth row). The blue shaded regions represent the $68\%$ (darker) and $95\%$ (lighter) confidence levels.}
    \label{fig:tarp}
\end{figure*}

\subsection{Training and testing on the same galaxy formation model}
\label{sec:results_same}

We start by considering the case in which we train the model on just one dataset---either \texttt{IllustrisTNG} or \texttt{SIMBA}---and we test the results self-consistently on realizations from the same dataset. We first perform a set of runs with both models and the three scale cuts reported above, performing inference of all the six cosmological and astrophysical parameters. We find that, regardless of the model and of the scale cuts that we choose, our inference framework picks clear information on the cosmological parameters, but the astrophysical ones converge to the prior mean during the chains. To stress test this hypothesis, we push the scale cuts to higher values ($k_{\rm max}=5.0~h~{\rm Mpc}^{-1}$ and $k_{\rm max}=9.0~h~{\rm Mpc}^{-1})$, which even being not being robustly converged in terms of predictions, consist of configurations that should be more sensitive to the underlying astrophysics. As will be shown in Appendix~\ref{app:large_kmax}, going to these large $k_{\rm max}$ results in a significant gain in the cosmological information. However, as shown in Appendix~\ref{app:joint}, the astrophysical parameters converge to the prior mean even for $k_{\rm max}=9.0~h~{\rm Mpc}^{-1}$ \citep[see also e.g.,][for other examples of SBI studies in which the astrophysical parameters show the same behavior]{VillaescusaNavaroo2022,Lovell2025}. 
This can have different explanations. First, it could be due to the intrinsic limited sensitivity of the Ly$\alpha$ fores to the variations of these specific parameters. In fact, on the one hand the cosmological volume embedded in the \texttt{CAMELS} simulations is quite small and hence it contains a small number of SMBH driving AGN feedback. Therefore, despite the fact that AGNs can affect large regions with thermal feedback large regions, the impact of AGN feedback may be negligible. On the other hand, supernovae are more ubiquitous and drive galactic winds, but the volume filling factor of these regions is still small \citep[see also e.g.,][]{Viel2013b}. Second, as already discussed in Section~\ref{sec:camels_cv}, the effect of cosmic variance could be dominating over the one of feedback.

Since there is no gain of information in explicitly using all the six parameters in training and testing phase, in the remainder of the paper we limit our inference to just $\Omega_m$ and $\sigma_8$. 

We then run the training and the testing procedures for both galaxy formation models and for all the $k$ cuts described above. Figure~\ref{fig:posteriors_om_s8_tngtng} shows the posterior distributions of the $\Omega_m$ and $\sigma_8$ parameters for two different test realizations, in the case in which both training and testing were performed on \texttt{IllustrisTNG}. The posteriors shown in the left column correspond to simulations having underlying parameters with intermediate values, while the ones shown in the right column correspond to extreme cases with values for both $\Omega_m$ and $\sigma_8$ at the edge of the priors. We display in each panel the posteriors for the three baseline scale cuts, in blue ($k_{\rm max}=1.5~h~{\rm Mpc}^{-1}$), orange ($k_{\rm max}=2.0~h~{\rm Mpc}^{-1}$), and green ($k_{\rm max}=3.0~h~{\rm Mpc}^{-1}$). A visual inspection reveals that the resulting constraints are unbiased and stable across the different scale cuts. 
Figure~\ref{fig:residuals_allk_tngtng} shows the predicted parameter values against the true ones in the top subpanel and the ratio of the two in the bottom subpanel, for the three different scale cuts and in the case in which both training and testing were performed on the \texttt{IllustrisTNG} simulations. We report the quantitative results in terms of percentages of data points with residuals $\rho$ within $5\%, 10\%$, and $20\%$, the average residuals $\hat{\rho}$, the median precision $\hat{\sigma}$, the Pearson correlation coefficient $R$ and the $R^2$ score in Table~\ref{tab:results}. We notice that the residuals and precision results are consistently accurate for all the three scale cuts (first, second, and third lines of each subtable) in this case. 
The correlation coefficients $R$ and $R^2$ for $\Omega_m$ are stable across the different scale cuts, but they increase for $\sigma_8$ for increasing $k_{\rm max}$. The top panel of Figure~\ref{fig:residuals_allmodels} shows the same results for $k_{\rm max}=3.0~h~{\rm Mpc}^{-1}$ and for the \texttt{SIMBA} model. 
Therefore, the results for \texttt{SIMBA} (reported in the fourth line of Table~\ref{tab:results}) are somewhat slightly worse than those for \texttt{IllustrisTNG}, potentially reflecting a stronger dependence on astrophysics as shown in Figures~\ref{fig:pk_comparison},~\ref{fig:pk_feedback},~\ref{fig:cosmic_variance}, and therefore a larger uncertainty on the cosmological parameters, but still quite accurate overall.
Figure~\ref{fig:posterior_coverage_tngtng} shows the posterior coverage for $\Omega_m$ (left) and $\sigma_8$ (right), in the $k_{\rm max}=1.5~h~{\rm Mpc}^{-1}$ (top), $k_{\rm max}=2.0~h~{\rm Mpc}^{-1}$ (mid), and $k_{\rm max}=3.0~h~{\rm Mpc}^{-1}$ (bottom). The resulting curves denote that both the univariate posteriors are well-calibrated, being slightly overconfident (below the diagonal) for the smaller scale cuts, but becoming unbiased at $k_{\rm max}=3.0~h~{\rm Mpc}^{-1}$. The top panel of Figure~\ref{fig:posterior_coverage_allmodels} shows the posterior coverage for the \texttt{SIMBA} model and $k_{\rm max}=3.0~h~{\rm Mpc}^{-1}$, also reflecting a good calibration of the posteriors. The same results are confirmed also by the TARP results, shown in Figure~\ref{fig:tarp}, where both \texttt{IllustrisTNG} (first row) and \texttt{SIMBA} (second row) are found to have unbiased posteriors at the various probed scale cuts (shown in different columns).

\begin{table*}[]
    \centering
    \begin{tabular}{lcccccccc}
    \toprule
         $\mathbf{\Omega_m}$ &  Sect. &  $\rho<5\%~[\%]$ & $\rho<10\%~[\%]$ & $\rho<20\%~[\%]$ & $\hat{\rho}~[\%]$ & $\hat{\sigma}~[\%]$ & $R$ & $R^2$ \\
         \midrule
         TNG-TNG $k_{\rm max}=1.0~h~{\rm Mpc}^{-1}$ & \ref{sec:results_same} & $42$ & $80$ & $98$ & $1.1\pm 8.3$ & $7.8\pm 1.3$ & $0.97$ & $0.95$ \\
         TNG-TNG $k_{\rm max}=2.0~h~{\rm Mpc}^{-1}$ & \ref{sec:results_same} & $54$ & $76$ & $98$ & $1.1\pm 8.3$ & $7.5\pm 1.5$ & $0.97$ & $0.95$ \\
         TNG-TNG $k_{\rm max}=3.0~h~{\rm Mpc}^{-1}$ & \ref{sec:results_same} & $54$ & $76$ & $100$ & $0.3\pm 8.0$ & $8.1\pm 1.5$ & $0.97$ & $0.96$ \\
         SIMBA-SIMBA $k_{\rm max}=3.0~h~{\rm Mpc}^{-1}$ & \ref{sec:results_same} & $36$ & $64$ & $94$ & $0.5\pm 11.2$ & $9.3\pm 1.4$ & $0.94$ & $0.89$ \\
         TNG-SIMBA $k_{\rm max}=3.0~h~{\rm Mpc}^{-1}$ & \ref{sec:results_diff} & $14$ & $48$ & $88$ & $1.4\pm 13.7$ & $8.9\pm 4.1$ & $0.91$ & $0.82$ \\
         Both-both $k_{\rm max}=3.0~h~{\rm Mpc}^{-1}$ & \ref{sec:results_mix} &  $48$ & $80$ & $96$ & $1.7\pm 8.5$ & $8.9\pm1.5$ & $0.96$ & $0.92$ \\
         Both-both $k_{\rm max}=3.0~h~{\rm Mpc}^{-1}$ (no resc.) & \ref{app:mix_noresc} & $34$ & $52$ & $74$ & $2.8\pm 14.8$ & $15.3\pm 5.0$ & $0.90$ & $0.79$ \\
         TNG-TNG $k_{\rm max}=3.0~h~{\rm Mpc}^{-1}$ MCMC calib. & \ref{sec:results_mcmc} & $48$ & $80$ & $100$ & $1.4\pm 7.9$ & $9.1\pm2.2$ & $0.98$ & $0.96$ \\
         TNG-TNG $k_{\rm max}=3.0~h~{\rm Mpc}^{-1}$ ${\rm SNR}=1$ & \ref{sec:results_noise} & $54$ & $74$ & $100$ & $1.1\pm 7.9$ & $7.4\pm 1.3$ & $0.98$ & $0.97$ \\
         TNG-TNG $k_{\rm max}=3.0~h~{\rm Mpc}^{-1}$ ${\rm SNR}=3$ & \ref{sec:results_noise} & $52$ & $78$ & $100$ & $1.7\pm 7.9$ & $7.5\pm 1.5$ & $0.98$ & $0.97$ \\
         TNG-TNG $k_{\rm max}=3.0~h~{\rm Mpc}^{-1}$ ${\rm SNR}=5$ & \ref{sec:results_noise} & $52$ & $76$ & $100$ & $1.3\pm 8.0$ & $7.3\pm 1.4$ & $0.98$ & $0.96$ \\
         TNG-TNG $k_{\rm max}=3.0~h~{\rm Mpc}^{-1}$ $\tau$ &  \ref{app:tau} & $66$ & $94$ & $100$ & $0.7\pm 4.9$ & $5.9\pm 1.2$ & $0.99$ & $0.99$ \\
         \midrule\midrule
         $\mathbf{\sigma_8}$& Sect. &  $\rho<5\%~[\%]$ & $\rho<10\%~[\%]$ & $\rho<20\%~[\%]$ & $\hat{\rho}~[\%]$ & $\hat{\sigma}~[\%]$ & $R$ & $R^2$ \\
         \midrule
         TNG-TNG $k_{\rm max}=1.5~h~{\rm Mpc}^{-1}$ & \ref{sec:results_same} & $56$ & $86$ & $98$ & $0.7\pm 7.3$ & $6.6\pm1.5$ & $0.84$ & $0.69$ \\
         TNG-TNG $k_{\rm max}=2.0~h~{\rm Mpc}^{-1}$ & \ref{sec:results_same} & $46$ & $86$ & $100$ & $0.1\pm 6.7$ & $5.8\pm1.1$ & $0.87$ & $0.75$ \\
         TNG-TNG $k_{\rm max}=3.0~h~{\rm Mpc}^{-1}$ & \ref{sec:results_same} & $60$ & $96$ & $100$ & $0.3\pm 5.7$ & $5.8\pm 1.2$ & $0.90$ & $0.82$ \\
         SIMBA-SIMBA $k_{\rm max}=3.0~h~{\rm Mpc}^{-1}$ & \ref{sec:results_same} & $50$ & $76$ & $100$ & $2.8\pm 8.0$ & $7.1\pm 0.9$ & $0.82$ & $0.63$ \\
         TNG-SIMBA $k_{\rm max}=3.0~h~{\rm Mpc}^{-1}$ & \ref{sec:results_diff} & $28$ & $56$ & $78$ & $9.2\pm 11.4$ & $6.4\pm 1.4$ & $0.68$ & $0.01$ \\
         Both-both $k_{\rm max}=3.0~h~{\rm Mpc}^{-1}$ & \ref{sec:results_mix} & $46$ & $90$ & $98$ & $0.6\pm 6.6$ & $7.0\pm 0.9$ & $0.88$ & $0.76$ \\
         Both-both $k_{\rm max}=3.0~h~{\rm Mpc}^{-1}$ (no resc.) & \ref{app:mix_noresc} & $40$ & $76$ & $100$ & $4.2\pm 7.6$ & $7.8\pm 1.4$ & $0.81$ & $0.57$ \\
         TNG-TNG $k_{\rm max}=3.0~h~{\rm Mpc}^{-1}$ MCMC calib. & \ref{sec:results_mcmc} & $44$ & $80$ & $100$ & $1.8\pm 7.0$ & $8.1\pm 1.9$ & $0.85$ & $0.70$ \\
         TNG-TNG $k_{\rm max}=3.0~h~{\rm Mpc}^{-1}$ ${\rm SNR}=1$ & \ref{sec:results_noise} & $62$ & $90$ & $100$ & $0.3\pm 5.7$ & $5.3\pm 0.9$ & $0.90$ & $0.81$ \\
         TNG-TNG $k_{\rm max}=3.0~h~{\rm Mpc}^{-1}$ ${\rm SNR}=3$ & \ref{sec:results_noise} & $56$ & $92$ & $100$ & $0.1\pm 6.0$ & $5.5\pm 1.0$ & $0.90$ & $0.79$ \\
         TNG-TNG $k_{\rm max}=3.0~h~{\rm Mpc}^{-1}$ ${\rm SNR}=5$ & \ref{sec:results_noise} & $60$ & $92$ & $100$ & $0.1\pm 6.0$ & $5.4\pm 0.8$ & $0.90$ & $0.80$ \\
         TNG-TNG $k_{\rm max}=3.0~h~{\rm Mpc}^{-1}$ $\tau$ & \ref{app:tau} & $78$ & $98$ & $100$ & $0.4\pm 4.3$ & $4.6\pm0.6$ & $0.95$ & $0.89$ \\
         \bottomrule
    \end{tabular}
    \caption{Numerical results underlying the inference tests for $\Omega_m$ (top subtable) and $\sigma_8$ (bottom subtable) discussed in this work. First column: the experiments the results correspond to. Second column: the section in which the tests are described and the results are discussed. Third, fourth, and fifth columns: percentages of data point with residuals $\rho$ within $5\%$, $10\%$, and $20\%$, respectively. Sixth column: median residuals $\hat{\rho}$. Seventh column: $\hat{\sigma}$. Eight and ninth columns: the Pearson correlation coefficients $R$ and the $R^2$ score, respectively.}
    \label{tab:results}
\end{table*}

\subsection{Training on \texttt{IllustrisTNG} and testing on \texttt{SIMBA}}
\label{sec:results_diff}

\begin{figure*}
    \centering
    \includegraphics[width=0.45\linewidth]{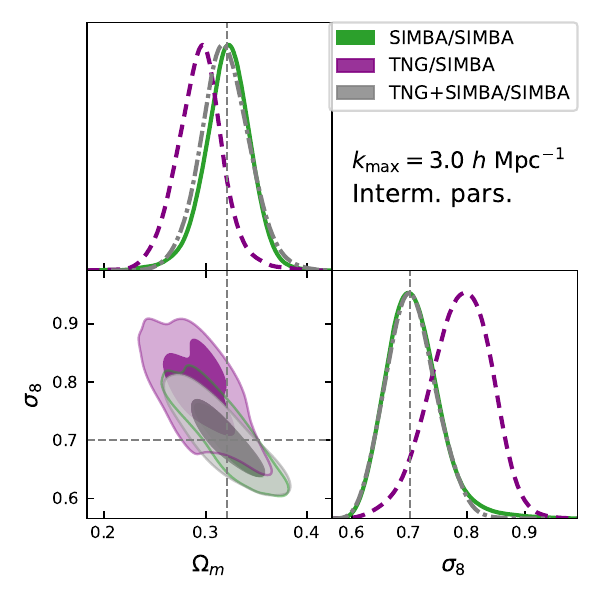}
    \includegraphics[width=0.45\linewidth]{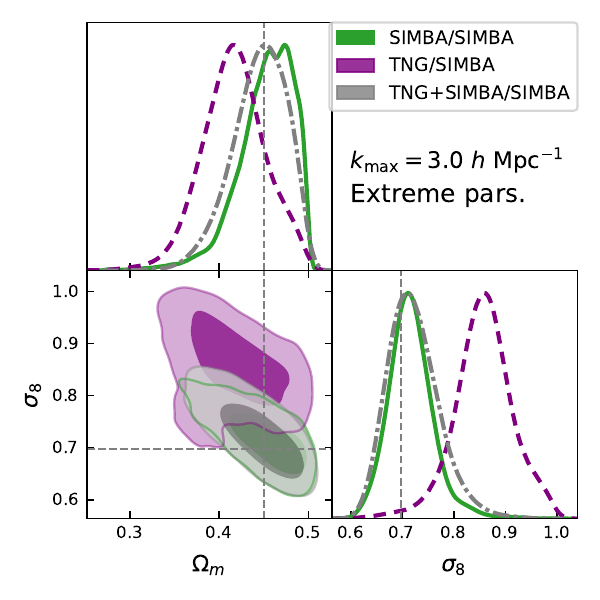}
    \caption{Posterior distributions of the $\Omega_m$ and $\sigma_8$ parameters for different test realizations, and $k_{\rm max}=3.0~h~{\rm Mpc}^{-1}$. Green: training and testing performed on \texttt{SIMBA}. Purple: training on \texttt{IllustrisTNG} and testing performed with on \texttt{SIMBA}. Gray: training on both models and testing on \texttt{SIMBA}. The posteriors shown on the left correspond to simulations having underlying parameters with intermediate values, while the ones shown on the right correspond to extreme cases, at the edge of the priors. The gray dashed lines indicate the true parameters underlying the considered realization.}
    \label{fig:posteriors_om_s8_tngsibamix}
\end{figure*}

We now pass to consider the case in which we perform training on one galaxy formation model and testing on another one. Specifically, here we train the networks on \texttt{IllustrisTNG} and validate them on \texttt{SIMBA}. As mentioned in Section \ref{sec:camels_lya}, when combining the two different galaxy formation models, we need to take into account the fact that predicted mean flux differ by $\sim 5-15\%$ between the two at $z\gtrsim 3$. We described in Appendix~\ref{app:tngsimba_noresc} the attempt that we made without rescaling the mean flux. Briefly, in such a case the inferred parameters values are severely biased, as expected by the fact that the normalization of the $P_{\rm 1D}(k)$ is sensitive both to the mean flux and the cosmological parameters, and an offset in the mean flux is degenerate with an offset in either of the cosmological parameters. Therefore, we proceeded by rescaling the mean flux as described in Section~\ref{sec:camels_lya}. The mid panels of Figure~\ref{fig:residuals_allmodels} shows the inferred $\Omega_m$ (left) and $\sigma_8$ (right) for $k_{\rm max}=3.0~h~{\rm Mpc}^{-1}$. The numerical results are reported in the fifth line of Table~\ref{tab:results}. 
The inferred values for $\Omega_m$ are therefore relatively unbiased, although with larger scatter with respect to the cases described previously. On the contrary, the prediction for $\sigma_8$ are biased towards too large values at the $\sim10\%$ level. In addition, both the posterior coverages (mid row in Figure~\ref{fig:posterior_coverage_allmodels}) and the TARP results (third row in Figure~\ref{fig:tarp}) show that the posteriors are overconfident, i.e. the inferred uncertainties too small compared to the expected ones. Finally, Figure~\ref{fig:posteriors_om_s8_tngsibamix} shows in purple the posteriors for this case, as in Figure~\ref{fig:posteriors_om_s8_tngtng}. If compared to the case in which both the training and the testing were performed with the \texttt{SIMBA} model, it turns out quite clearly that in this case the derived posteriors are either biased and/or provide much looser constraints on the parameter values.

These results represent a typical example of cross-generalization problem that arises when performing training and testing across different datasets, due to the intrinsic lack of convergence in the predictions of different galaxy formation models and codes \citep[e.g.,][]{VillaescusaNavaroo2022,Angeloudi2023,Gluck2024}.

\subsection{Training on both \texttt{IllustrisTNG} and \texttt{SIMBA}}
\label{sec:results_mix}

In the previous section, we showed that performing training and testing on two different galaxy formation models poses important cross-generalization challenges in this inference framework. How can we tackle this problem, in order to ensure the reliability of the results when we will perform inference on real data (not used for training)? Among the possible solutions---referred to as `domain adaptation' \citep[e.g.,][]{Ciprjanovic2021,Ciprjanovic2022,Roncoli2023,Akhmetzhanova2024,Gilda2024,Ntampaka2025}---we adopt here multi-domain training and train our neural networks on the combination of \texttt{IllustrisTNG} and \texttt{SIMBA} simulations, leaving more complex techniques for future studies. The rationale behind this choice is that the neural network should be able to learn shared patterns between different models and marginalize over galaxy formation models' uncertainties. In this way, one would obtain an increasingly more representative sample of Ly$\alpha$ forest spectra the larger the number of models using in the training. 

We therefore train the neural networks on $950$ simulations ($475$ from \texttt{IllustrisTNG}, $475$ from \texttt{SIMBA}, selected with a uniform random sampling), and perform testing on $50$ simulations ($25$ from \texttt{IllustrisTNG}, $25$ from \texttt{SIMBA}). While we could in principle use the full dataset and get a more complete training, we decided to use the same number of simulations to make the comparison with the cases discussed previously easier. We discuss here the case in which we perform the flux rescaling, and discuss the case without rescaling in Appendix~\ref{app:mix_noresc}.

From the bottom panel of Figure~\ref{fig:residuals_allmodels}, one can see that both $\Omega_m$ and $\sigma_8$ are correctly recovered. The numerical results are reported in the sixth line of Table~\ref{tab:results}.
Both the posterior coverages (bottom row of Figure~\ref{fig:posterior_coverage_allmodels}) and the TARP results (fourth row of Figure~\ref{fig:tarp}) show that the posteriors are robustly unbiased. A visual inspection of the posteriors in Figure~\ref{fig:posteriors_om_s8_tngsibamix} (gray versus green) confirms that the multi-domain training recovers unbiased values for the parameters, with comparable levels of scatter in the residuals and of precision in the derived parameter constraints. Therefore, the impact of the multi-domain training is small in the studied case.

We acknowledge that the tests shown in this Section represent a first step towards tackling the cross-generalization issue, but that more work should be done to conclusively addressing it. We defer this to future work. 

\subsection{Comparison with explicit-likelihood inference}
\label{sec:results_mcmc}
 
In this section, we test SBI against a traditional MCMC explicit-likelihood inference. To do so, we maximize a Gaussian likelihood $\mathcal{L}$ of the form: 

\begin{equation}
    \mathcal{L} = \left(P_{1D}(k)-{P}^\dagger_{1D}(k)\right)^T\mathcal{C}^{-1}\left(P_{1D}(k)-{P}^\dagger_{1D}(k)\right) 
\end{equation}

where $\mathcal{C}$ is the covariance matrix and ${P}^\dagger_{1D}(k)$ is the theory prediction, both of which need to be estimated.

To obtain an estimate for the covariance, we cannot rely on the \texttt{CAMELS} LH suite, as it varies the cosmology and has only one realization (i.e., only one random seed) per each cosmology. Therefore we use the \texttt{CAMELS} CV suite, that includes $27$ realizations at fixed cosmology and different random seeds. Due to the low number of simulations available compared to the degrees of freedom, the covariance estimated from the CV suite is initially singular and therefore not invertible. We addressed this issue by applying a regularization based on the Ledoit-Wolf shrinkage technique \citep[][]{LedoitWolf2004}. In addition, the covariance obtained from the CV suite includes the cosmic variance contribution, but does not account for the fact that \texttt{CAMELS} spans a wide priors on the cosmological parameters. In this way, the covariance matrix estimated at the fixed fiducial cosmology may not be accurate when used in a MCMC step proposing a set of cosmological parameters that are significantly different from the fiducial ones. A preliminary MCMC test with the fixed-cosmology produced unrealistically narrow posterior distributions. To overcome this issue, at each likelihood evaluation we rescale the covariance by a factor $r^2$, where $r=P^\dagger_{\rm 1D}(k)/P^\dagger_{{\rm 1D,fid}}(k)$, i.e. the ratio of the $P_{\rm 1D}(k)$ evaluated with the current cosmology and the fiducial-cosmology $P_{\rm 1D}(k)$, averaged over all the $k$ bins. We notice that it would be more accurate rescaling each element $\mathcal{C}_{ij}$ by factor $r^2_{ij}=\left(P^\dagger_{i,{\rm 1D}}(k)/P^\dagger_{i,{\rm 1D, fid}}(k)\right)\left(P^\dagger_{j,{\rm 1D}}(k)/P^\dagger_{j,{\rm 1D,fid}}(k)\right)$, where the indeces $\{i,j\}$ run over the $k$ bins, to consistently capture the scale dependence of the correction. However, the significant cosmic variance contribution in the \texttt{CAMELS} $P_{\rm 1D}(k)$ (see Section~\ref{sec:camels_cv}) introduces a large noise contribution in the correction, therefore we stick here to the simpler scalar correction described above.

To obtain a theoretical prediction for the $P_{\rm 1D}(k)$, instead, we interpolate in between predictions from the available LH simulations. Specifically, we use a radial basis function (RBF) interpolation. In practice, we use the \texttt{scipy.interpolate.RBFInterpolator} routine, with smoothing parameter $\epsilon=0.1$. We have explicitly verified that the specific choice for the smoothing parameter does not affect the results, provided that it stay within a range $\epsilon=10^{-3}-1$. Setting $\epsilon=0$ produces ill-defined posteriors. This interpolation scheme can cope with the high dimensionality of the parameters space ($n=6$, despite we only sample the two cosmological parameters) more easily than other simpler approaches such as e.g. a Delaunay triangulation.

We adopt the affine-invariant MCMC sampler \texttt{emcee} \citep[][]{ForemanMackey2013}, using the same scale cut ($k_{\rm max}=3.0~h~{\rm Mpc}^{-1}$), priors, number of samples, and training/testing splitting as for the baseline SBI analysis. Specifically, we use the $950$ \texttt{CAMELS} realizations used for the SBI training as interpolation nodes, and run explicit-likelihood MCMC on the $50$ testing realizations. We use the full redshift range, stacking the model predictions $P^\dagger_{1D}(k)$ for the $8$ redshift snapshots in one single data vector. The corresponding full covariance matrix is built in a block-diagonal form, with different blocks corresponding to the covariances at distinct redshifts. We assume zero cross-covariance in between different redshift bins. When running the explicit-likelihood inference this way, a set of posterior coverage tests (whose results are shown later in this Section) reveal that the resulting posteriors are severely overconfident, and hence, the uncertainties associated to the estimated parameters underestimated. This issue may have different origins, but we generally ascribe it to the insufficiency of the covariance computed as described above to capture all the sources of uncertainties. To overcome this problem, we remap the covariance $\mathcal{C}$ to by further rescaling it as $\lambda\mathcal{C}$, where $\lambda$ is an additional factor used to enforce the correct posterior calibration. We find $\lambda=2.8$ heuristically as the value that ensures the best posteriors calibration when looking simultaneously at the univariate posterior coverages and at TARP.

\begin{figure*}
    \centering
    \includegraphics[width=0.45\linewidth]{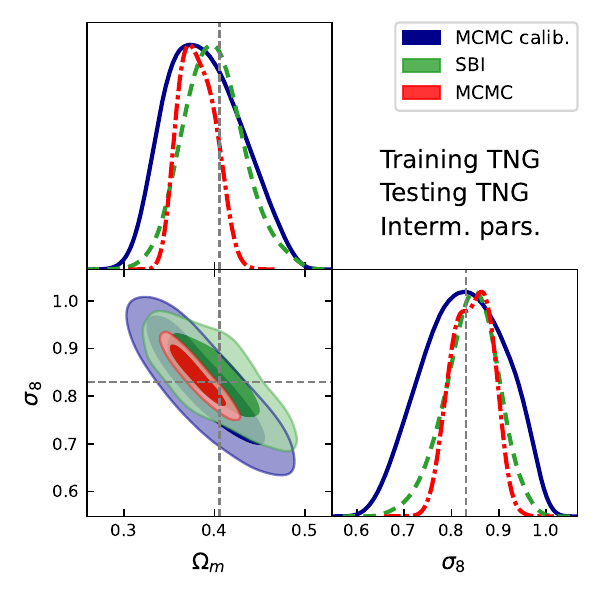}
    \includegraphics[width=0.45\linewidth]{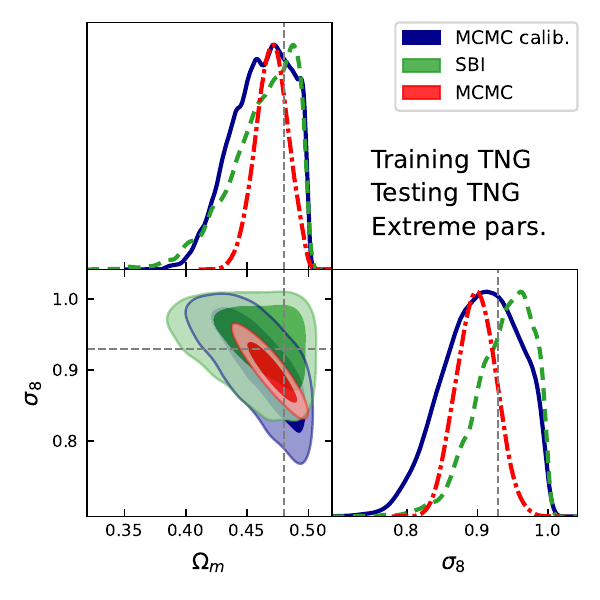}
    \caption{Posterior distributions of the $\Omega_m$ and $\sigma_8$ parameters for different test realizations, $k_{\rm max}=3.0~h~{\rm Mpc}^{-1}$, and obtained through SBI (green), explicit-likelihood MCMC inference (light orange), and MCMC inference with additional posteriors calibration (light blue). Training and testing were performed using simulations run with the \texttt{IllustrisTNG} model. The posteriors shown on the left correspond to simulations having underlying parameters with intermediate values, while the ones shown on the right correspond to extreme cases, at the edge of the priors. The gray dashed lines indicate the true parameters underlying the considered realization.}
    \label{fig:posteriors_om_s8_tngtng_likelihood}
\end{figure*}

\begin{figure*}
    \centering
    \includegraphics[width=0.9\linewidth]{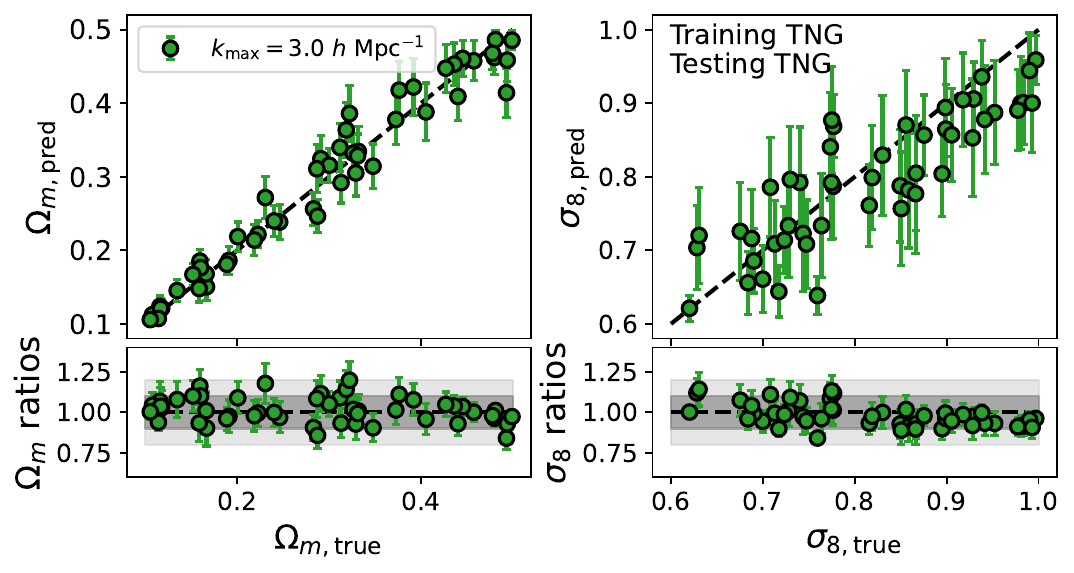 }
    \caption{Predicted values for $\Omega_m$ (left column) and $\sigma_8$ (right column) against the true parameters values, for $k_{\rm max}=3.0~h~{\rm Mpc}^{-1}$. In this case the predicted values were obtained through an explicit-likelihood analysis and not through SBI. Both training and testing were performed on simulations from the \texttt{IllustrisTNG} suite. The top panels show the 1:1 comparison. The bottom panels display the ratios between the predicted and the true values, and the gray shaded regions stand for $10\%$ (darker) and $20\%$ (lighter) deviations.}
    \label{fig:residuals_tngntg_likelihood}
\end{figure*}

\begin{figure*}
    \centering
    \includegraphics[width=0.8\linewidth]{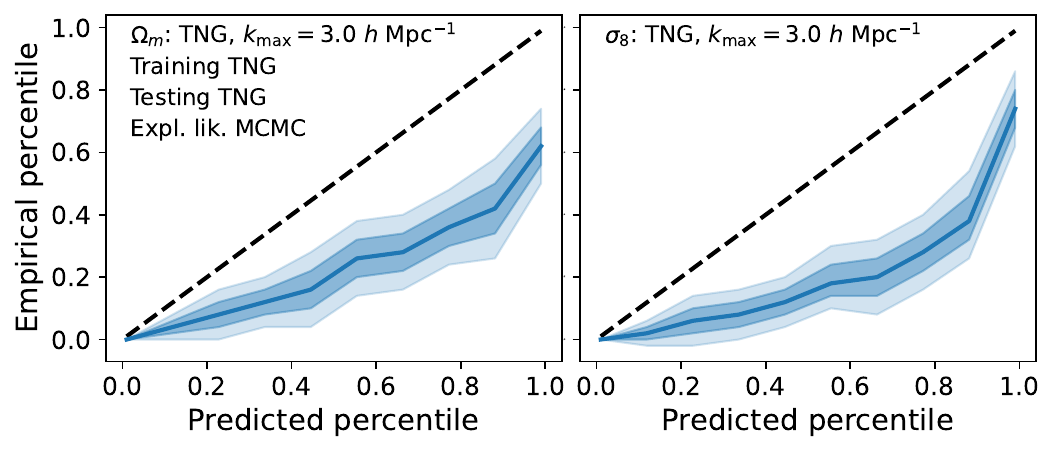}
    \includegraphics[width=0.8\linewidth]{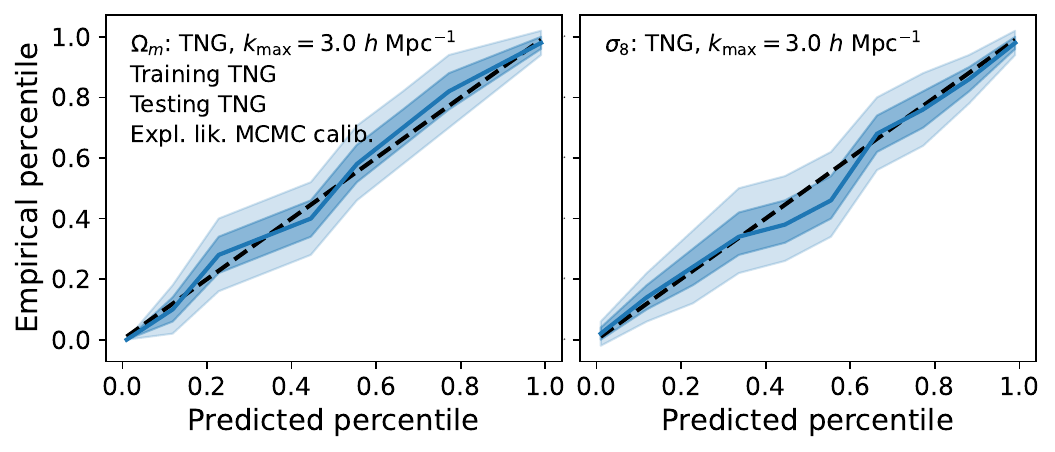}
    \caption{Univariate posterior coverage tests, showing the empirical versus predicted percentiles for the $\Omega_m$ (left) and $\sigma_8$ (right) posteriors, for $k_{\rm max}=3.0~h~{\rm Mpc}^{-1}$, in the case in which both training and testing were performed on \texttt{IllustrisTNG} and for the explicit-likelihood MCMC inference without (top) and with (bottom) posteriors calibration. The blue shaded regions represent the $68\%$ (darker) and $95\%$ (lighter) confidence levels.}
    \label{fig:posterior_coverage_likelihood}
\end{figure*}

\begin{figure*}
    \centering
    \includegraphics[width=0.85\linewidth]{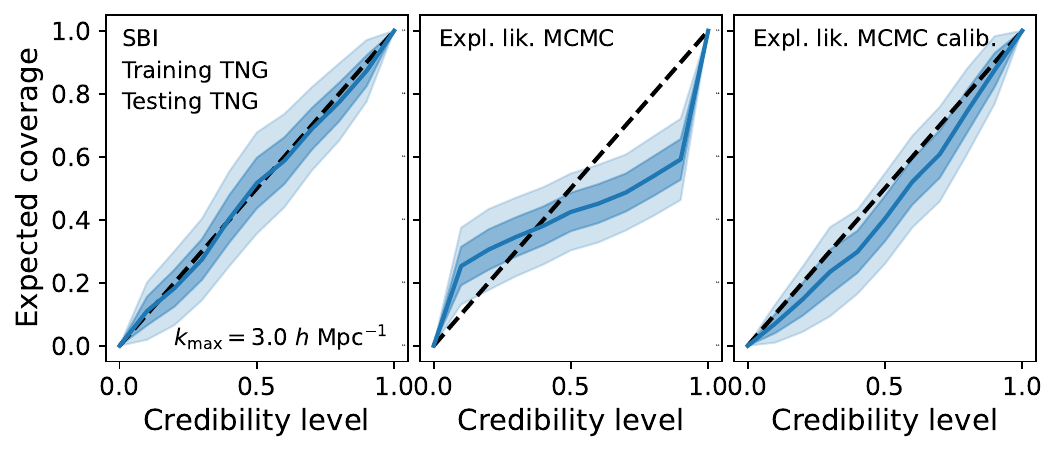}
    \caption{Expected coverage versus credibility level results from TARP, for $k_{\rm max}=3.0~h~{\rm Mpc}^{-1}$ and with both training and testing performed on \texttt{IllustrisTNG}. The different panels show the TARP results for SBI (left), and explicit-likelihood MCMC without (mid) and with (right) posteriors calibration. The blue shaded regions represent the $68\%$ (darker) and $95\%$ (lighter) confidence levels.}
    \label{fig:tarp_likelihood}
\end{figure*}

Figure~\ref{fig:posteriors_om_s8_tngtng_likelihood} shows the posterior distributions of the $\Omega_m$ and $\sigma_8$ parameters for two different test realizations, in the case in which both training and testing were performed on \texttt{IllustrisTNG} and for $k_{\rm max}=3.0~h~{\rm Mpc}^{-1}$. The posteriors shown in the left column correspond to simulations having underlying parameters with intermediate values, while the ones shown in the right column correspond to extreme cases with values for both $\Omega_m$ and $\sigma_8$ at the edge of the priors. We display in green the posteriors obtained through SBI, and in dark blue and red the posteriors yielded by the explicit-likelihood MCMC analysis with and without calibration, respectively. A visual inspection reveals that the calibrated MCMC posteriors agree well with the SBI ones, while being slightly less precise, and the posteriors from MCMC without calibration are clearly too narrow. 
Figure~\ref{fig:residuals_tngntg_likelihood} shows the predicted parameter values against the true ones in the top subpanel and the ratio of the two in the bottom subpanel, for the three different scale cuts and in the case in which both training and testing were performed on the \texttt{IllustrisTNG} simulations, and for $k_{\rm max}=3.0~h~{\rm Mpc}^{-1}$. 
The numerical results are reported in the eight line of Table~\ref{tab:results}. Compared to the SBI results, the explicit likelihood analysis achieves a factor $\sim 4.7$ and $\sim 6$ lower accuracy in the $\Omega_m$ and $\sigma_8$ residuals, respectively. The scatter is comparable for $\Omega_m$ across the two methods, while for $\sigma_8$ it is $\sim 25\%$ larger for the explicit likelihood inference. Finally, the calibrated MCMC achieves $\sim 12\%$ and $\sim40\%$ lower median precision than SBI in $\Omega_m$ and $\sigma_8$, respectively. Figure~\ref{fig:posterior_coverage_likelihood} shows the posterior coverage for $\Omega_m$ (left) and $\sigma_8$ (right), for $k_{\rm max}=3.0~h~{\rm Mpc}^{-1}$. The top panels show the results for the explicit-likelihood MCMC inference (no calibration), while the bottom panels display the ones for the calibrated MCMC analysis. As anticipated, the former case results in poorly calibrated posteriors, found to be quite overconfident. The latter, instead, are well-calibrated by construction, given that the choice of the values for $\lambda$ was done by ensuring the correct behavior of these curves. The same results are confirmed also by the TARP results, shown in Figure~\ref{fig:tarp_likelihood}, which displays the SBI TARP curve in the left panel, and the explicit-likelihood MCMC TARP curves without calibration in the mid panel and with calibration in the right panel. 

Therefore, overall the SBI procedure is here shown to yield robust and well-calibrated posteriors, with an overall comparable and even superior performance compared a traditional explicit-likelihood analysis.

As a final point, we note that since the results from an explicit-likelihood analysis relying on obtaining a $P_{\rm 1D}(k)$ prediction agree with the results from SBI, the former is as likely to incur into the model mis-specification issue as the latter. While state-of-art analyses have investigated this issue by cross-generalizing the model learnt from one simulation suite to another set of testing simulations and found no evident biases \citep[see e.g.,][]{ChavesMontero2026}, the situation may change in the future as the data become more and more precise. In this sense, we point out that SBI constitutes a natural framework to include predictions from different galaxy formation model in the inference. In fact, different models maps onto one set of cosmological parameters. The resulting transform is therefore not bijective and, consequently, not invertible. A traditional explicit-likelihood inference, which relies on $P_{\rm 1D}(k)$ predictions starting form the cosmological parameters through emulators,  is hence not able to invert the mapping and is restricted to only one model per training, unless the different models are accounted for at the likelihood level. In this sense, SBI represent a natural way forward to account for the uncertainty on baryon feedback at the training level (as done herein using a multi-domain training approach) and marginalize over it.

\subsection{Noise injection}
\label{sec:results_noise}

In this section, we test the impact of noise in the recovery of the cosmological parameters. We test herein a Gaussian noise model, that we inject directly into the skewers. In this regime, SBI should yield consistent results as a traditional explicit-likelihood analysis employing a Gaussian likelihood. In future work, we will explore more complex noise models beyond Gaussian.

We model the noise as follows. We first set the desired per-sightline signal-to-noise (SNR). Afterwards, we randomly-sample the noise per each skewer from a Gaussian distribution with zero mean and standard deviation $\sigma=\bar{F}/{\rm SNR}$, where $\bar{F}$ is the mean flux computed from each skewer. We repeat this procedure for ${\rm SNR}=1,3,5$, compute the resulting $P_{\rm 1D}(k)$, and run the SBI framework over them.

\begin{figure*}
    \centering
    \includegraphics[width=0.45\linewidth]{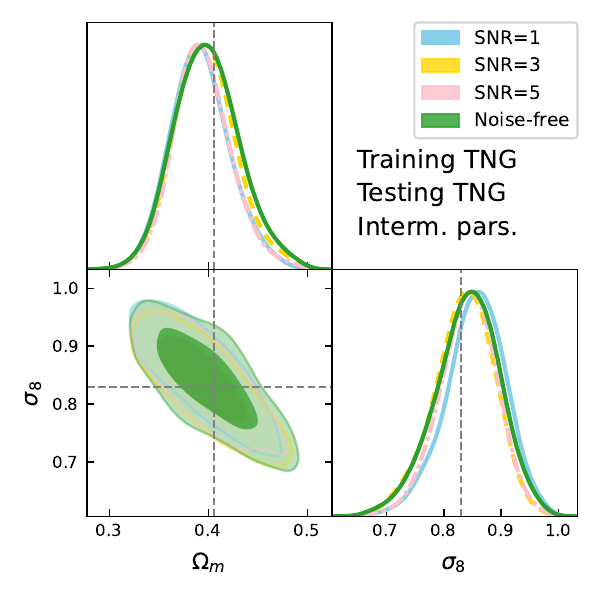}
    \includegraphics[width=0.45\linewidth]{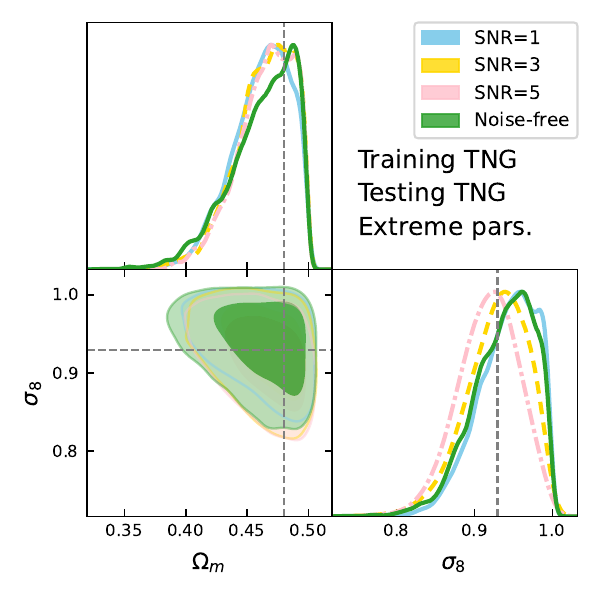}
    \caption{Posterior distributions of the $\Omega_m$ and $\sigma_8$ parameters for different test realizations, $k_{\rm max}=3.0~h~{\rm Mpc}^{-1}$, and obtained by injecting Gaussian noise int the skewers with different SNR: ${\rm SNR=1}$ (light blue), ${\rm SNR=3}$ (gold), ${\rm SNR=5}$ (pink), and no noise (green). Training and testing were performed using simulations run with the \texttt{IllustrisTNG} model. The posteriors shown on the left correspond to simulations having underlying parameters with intermediate values, while the ones shown on the right correspond to extreme cases, at the edge of the priors. The gray dashed lines indicate the true parameters underlying the considered realization.}
    \label{fig:posteriors_noisy}
\end{figure*}

We find that injecting noise with the aforementioned model has negligible impact compared to the noise-free case. We report the numerical values quantifying the accuracy and precision of this experiment in the ninth (${\rm SNR}=1$), tenth (${\rm SNR}=3$), and eleventh (${\rm SNR}=5$) lines of  Table~\ref{tab:results}. Figure~\ref{fig:posteriors_noisy} shows the comparison between the posterior distributions of $\Omega_m$ and $\sigma_8$, and obtained by injecting Gaussian noise with different SNR: ${\rm SNR=1}$ (light blue), ${\rm SNR=3}$ (gold), ${\rm SNR=5}$ (pink), and no noise (green). This comparison indeed confirms an excellent agreement between the posteriors.

As anticipated, the posteriors may change, either becoming broader/narrower or suffering a shift, if a different noise model is assumed. We will further investigate this point in future publications.

\section{Conclusions}
\label{sec:conclusions}

In this work, we have presented a simulation-based inference framework from the Ly$\alpha$ forest $P_{\rm 1D}(k)$. In particular, we rely on the publicly-available Lya forest skewers at $2<z<3.5$ from the \texttt{CAMELS} cosmological hydrodynamic simulations, covering $V=(25~h^{-1}~{\rm Mpc})^3$ and sampling two cosmological parameters ($\Omega_m$ and $\sigma_8$) and four astrophysical parameters ($A_{\rm SN1}$, $A_{\rm AGN1}$, $A_{\rm SN2}$, $A_{\rm AGN2}$) over broad priors. We adopt the neural posterior estimation technique to approximate the posterior distributions using a normalizing flow, and use it to perform inference of the model parameters given the skewers as input. We consider two galaxy formation models---\texttt{IllustrisTNG} and \texttt{SIMBA}---each of which was used to run $1,000$ LH simulations within \texttt{CAMELS}.

We find that in general this inference framework is able to constrain well the cosmological parameters, but not the astrophysical ones, most likely due to volume effects. For this reason, in the remainder of the work we perform inference only of the cosmological parameters, leaving the astrophysics for future work relying on larger-volume simulations.

When performing training and testing on the \texttt{IllustrisTNG} model, we find that our inference scheme predicts unbiased parameter values, within $10\%(20\%)$ deviations in $\gtrsim 75\%~(100\%)$ of the cases for $\Omega_m$ and in $\gtrsim 90\%~(100\%)$ of the cases for $\sigma_8$, and achieving $\sim 8\%$ and $\sim 6\%$ precision on $\Omega_m$ and $\sigma_8$, respectively. The results are slightly worse when training and testing on \texttt{SIMBA}, tentatively due to the stronger dependency of the $P_{\rm 1D}(k)$ prediction on the astrophysical parameters which reflects into a larger uncertainty. Nonetheless, the results are qualitatively unchanged and the results unbiased also in this case. When however we perform training on \texttt{IllustrisTNG} and testing on \texttt{SIMBA}, we report a significant worsening of the performance. Specifically, just $50\%$ of the cases are found to be within $10\%$ deviations both for $\Omega_m$ and $\sigma_8$, with an overall $\sim 10\%$ positive bias in the $\sigma_8$ predictions. The achieved precision is comparable to the one achieved in the previous cases, but the posteriors are generally found to be severely overconfident, and hence, the posterior
uncertainties are underestimated. Finally, when performing multi-domain training, the results are found to be unbiased and overall comparable with the ones achieved when training on one single model and the posteriors to be statistically robust, supporting this technique as an effective potential way forward to deal with the cross-generalization issue.

An important point to be stressed is that the \texttt{CAMELS} LH suite used herein might not be ideal to test the impact of baryon physics on the Ly$\alpha$ forest. The $4$ feedback parameters are in fact related to galactic feedback and might not allow a full exploration of more physical thermal histories \citep[see e.g.,][]{Puchwein2023}. In this way, the derived posteriors on cosmological parameters are likely to be neglecting at least partly some of the correlations with the astrophysical parameters. Therefore, combining \texttt{CAMELS} with other sets of simulations such as \texttt{PRIYA} \citep{Bird2023} or \texttt{Sherwood-Relics} \citep{Puchwein2023} might improve the accuracy and the precision of the posteriors.

In future studies, we will expand this framework to the next generations of \texttt{CAMELS} larger-volume simulations, which will allow to probe larger scales and potentially achieve a better statistical precision on the cosmological parameters, as well as constrain the astrophysics. On another line of research, we will apply this framework to other simulation suites including different dark matter physics such as \texttt{DREAMS} \citep{Rose2025}. Both extensions will be leveraged to perform inference from real state-of-the-art $P_{\rm 1D}(k)$ measurements from e.g. DESI \cite{Karacayli2025,Ravoux2025} and KODIAQ-SQUAD \citep{Karacayli2022}. 

Specifically, to perform inference from real data, we will need to address two important issues:
\begin{itemize}
    \item Modeling the main observational effects in the training dataset: in addition to instrumental noise (addressed in this work in Section~\ref{sec:results_noise}), other major contaminants for the Ly$\alpha$ forest are metals and HCDs. HCDs may be partly included in the CAMELS simulations, but the relatively low resolution and the small volume that they encompass make the HCDs modeling likely to be incomplete. The injection of systematics can be done either at the level of individual skewers, or directly at the level of the $P_{\rm 1D}(k)$. The former implies a significant modeling effort, but allows us to study the impact of going beyond a Gaussian likelihood. The latter provides an easier route and can be achieved by sampling noise terms corresponding to the systematic uncertainties directly from the publicly-available covariance matrix of the data \citep[see e.g.,][and references therein]{Hahn2022,IglesiasNavarro2025}, but it preserves the Gaussianity of the likelihood. We will study these points in future work.
    \item Minimizing the impact of cross-generalization when applying the networks trained on simulations to real data: this aspect can be tackled in a multi-domain training fashion as done in this work by including as many models as possible in the training phase and let the network learn the shared patterns and marginalize over the astrophysical uncertainties, in combination with other domain-adaptation techniques \citep[e.g.,][]{Ciprjanovic2021,Ciprjanovic2022,Roncoli2023,Akhmetzhanova2024,Gilda2024,Ntampaka2025}.
\end{itemize} 
We will explore these avenues in future works.

We conclude that this study represents a promising step forward in the estimation of cosmological parameters via simulation-based inference dealing with the complex problem of the lack of convergence in the predictions from different baryon physics models, and paves the way towards constraining fundamental physics with the Lyman-$\alpha$ forest.

\vspace{0.5cm}
\begin{acknowledgments}
F.S. is grateful to Francisco-Shu Kitaura, Kentaro Nagamine, and ChangHoon Hahn for useful discussions. F.S. acknowledges the support from the postdoctoral fellowship scheme of the {\it Institute for Fundamental Physics of the Universe}. P.I.N. thanks the LSST-DA Data Science Fellowship Program, which is funded by LSST-DA, the Brinson Foundation, the WoodNext Foundation, and the Research Corporation for Science Advancement Foundation; her participation in the program has benefited this work. P.I.N. acknowledges financial support from the State Research Agency of the Spanish Ministry of Science and Innovation (AEI-MCINN) under the grants ``Galaxy Evolution with Artificial Intelligence'' with reference PGC2018-100852-A-I00 and ``BASALT'' with reference PID2021-126838NB-I00. M.V. is partly supported by INFN INDARK and SISSA IDEAS grants, and by the INAF Theory Grant "Cosmological investigation of the cosmic web". M.V. acknowledges the funding by the European Union - NextGenerationEU, in the framework of the HPC project – “National Centre for HPC, Big Data and Quantum Computing” (PNRR - M4C2 - I1.4 - CN00000013 – CUP J33C22001170001).
\end{acknowledgments}

\appendix


\appendix
\section{Analysis with larger $k_{\rm max}$}\label{app:large_kmax}

\begin{figure*}
    \centering
    \includegraphics[width=0.45\linewidth]{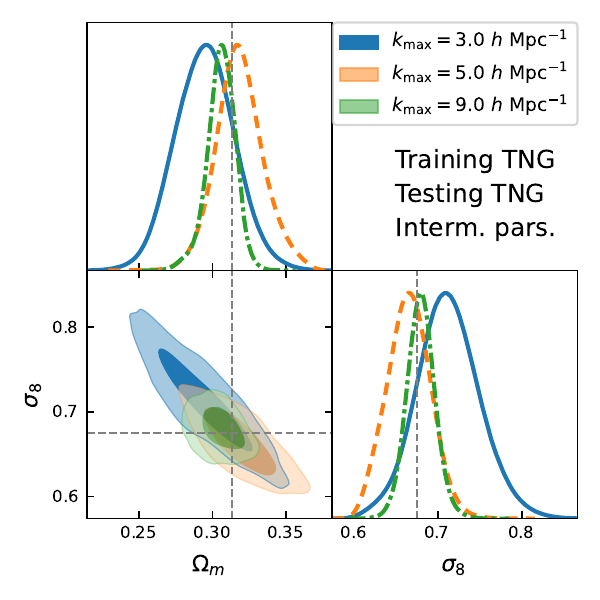}
    \includegraphics[width=0.45\linewidth]{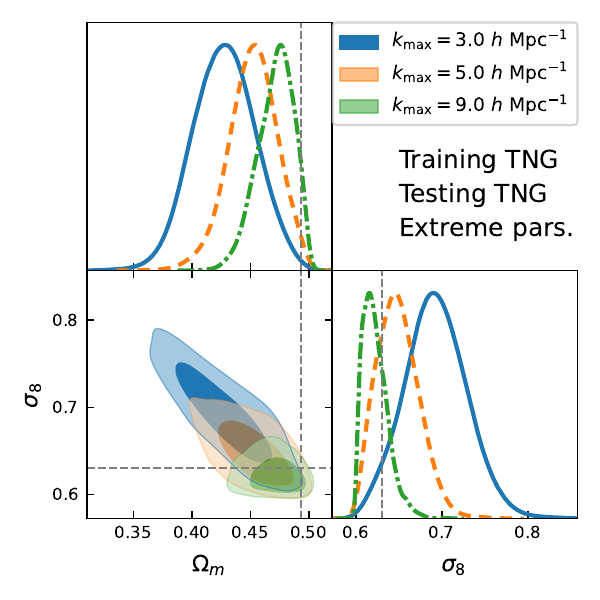}
    
    \caption{Posterior distributions of the $\Omega_m$ and $\sigma_8$ parameters for different test realizations. Training and testing performed with simulations run with the \texttt{IllustrisTNG} model. The posteriors shown in the left column correspond to simulations having underlying parameters with intermediate values, while the ones shown in the right column correspond to extreme cases with $\Omega_m$ values at the edge of the priors. The neural networks were trained setting $k_{\rm max}=3.0~h~{\rm Mpc}^{-1}$ (blue), $k_{\rm max}=5.0~h~{\rm Mpc}^{-1}$ (orange), and  $k_{\rm max}=9.0~h~{\rm Mpc}^{-1}$ (green). The gray dashed lines indicate the true parameters underlying the considered realization.}
    \label{fig:posteriors_om_s8_highk}
\end{figure*}

In the baseline analysis underlying this work, we have tested the performance of the SBI framework presented herein adopting conservative scale cuts due to the limited resolution of the training simulations. It is however interesting to test also cases in which we exploit the larger amount of information coming from smaller scales. In particular, we compare here three cases: $k_{\rm max}=3.0~h~{\rm Mpc}^{-1}$, $k_{\rm max}=5.0~h~{\rm Mpc}^{-1}$, and $k_{\rm max}=9.0~h~{\rm Mpc}^{-1}$. While a quantitative detailed diagnostics of the results as the one presented for the baseline analysis would not be very reliable given the fact that the $P_{\rm 1D}(k)$ predictions are not fully converged on the scale, it is still informative to look at the posteriors to develop an intuitive understanding of the potential information gain. Figure~\ref{fig:posteriors_om_s8_highk} shows the posterior distributions of the $\Omega_m$ and $\sigma_8$ in the case in which training and testing were performed on \texttt{IllustrisTNG}. One clearly sees that there is a very significant gain in precision when going towards larger $k$s and that the posteriors remain fully unbiased. In the case with extreme parameter values, the skewness of the marginalized posteriors for both $\Omega_m$ and $\sigma_8$  captures significantly the true parameters values, compared to the more Gaussian posteriors yielded by $k_{\rm max}=3.0~h~{\rm Mpc}^{-1}$. As will be shown in Appendix~\ref{app:joint}, choosing larger $k$s do not provide any larger constraining power on the astrophysical parameters. Nonetheless, while not conclusive, the potential cosmological information gain in extending the analysis to these smaller scales is remarkable, and motivates the need for future higher-resolution SBI-oriented suites of simulations.

\section{Dependence on redshift}\label{app:redshift}

\begin{figure*}
    \centering
    \includegraphics[width=0.45\linewidth]{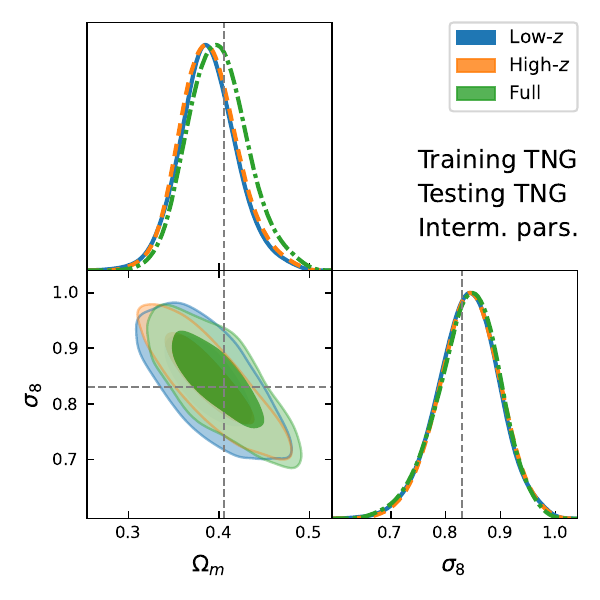}
    \includegraphics[width=0.45\linewidth]{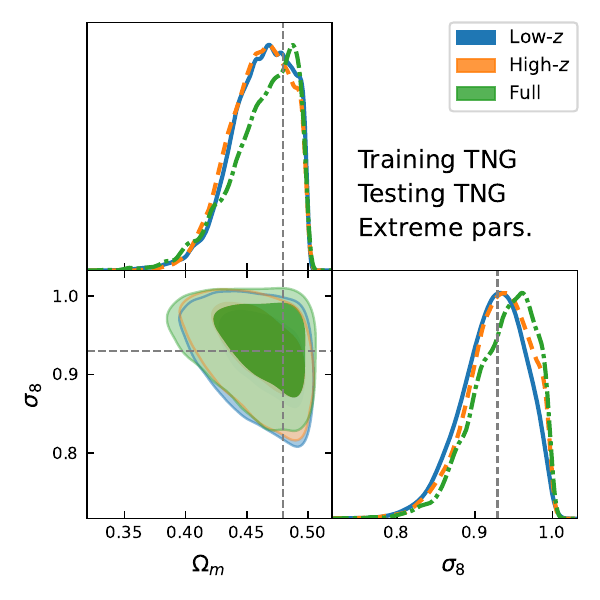}
    \caption{Posterior distributions of the $\Omega_m$ and $\sigma_8$ parameters for different test realizations. Training and testing performed with simulations run with the \texttt{IllustrisTNG} model. The posteriors shown in the left column correspond to simulations having underlying parameters with intermediate values, while the ones shown in the right column correspond to extreme cases with $\Omega_m$ values at the edge of the priors. The neural networks on the low-$z$ sample $z<2.5$ (blue), on the high-$z$ $z>2.5$ (orange), and on the full redshift range (green). The gray dashed lines indicate the true parameters underlying the considered realization.}
    \label{fig:posteriors_redshift}
\end{figure*}

While the baseline analysis was performed including all the $8$ redshift bins listed in Section~\ref{sec:camels}, we test here the inference framework by splitting the sample into two subsamples: a low-$z$ subsample at $z<2.5$ ($z=\{2.00,~2.15,~2.30,~2.46\}$), and a high-$z$ subsample at $z>2.5$ ($z=\{2.63,~2.80,~3.01,~3.49\}$). We find no significant difference with respect to the baseline analysis, neither in accuracy nor in precision. Figure~\ref{fig:posteriors_redshift} shows the comparison of the posterior distributions for the same realizations as in Figure~\ref{fig:posteriors_om_s8_tngtng}, for the low-$z$ sample (blue), the high-$z$ sample (orange), and the full sample (green). In the intermediate parameters case, the posteriors are in excellent agreement, with the full sample having slightly higher accuracy with respect to the other subsamples. In the extreme parameters case, the posteriors associated with the full sample are more skewed, better capturing the deviation from a Gaussian distribution. However, there is no significant difference in the recovery of the cosmological parameters when using just the low$z$ or high-$z$ subsample, coming e.g. from the fact that feedback effects have stronger impact at lower redshift.
 
\section{Posterior distributions for the cosmological and astrophysical inference}\label{app:joint}

\begin{figure*}
    \centering
    \includegraphics[width=\linewidth]{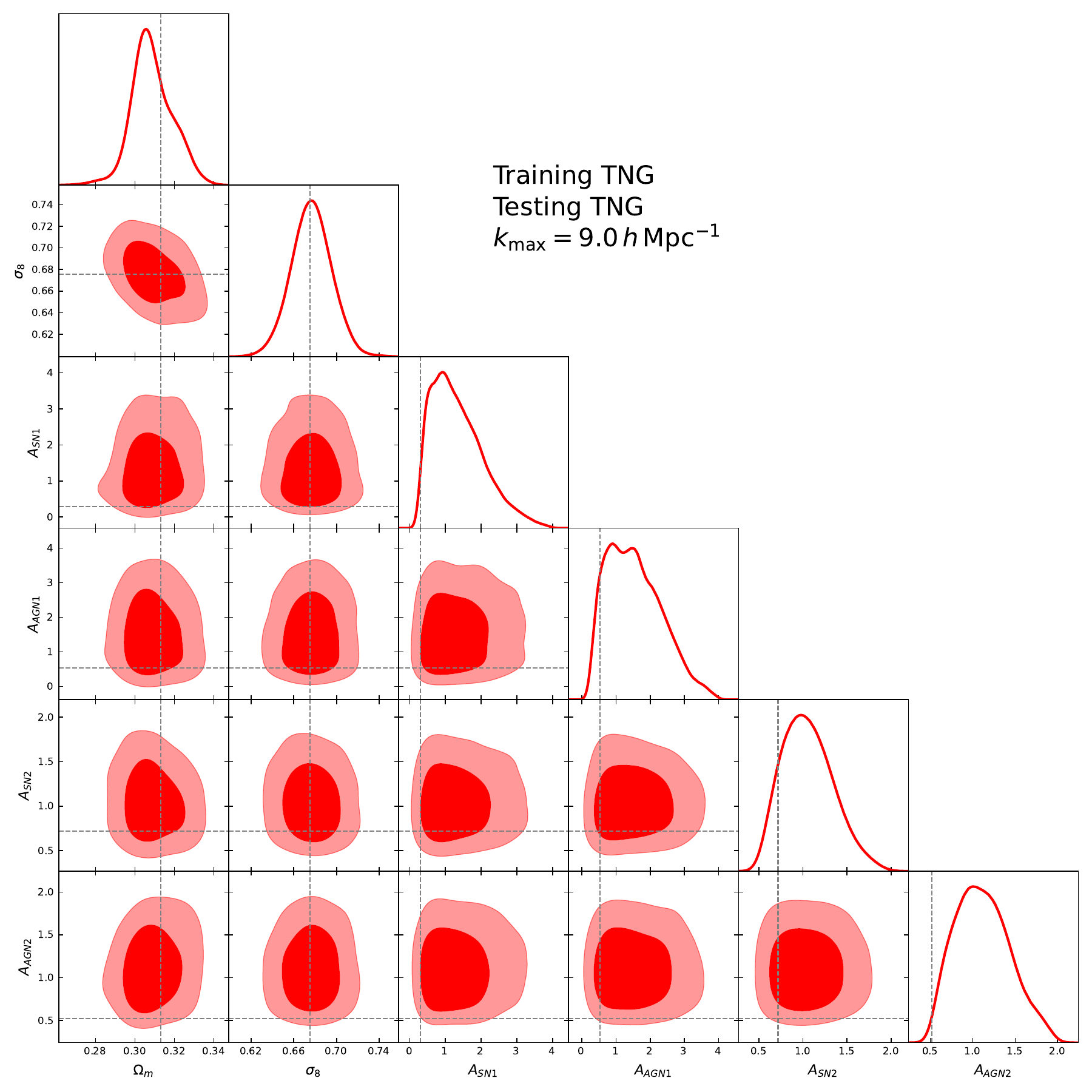}

    \caption{Posterior distributions of all the model parameters, for a test realization from the \texttt{IllustrisTNG} model. The neural networks were trained using the \texttt{IllustrisTNG} model and setting $k_{\rm max}=9.0~h~{\rm Mpc}^{-1}$. The gray dashed lines indicate the true parameters underlying the considered realization.}
    \label{fig:posteriors_all_tng_tng}
\end{figure*}

\begin{figure*}
    \centering
    \includegraphics[width=0.8\linewidth]{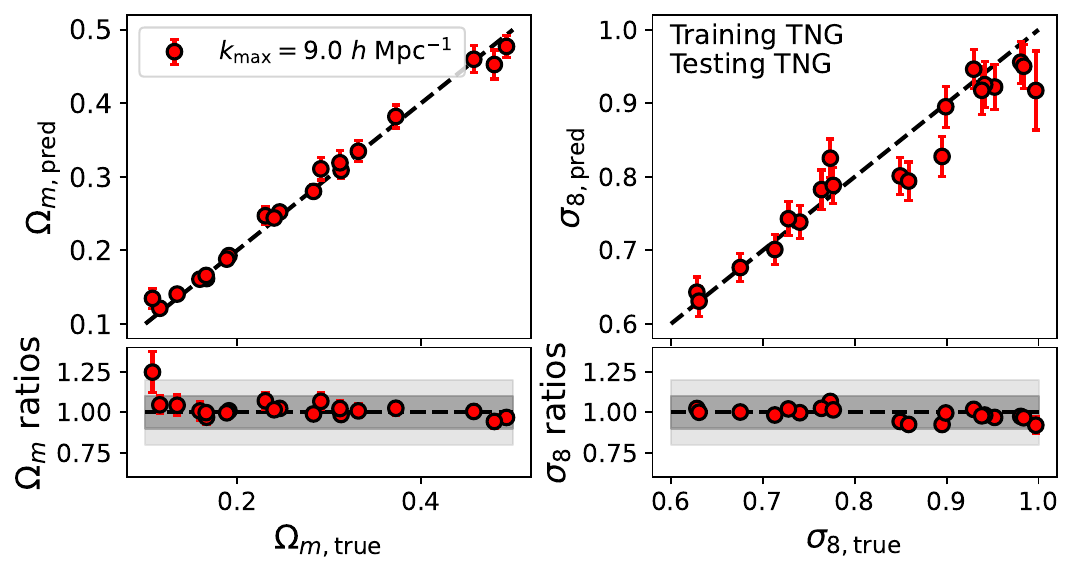}
    \includegraphics[width=0.8\linewidth]{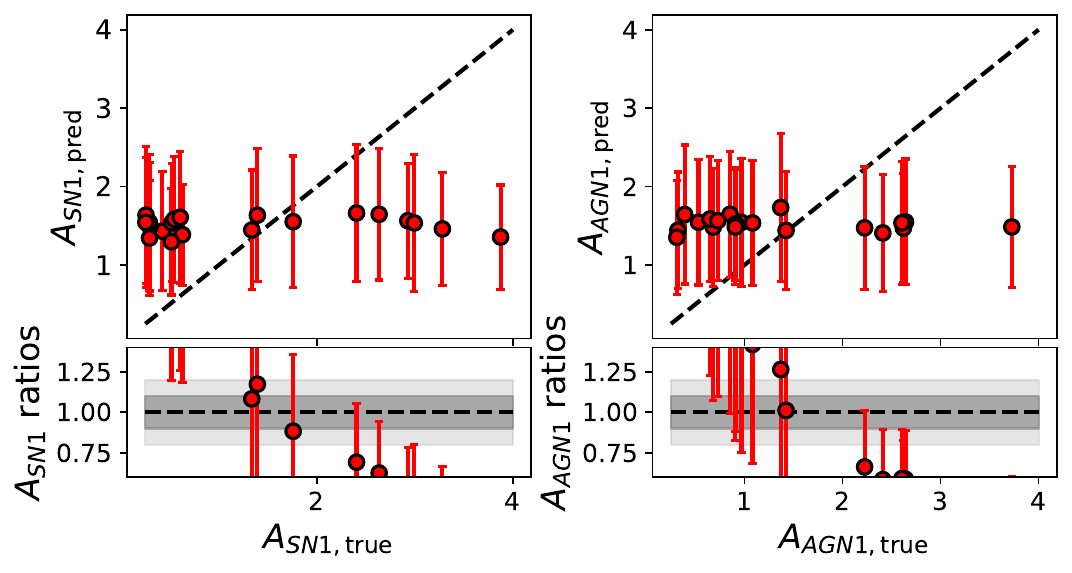}
    \includegraphics[width=0.8\linewidth]{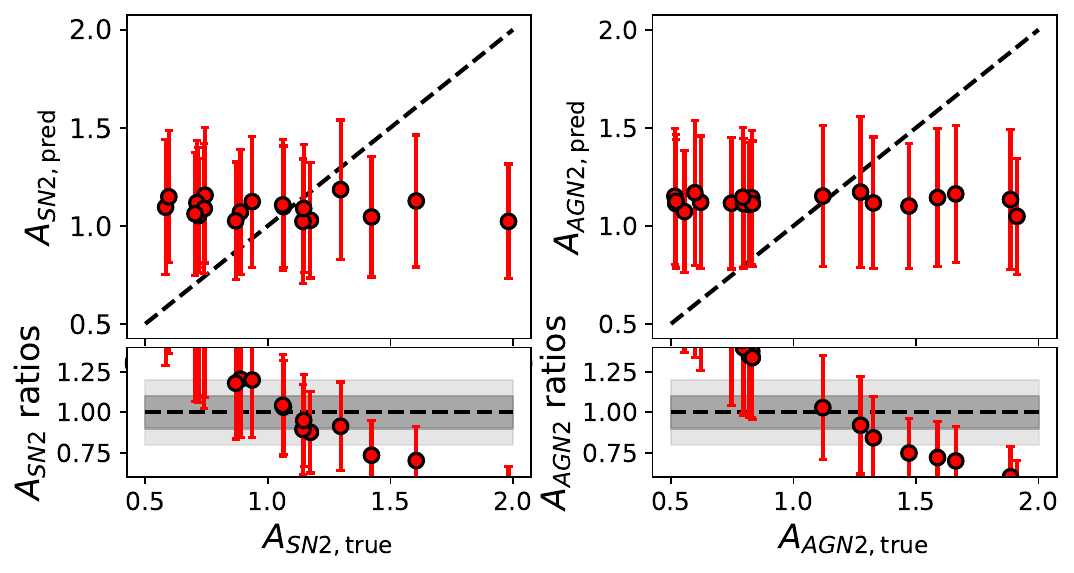}
    \caption{Predicted values for $\Omega_m$ (top left), $\sigma_8$ (top right), $A_{\rm SN1}$ (mid left), $A_{\rm AGN1}$ (mid right), $A_{\rm SN2}$ (bottom left), and $A_{\rm AGN2}$ (bottom right), against the true parameters values, for $k_{\rm max}=9.0~h~{\rm Mpc}^{-1}$. Both training and testing were performed on simulations from the \texttt{IllustrisTNG} suite. The top panels show the 1:1 comparison. The bottom panels display the ratios between the predicted and the true values, and the gray shaded regions stand for $10\%$ (darker) and $20\%$ (lighter) deviations.}
    \label{fig:residuals_allpars_tng_tng}
\end{figure*}

To address whether the neural network is able to constrain the astrophysical parameters, we perform an experiment performing inference of all the six parameters which are varied within the \texttt{CAMELS} LH suite. In this case, to make the training more robust and the testing quicker (given that we have to sample from the posteriors of six parameters and not only two), we use $980$ simulations for training and $20$ for testing, the latter being sufficient to prove the point that we want to make in this section. As anticipated in Section~\ref{sec:results_same}, we perform both training and testing on the \texttt{IllustrisTNG} dataset, adopting different scale cuts. We show in what follows only the results for $k_{\rm max}=9.0~h~{\rm Mpc}^{-1}$---the case that has the largest constraining power on the astrophysical parameters---but the other configurations yield equivalent results. 
Figure~\ref{fig:posteriors_all_tng_tng} show the posteriors for \texttt{IllustrisTNG} and $k_{\rm max}=9.0~h~{\rm Mpc}^{-1}$ case, for one test realization (i.e., not included in the training). One clearly sees that the two cosmological parameters---$\Omega_m$ and $\sigma_8$---are well-constrained, while the four astrophysical parameters have broad posteriors, spanning the whole ranges covered by the priors. To better understand these results, we compared the predicted parameter values against the true ones with which the simulations were run. Figure~\ref{fig:residuals_allpars_tng_tng} shows in each panel the predicted parameter values against the true ones in the top subpanel and the ratio of the two in the bottom subpanel, for all the six parameters. This figure clearly shows that we reproduce with excellent accuracy and precision the true cosmological parameters, while the networks seem to be insensitive to the values of the astrophysical parameters. A qualitatively similar result is found also using the \texttt{SIMBA} model. This finding is not so surprising considering that the cosmological volume embedded in the \texttt{CAMELS} simulations is quite small and the effect of cosmic variance is larger than the one induced by feedback, as discussed in Sections~\ref{sec:camels_feedback} and \ref{sec:camels_cv}. 

\section{Inference on the optical depth $P_{\rm 1D}(k)$}\label{app:tau}

\begin{figure*}
    \centering
    \includegraphics[width=0.9\linewidth]{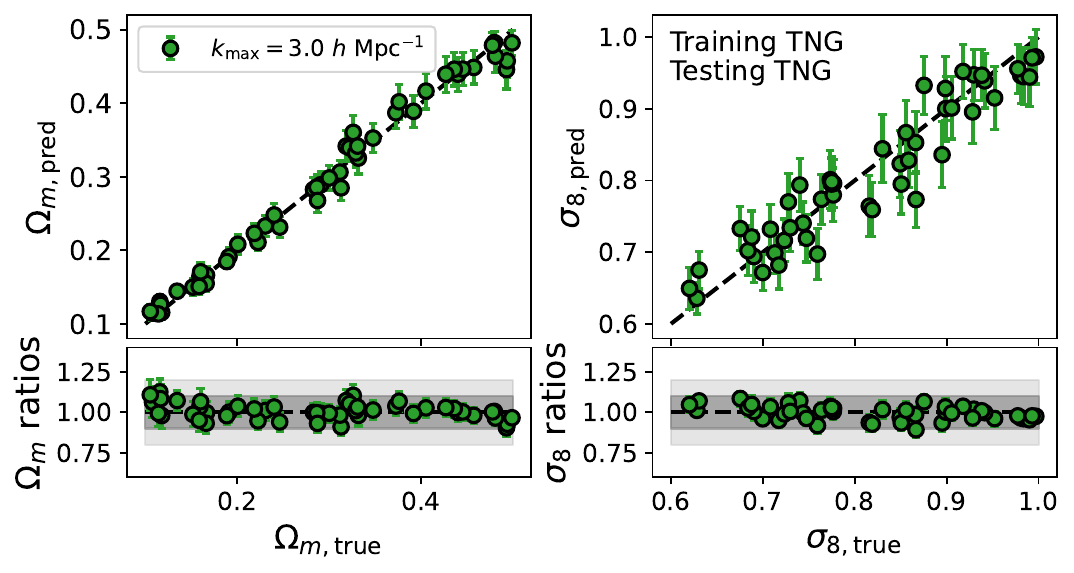}
    \caption{Predicted values for $\Omega_m$ (left) and $\sigma_8$ (right) against the true parameters values, for $k_{\rm max}=3.0~h~{\rm Mpc}^{-1}$. The training was performed on \texttt{IllustrisTNG} and testing on \texttt{SIMBA}, based on the optical field $\tau$. The top panels show the 1:1 comparison. The bottom panels display the ratios between the predicted and the true values, and the gray shaded regions stand for $10\%$ (darker) and $20\%$ (lighter) deviations.}
    \label{fig:residuals_tngtng_tau}
\end{figure*}

\begin{figure*}
    \centering
    \includegraphics[width=0.45\linewidth]{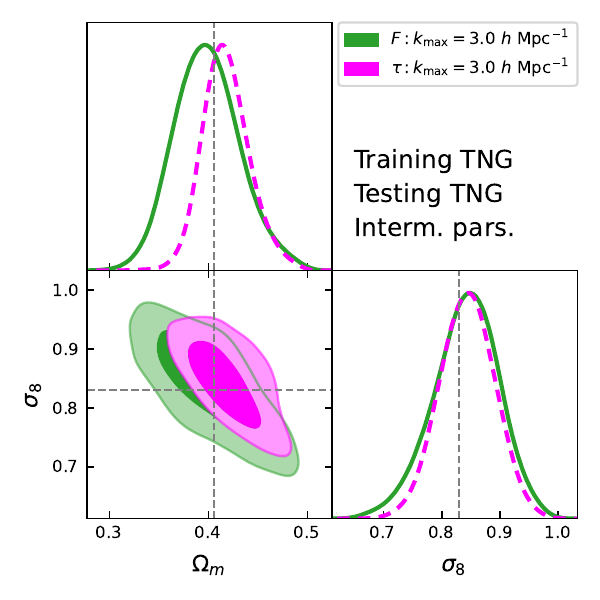}
    \includegraphics[width=0.45\linewidth]{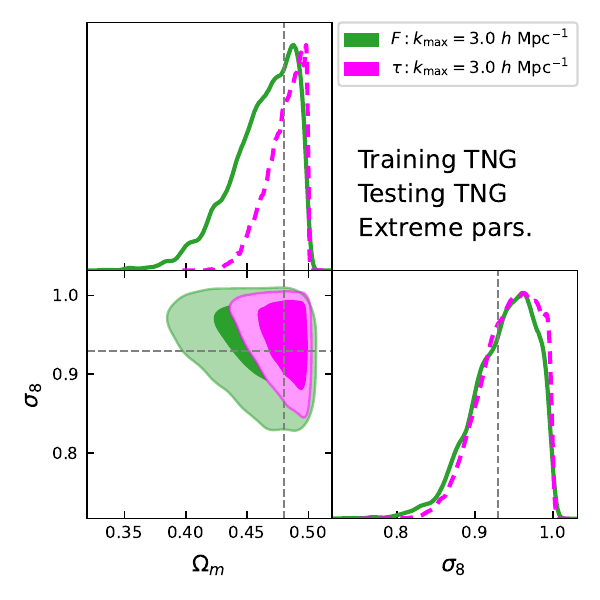}

    \caption{Posterior distributions of the $\Omega_m$ and $\sigma_8$ parameters for different test realizations, where both training and testing were performed on \texttt{IllustrisTNG} and with $k_{\rm max}=3.0~h~{\rm Mpc}^{-1}$. Green: posterior distribution obtained from performing inference on the flux $P_{\rm 1D}(k)$. Magenta: posterior distribution obtained from performing inference on the optical depth $P_{\rm 1D}(k)$. The posteriors shown on the left correspond to simulations having underlying parameters with intermediate values, while the ones shown on the right correspond to extreme cases, at the edge of the priors. The gray dashed lines indicate the true parameters underlying the considered realization.}
    \label{fig:posteriors_tngtng_tau}
\end{figure*}

As a variation around the baseline analysis, we test in this section the inference on the $P_{\rm 1D}(k)$ computed from the optical depth $\tau$ instead of from the flux $F=\exp(-\tau)$. In fact, the exponential mapping quickly saturates the flux to $F\sim 0$ at increasing $\tau$, potentially losing sensitivity to the information coming from the intermediate- and high-density regions. We first perform a first inference run of all the six parameters. However, as in the previous case shown in Appendix~\ref{app:joint}, the astrophysical parameters converge to the prior mean. Therefore, we repeat the inference restricting the analysis to only the two cosmological parameters. Figure~\ref{fig:residuals_tngtng_tau} shows the predicted parameter values against the true ones in the top subpanel and the ratio of the two in the bottom subpanel. In this case, $66\%$, $94\%$, and $100\%$ of the $\Omega_m$ data points and $78\%$, $98\%$, and $100\%$ of the $\sigma_8$ data points are within $5\%$, $10\%$, and $20\%$ deviations. The average residuals for $\Omega_m$ and $\sigma_8$ are $(0.7\pm 4.9)\%$ and $(0.3\pm 4.3)\%$, respectively. The median precision for $\Omega_m$ and $\sigma_8$ is $5.8\%$ and $4.6\%$, respectively. Overall, we obtain better results (reported in the twelfth line of Table~\ref{tab:results}) than in the case in which we perform inference on the flux field: the average of the $50$ data points is within $1\%$ both for $\Omega_m$ and $\sigma_8$, the scatter is reduced from $\sim 8\%$ to $\sim 5\%$ for both parameters, and the median precision on the single data point passes from $\sim 8.1\%$ to $\sim 5.8\%$ for $\Omega_m$ and from $\sim 5.8\%$ to $\sim 4.6\%$ for $\sigma_8$. The improvement can be clearly appreciated by looking at the posterior distributions shown in Figure~\ref{fig:posteriors_tngtng_tau}, in which we display the results from the inference on the flux and optical depth power spectra in green and magenta, respectively. In both the cases with intermediate and extreme parameters, training on the optical depth $P_{\rm 1D}(k)$ yields a clear improvement in the precision of both parameters (especially of $\Omega_m$), while still preserving unbiased results. This result has important implications. While observations yield the flux field, one can always transform it back to optical depth $\tau$ by inverting the exponential mapping, or more in general apply an arbitrary nonlinear transform that maximizes the performance of the inference framework (albeit the noise treatment becomes nontrivial when dealing with real observations). We will explore such possibilities in future works.    

\section{Training on \texttt{IllustrisTNG} and testing on \texttt{SIMBA} without flux rescaling}\label{app:tngsimba_noresc}

\begin{figure*}
    \centering
    \includegraphics[width=0.9\linewidth]{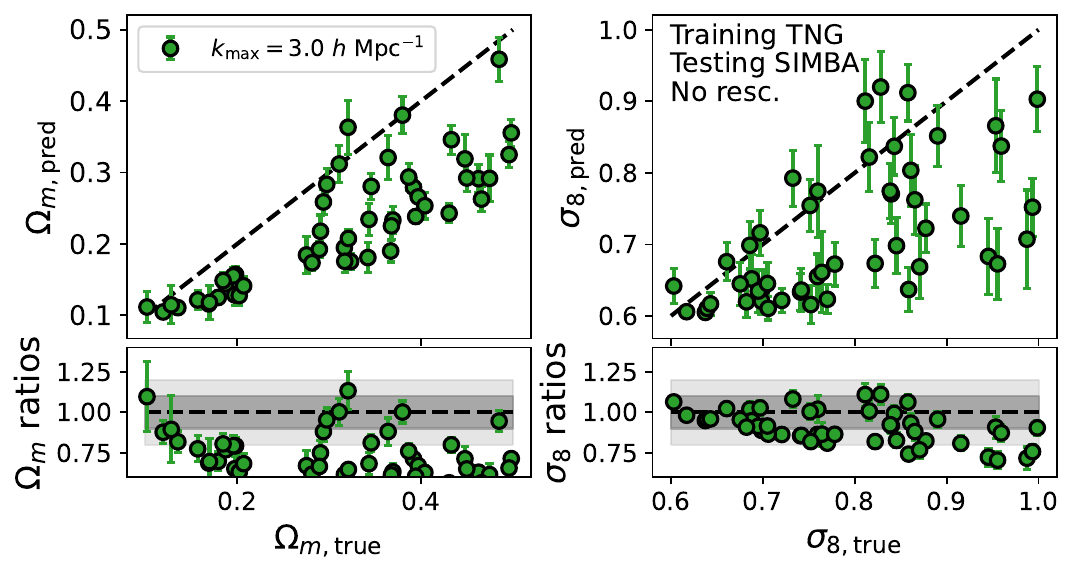}
    \caption{Predicted values for $\Omega_m$ (left) and $\sigma_8$ (right) against the true parameters values, for $k_{\rm max}=3.0~h~{\rm Mpc}^{-1}$. The training was performed on \texttt{IllustrisTNG} and testing on \texttt{SIMBA}, without mean flux rescaling. The top panels show the 1:1 comparison. The bottom panels display the ratios between the predicted and the true values, and the gray shaded regions stand for $10\%$ (darker) and $20\%$ (lighter) deviations.}
    \label{fig:residuals_tngsimba_noresc}
\end{figure*}

In this section, we present the results related to the cases in which we perform the training on \texttt{IllustrisTNG} and the testing on \texttt{SIMBA}, without applying the mean flux rescaling correction described in Section~\ref{sec:camels_lya}. Figure~\ref{fig:residuals_tngsimba_noresc} shows the predicted parameter values against the true ones in the top subpanel and the ratio of the two in the bottom subpanel. One clearly sees that the inference results in severe biases both in $\Omega_m$ and in $\sigma_8$ in this case. For this reason, we do not diagnose it further. In Section~\ref{sec:results_diff}, we test the same case but including the flux rescaling, and show that despite still some residual bias, the results significantly improve.


\section{Training on both \texttt{IllustrisTNG} and \texttt{SIMBA} without flux rescaling}\label{app:mix_noresc}

\begin{figure*}
    \centering
    \includegraphics[width=0.9\linewidth]{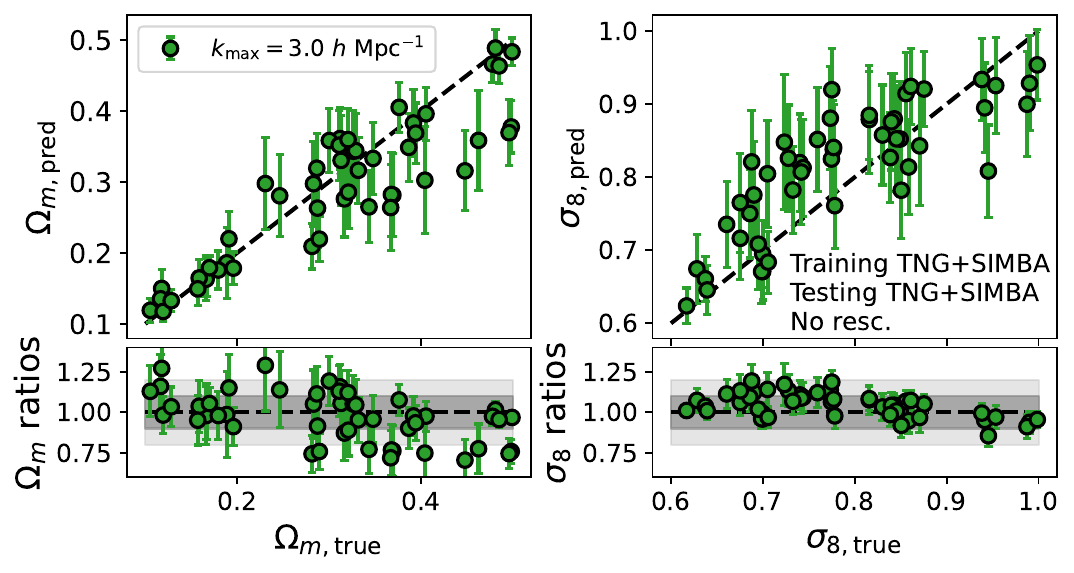}

    \caption{Predicted values for $\Omega_m$ (left) and $\sigma_8$ (right) against the true parameters values, for $k_{\rm max}=3.0~h~{\rm Mpc}^{-1}$. Both training and testing were performed on the combination of \texttt{IllustrisTNG} and \texttt{SIMBA}, without mean flux rescaling. The top panels show the 1:1 comparison. The bottom panels display the ratios between the predicted and the true values, and the gray shaded regions stand for $10\%$ (darker) and $20\%$ (lighter) deviations.}
    \label{fig:residuals_mixix_noresc}
\end{figure*}

\begin{figure*}
    \centering
    \includegraphics[width=0.9\linewidth]{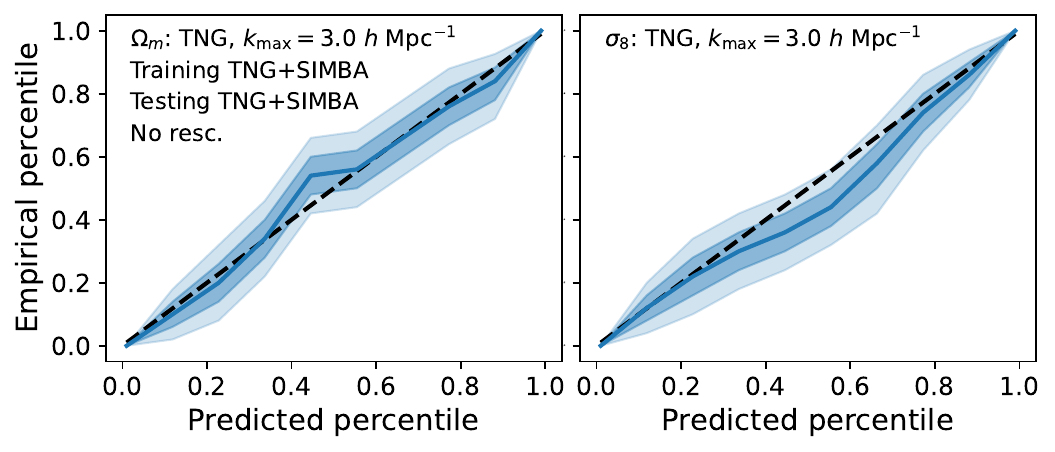}
     \includegraphics[width=0.5\linewidth]{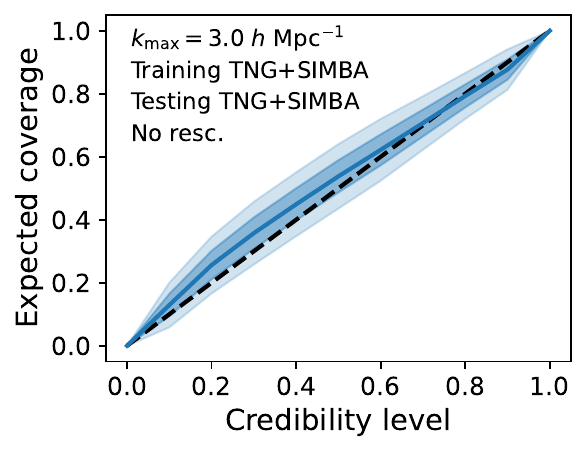}

    \caption{Posteriors coverage test results for the case in which both training and testing were performed on the combination of \texttt{IllustrisTNG} and \texttt{SIMBA}, without mean flux rescaling and $k_{\rm max}=3.0~h~{\rm Mpc}^{-1}$. Top: Univariate posterior coverage tests, showing the empirical versus predicted percentiles for the $\Omega_m$ (left) and $\sigma_8$ (right) posteriors. Bottom: Expected coverage versus credibility level results from TARP. The blue shaded regions represent the $68\%$ (darker) and $95\%$ (lighter) confidence levels.}
    \label{fig:cov_mixix_noresc}
\end{figure*}

\begin{figure*}
    \centering
    \includegraphics[width=0.45\linewidth]{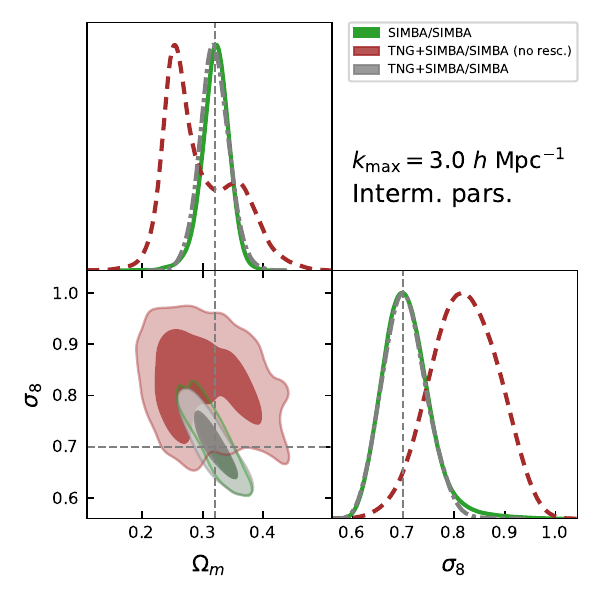}
    \includegraphics[width=0.45\linewidth]{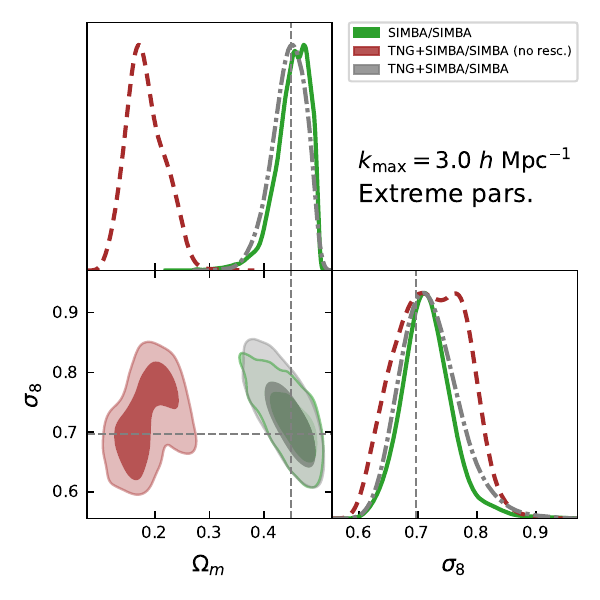}

    \caption{Posterior distributions of the $\Omega_m$ and $\sigma_8$ parameters for different test realizations, and $k_{\rm max}=3.0~h~{\rm Mpc}^{-1}$. Green: training and testing performed on \texttt{SIMBA}. Brown: training on \texttt{IllustrisTNG} and testing performed on \texttt{SIMBA}, without mean flux rescaling. Gray: training on both models and testing on \texttt{SIMBA}. The posteriors shown on the left correspond to simulations having underlying parameters with intermediate values, while the ones shown on the right correspond to extreme cases, at the edge of the priors. The gray dashed lines indicate the true parameters underlying the considered realization.}
    \label{fig:cov_mixmix_noresc}
\end{figure*}

In this section, we present the results related to the cases in which we perform the multi-domain training on \texttt{IllustrisTNG} \texttt{SIMBA}, but without applying the mean flux rescaling correction described in Section~\ref{sec:camels_lya}. Figure~\ref{fig:residuals_mixix_noresc} shows the predicted parameter values against the true ones in the top subpanel and the ratio of the two in the bottom subpanel. The numerical results are reported in the seventh line of Table~\ref{tab:results}.
The performance is clearly worse than the case in which we perform the mean flux rescaling (Section~\ref{sec:results_mix}), both in accuracy and precision. Figure~\ref{fig:cov_mixmix_noresc} offers a clear picture of this. The case in which the training on the combined dataset was performed with flux rescaling well reproduces the posteriors yielded by the training performed on a single dataset, while the training performed on the combined dataset without rescaling the mean flux results in looser and biased posteriors. In the case shown on the left, the marginalized posterior for $\Omega_m$ is even bimodal, reflecting the presence of two different datasets with distinct mean flux normalizations, which are misinterpreted by the network as two possible solutions for $\Omega_m$. In summary, both this section and Appendix~\ref{app:tngsimba_noresc} demonstrate that the networks are quite sensitive to the degeneracy in the power spectrum normalization between the mean flux and the cosmological parameter, and support the need to break this degeneracy by manually rescaling the mean flux to the same reference system.


\bibliography{main}

@ARTICLE{Akeret2015,
       author = {{Akeret}, Jo{\"e}l and {Refregier}, Alexandre and {Amara}, Adam and {Seehars}, Sebastian and {Hasner}, Caspar},
        title = "{Approximate Bayesian computation for forward modeling in cosmology}",
      journal = {\jcap},
         year = 2015,
        month = aug,
       volume = {2015},
       number = {8},
        pages = {043-043},
          doi = {10.1088/1475-7516/2015/08/043},
archivePrefix = {arXiv},
       eprint = {1504.07245},
 primaryClass = {astro-ph.CO},
       adsurl = {https://ui.adsabs.harvard.edu/abs/2015JCAP...08..043A}
}

@ARTICLE{Akhmetzhanova2024,
       author = {{Akhmetzhanova}, Aizhan and {Mishra-Sharma}, Siddharth and {Dvorkin}, Cora},
        title = "{Data compression and inference in cosmology with self-supervised machine learning}",
      journal = {\mnras},
         year = 2024,
        month = jan,
       volume = {527},
       number = {3},
        pages = {7459-7481},
          doi = {10.1093/mnras/stad3646},
archivePrefix = {arXiv},
       eprint = {2308.09751},
 primaryClass = {astro-ph.CO},
       adsurl = {https://ui.adsabs.harvard.edu/abs/2024MNRAS.527.7459A}
}

@ARTICLE{Alsing2018,
       author = {{Alsing}, Justin and {Wandelt}, Benjamin and {Feeney}, Stephen},
        title = "{Massive optimal data compression and density estimation for scalable, likelihood-free inference in cosmology}",
      journal = {\mnras},
         year = 2018,
        month = jul,
       volume = {477},
       number = {3},
        pages = {2874-2885},
          doi = {10.1093/mnras/sty819},
archivePrefix = {arXiv},
       eprint = {1801.01497},
 primaryClass = {astro-ph.CO},
       adsurl = {https://ui.adsabs.harvard.edu/abs/2018MNRAS.477.2874A}
}

@ARTICLE{Alsing2019,
       author = {{Alsing}, Justin and {Charnock}, Tom and {Feeney}, Stephen and {Wandelt}, Benjamin},
        title = "{Fast likelihood-free cosmology with neural density estimators and active learning}",
      journal = {\mnras},
         year = 2019,
        month = sep,
       volume = {488},
       number = {3},
        pages = {4440-4458},
          doi = {10.1093/mnras/stz1960},
archivePrefix = {arXiv},
       eprint = {1903.00007},
 primaryClass = {astro-ph.CO},
       adsurl = {https://ui.adsabs.harvard.edu/abs/2019MNRAS.488.4440A}
}

@ARTICLE{Angeloudi2023,
       author = {{Angeloudi}, Eirini and {Falc{\'o}n-Barroso}, Jes{\'u}s and {Huertas-Company}, Marc and {Sarmiento}, Regina and {Pillepich}, Annalisa and {Walo-Mart{\'\i}n}, Daniel and {Eisert}, Lukas},
        title = "{ERGO-ML: towards a robust machine learning model for inferring the fraction of accreted stars in galaxies from integral-field spectroscopic maps}",
      journal = {\mnras},
         year = 2023,
        month = aug,
       volume = {523},
       number = {4},
        pages = {5408-5429},
          doi = {10.1093/mnras/stad1669},
archivePrefix = {arXiv},
       eprint = {2306.01056},
 primaryClass = {astro-ph.GA},
       adsurl = {https://ui.adsabs.harvard.edu/abs/2023MNRAS.523.5408A}
}

@ARTICLE{Angulo2016,
       author = {{Angulo}, Raul E. and {Pontzen}, Andrew},
        title = "{Cosmological N-body simulations with suppressed variance}",
      journal = {\mnras},
         year = 2016,
        month = oct,
       volume = {462},
       number = {1},
        pages = {L1-L5},
          doi = {10.1093/mnrasl/slw098},
archivePrefix = {arXiv},
       eprint = {1603.05253},
 primaryClass = {astro-ph.CO},
       adsurl = {https://ui.adsabs.harvard.edu/abs/2016MNRAS.462L...1A}
}

@ARTICLE{Becker2011,
       author = {{Becker}, George D. and {Bolton}, James S. and {Haehnelt}, Martin G. and {Sargent}, Wallace L.~W.},
        title = "{Detection of extended He II reionization in the temperature evolution of the intergalactic medium}",
      journal = {\mnras},
         year = 2011,
        month = jan,
       volume = {410},
       number = {2},
        pages = {1096-1112},
          doi = {10.1111/j.1365-2966.2010.17507.x},
archivePrefix = {arXiv},
       eprint = {1008.2622},
 primaryClass = {astro-ph.CO},
       adsurl = {https://ui.adsabs.harvard.edu/abs/2011MNRAS.410.1096B}
}

@ARTICLE{Bird2015,
       author = {{Bird}, Simeon and {Haehnelt}, Martin and {Neeleman}, Marcel and {Genel}, Shy and {Vogelsberger}, Mark and {Hernquist}, Lars},
        title = "{Reproducing the kinematics of damped Lyman {\ensuremath{\alpha}} systems}",
      journal = {\mnras},
         year = 2015,
        month = feb,
       volume = {447},
       number = {2},
        pages = {1834-1846},
          doi = {10.1093/mnras/stu2542},
archivePrefix = {arXiv},
       eprint = {1407.7858},
 primaryClass = {astro-ph.GA},
       adsurl = {https://ui.adsabs.harvard.edu/abs/2015MNRAS.447.1834B}
}

@software{Bird2017,
       author = {{Bird}, Simeon},
        title = "{FSFE: Fake Spectra Flux Extractor}",
 howpublished = {Astrophysics Source Code Library, record ascl:1710.012},
         year = 2017,
        month = oct,
          eid = {ascl:1710.012},
archivePrefix = {ascl},
       eprint = {1710.012},
       adsurl = {https://ui.adsabs.harvard.edu/abs/2017ascl.soft10012B}
}

@article{Bird2019,
  author       = {Bird, Simeon and Rogers, Keir K. and Peiris, Hiranya V. and Verde, Licia and Font-Ribera, Andreu and Pontzen, Andrew},
  title        = {An Emulator for the Lyman-Alpha Forest},
  journal      = {Journal of Cosmology and Astroparticle Physics},
  volume       = {2019},
  number       = {02},
  pages        = {050},
  year         = {2019},
  doi          = {10.1088/1475-7516/2019/02/050},
  eprint       = {1812.04654},
  archivePrefix= {arXiv},
  primaryClass = {astro-ph.CO}
}

@ARTICLE{Bird2023,
       author = {{Bird}, Simeon and {Fernandez}, Martin and {Ho}, Ming-Feng and {Qezlou}, Mahdi and {Monadi}, Reza and {Ni}, Yueying and {Chen}, Nianyi and {Croft}, Rupert and {Di Matteo}, Tiziana},
        title = "{PRIYA: a new suite of Lyman-{\ensuremath{\alpha}} forest simulations for cosmology}",
      journal = {\jcap},
         year = 2023,
        month = oct,
       volume = {2023},
       number = {10},
          eid = {037},
        pages = {037},
          doi = {10.1088/1475-7516/2023/10/037},
archivePrefix = {arXiv},
       eprint = {2306.05471},
 primaryClass = {astro-ph.CO},
       adsurl = {https://ui.adsabs.harvard.edu/abs/2023JCAP...10..037B}
}

@article{BoeltsDeistler_2025,
  doi = {10.21105/joss.07754},
  url = {https://doi.org/10.21105/joss.07754},
  year = {2025},
  publisher = {The Open Journal},
  volume = {10},
  number = {108},
  pages = {7754},
  author = {Jan Boelts and Michael Deistler and Manuel Gloeckler and Álvaro Tejero-Cantero and Jan-Matthis Lueckmann and Guy Moss and Peter Steinbach and Thomas Moreau and Fabio Muratore and Julia Linhart and Conor Durkan and Julius Vetter and Benjamin Kurt Miller and Maternus Herold and Abolfazl Ziaeemehr and Matthijs Pals and Theo Gruner and Sebastian Bischoff and Nastya Krouglova and Richard Gao and Janne K. Lappalainen and Bálint Mucsányi and Felix Pei and Auguste Schulz and Zinovia Stefanidi and Pedro Rodrigues and Cornelius Schröder and Faried Abu Zaid and Jonas Beck and Jaivardhan Kapoor and David S. Greenberg and Pedro J. Gonçalves and Jakob H. Macke},
  title = {sbi reloaded: a toolkit for simulation-based inference workflows},
  journal = {Journal of Open Source Software}
}

@ARTICLE{Boera2019,
       author = {{Boera}, Elisa and {Becker}, George D. and {Bolton}, James S. and {Nasir}, Fahad},
        title = "{Revealing Reionization with the Thermal History of the Intergalactic Medium: New Constraints from the Ly{\ensuremath{\alpha}} Flux Power Spectrum}",
      journal = {\apj},
         year = 2019,
        month = feb,
       volume = {872},
       number = {1},
          eid = {101},
        pages = {101},
          doi = {10.3847/1538-4357/aafee4},
archivePrefix = {arXiv},
       eprint = {1809.06980},
 primaryClass = {astro-ph.CO},
       adsurl = {https://ui.adsabs.harvard.edu/abs/2019ApJ...872..101B}
}

@ARTICLE{Bolton2014,
       author = {{Bolton}, James S. and {Becker}, George D. and {Haehnelt}, Martin G. and {Viel}, Matteo},
        title = "{A consistent determination of the temperature of the intergalactic medium at redshift z = 2.4}",
      journal = {\mnras},
         year = 2014,
        month = mar,
       volume = {438},
       number = {3},
        pages = {2499-2507},
          doi = {10.1093/mnras/stt2374},
archivePrefix = {arXiv},
       eprint = {1308.4411},
 primaryClass = {astro-ph.CO},
       adsurl = {https://ui.adsabs.harvard.edu/abs/2014MNRAS.438.2499B}
}

@ARTICLE{Bolton2017,
       author = {{Bolton}, James S. and {Puchwein}, Ewald and {Sijacki}, Debora and {Haehnelt}, Martin G. and {Kim}, Tae-Sun and {Meiksin}, Avery and {Regan}, John A. and {Viel}, Matteo},
        title = "{The Sherwood simulation suite: overview and data comparisons with the Lyman {\ensuremath{\alpha}} forest at redshifts 2 {\ensuremath{\leq}} z {\ensuremath{\leq}} 5}",
      journal = {\mnras},
         year = 2017,
        month = jan,
       volume = {464},
       number = {1},
        pages = {897-914},
          doi = {10.1093/mnras/stw2397},
archivePrefix = {arXiv},
       eprint = {1605.03462},
 primaryClass = {astro-ph.CO},
       adsurl = {https://ui.adsabs.harvard.edu/abs/2017MNRAS.464..897B}
}

@ARTICLE{Borde2014,
       author = {{Borde}, Arnaud and {Palanque-Delabrouille}, Nathalie and {Rossi}, Graziano and {Viel}, Matteo and {Bolton}, James S. and {Y{\`e}che}, Christophe and {LeGoff}, Jean-Marc and {Rich}, Jim},
        title = "{New approach for precise computation of Lyman-{\ensuremath{\alpha}} forest power spectrum with hydrodynamical simulations}",
      journal = {\jcap},
         year = 2014,
        month = jul,
       volume = {2014},
       number = {7},
          eid = {005},
        pages = {005},
          doi = {10.1088/1475-7516/2014/07/005},
archivePrefix = {arXiv},
       eprint = {1401.6472},
 primaryClass = {astro-ph.CO},
       adsurl = {https://ui.adsabs.harvard.edu/abs/2014JCAP...07..005B}
}

@ARTICLE{Brehmer2018,
       author = {{Brehmer}, Johann and {Louppe}, Gilles and {Pavez}, Juan and {Cranmer}, Kyle},
        title = "{Mining gold from implicit models to improve likelihood-free inference}",
      journal = {arXiv e-prints},
         year = 2018,
        month = may,
          eid = {arXiv:1805.12244},
        pages = {arXiv:1805.12244},
          doi = {10.48550/arXiv.1805.12244},
archivePrefix = {arXiv},
       eprint = {1805.12244},
 primaryClass = {stat.ML},
       adsurl = {https://ui.adsabs.harvard.edu/abs/2018arXiv180512244B}
}

@article{Cabayol2023,
  author       = {Cabayol-Garcia, Laura and Chaves-Montero, Jon{\'a}s and Font-Ribera, Andreu and Pedersen, Christian},
  title        = {A Neural Network Emulator for the Lyman-\ensuremath{\alpha} Forest 1D Flux Power Spectrum},
  journal      = {Monthly Notices of the Royal Astronomical Society},
  volume       = {525},
  number       = {3},
  pages        = {3499--3515},
  year         = {2023},
  doi          = {10.1093/mnras/stad2512},
  eprint       = {2305.19064},
  archivePrefix= {arXiv},
  primaryClass = {astro-ph.CO}
}

@article{ChavesMontero2024,
  author       = {Chaves-Montero, Jon{\'a}s and Cabayol-Garcia, Laura and Lokken, M. and Font-Ribera, Andreu and others},
  title        = {ForestFlow: Cosmological Emulation of Lyman-\ensuremath{\alpha} Forest Clustering from Linear to Nonlinear Scales},
  journal      = {arXiv e-prints},
  year         = {2024},
  eprint       = {2409.05682},
  archivePrefix= {arXiv},
  primaryClass = {astro-ph.CO}
}

@ARTICLE{Chabanier2019,
       author = {{Chabanier}, Sol{\`e}ne and {Palanque-Delabrouille}, Nathalie and {Y{\`e}che}, Christophe and {Le Goff}, Jean-Marc and {Armengaud}, Eric and {Bautista}, Julian and {Blomqvist}, Michael and {Busca}, Nicolas and {Dawson}, Kyle and {Etourneau}, Thomas and {Font-Ribera}, Andreu and {Lee}, Youngbae and {du Mas des Bourboux}, H{\'e}lion and {Pieri}, Matthew and {Rich}, James and {Rossi}, Graziano and {Schneider}, Donald and {Slosar}, An{\v{z}}e},
        title = "{The one-dimensional power spectrum from the SDSS DR14 Ly{\ensuremath{\alpha}} forests}",
      journal = {\jcap},
         year = 2019,
        month = jul,
       volume = {2019},
       number = {7},
          eid = {017},
        pages = {017},
          doi = {10.1088/1475-7516/2019/07/017},
archivePrefix = {arXiv},
       eprint = {1812.03554},
 primaryClass = {astro-ph.CO},
       adsurl = {https://ui.adsabs.harvard.edu/abs/2019JCAP...07..017C}
}

@ARTICLE{ChavesMontero2026,
       author = {{Chaves-Montero}, J. and {Font-Ribera}, A. and {McDonald}, P. and {Armengaud}, E. and {Chebat}, D. and {Garcia-Quintero}, C. and {Kara{\c{c}}ayl{\i}}, N.~G. and {Ravoux}, C. and {Satyavolu}, S. and {Sch{\"o}neberg}, N. and {Walther}, M. and {Aguilar}, J. and {Ahlen}, S. and {Bailey}, S. and {Bianchi}, D. and {Brooks}, D. and {Claybaugh}, T. and {Cuceu}, A. and {de la Macorra}, A. and {Doel}, P. and {Ferraro}, S. and {Forero-Romero}, J.~E. and {Gazta{\~n}aga}, E. and {Gontcho}, S. Gontcho A and {Gonzalez-Morales}, A.~X. and {Gutierrez}, G. and {Guy}, J. and {Hahn}, C. and {Herrera-Alcantar}, H.~K. and {Honscheid}, K. and {Ishak}, M. and {Joyce}, R. and {Juneau}, S. and {Kirkby}, D. and {Kremin}, A. and {Lahav}, O. and {Lamman}, C. and {Landriau}, M. and {Le Goff}, J.~M. and {Le Guillou}, L. and {Leauthaud}, A. and {Levi}, M.~E. and {Manera}, M. and {Martini}, P. and {Meisner}, A. and {Miquel}, R. and {Moustakas}, J. and {Nadathur}, S. and {Niz}, G. and {Palanque-Delabrouille}, N. and {Percival}, W.~J. and {Prada}, F. and {P{\'e}rez-R{\`a}fols}, I. and {Rossi}, G. and {Sanchez}, E. and {Schlegel}, D. and {Schubnell}, M. and {Seo}, H. and {Silber}, J. and {Sprayberry}, D. and {Tan}, T. and {Tarl{\'e}}, G. and {Weaver}, B.~A. and {Y{\`e}che}, C. and {Zhou}, R. and {Zou}, H.},
        title = "{Cosmological analysis of the DESI DR1 Lyman alpha 1D power spectrum}",
      journal = {arXiv e-prints},
         year = 2026,
        month = jan,
          eid = {arXiv:2601.21432},
        pages = {arXiv:2601.21432},
          doi = {10.48550/arXiv.2601.21432},
archivePrefix = {arXiv},
       eprint = {2601.21432},
 primaryClass = {astro-ph.CO},
       adsurl = {https://ui.adsabs.harvard.edu/abs/2026arXiv260121432C}
}

@ARTICLE{Charnock2018,
       author = {{Charnock}, Tom and {Lavaux}, Guilhem and {Wandelt}, Benjamin D.},
        title = "{Automatic physical inference with information maximizing neural networks}",
      journal = {\prd},
         year = 2018,
        month = apr,
       volume = {97},
       number = {8},
          eid = {083004},
        pages = {083004},
          doi = {10.1103/PhysRevD.97.083004},
archivePrefix = {arXiv},
       eprint = {1802.03537},
 primaryClass = {astro-ph.IM},
       adsurl = {https://ui.adsabs.harvard.edu/abs/2018PhRvD..97h3004C}
}

@ARTICLE{Ciprjanovic2021,
       author = {{{\'C}iprijanovi{\'c}}, A. and {Kafkes}, D. and {Downey}, K. and {Jenkins}, S. and {Perdue}, G.~N. and {Madireddy}, S. and {Johnston}, T. and {Snyder}, G.~F. and {Nord}, B.},
        title = "{DeepMerge - II. Building robust deep learning algorithms for merging galaxy identification across domains}",
      journal = {\mnras},
         year = 2021,
        month = sep,
       volume = {506},
       number = {1},
        pages = {677-691},
          doi = {10.1093/mnras/stab1677},
archivePrefix = {arXiv},
       eprint = {2103.01373},
 primaryClass = {astro-ph.IM},
       adsurl = {https://ui.adsabs.harvard.edu/abs/2021MNRAS.506..677C}
}

@ARTICLE{Ciprjanovic2022,
       author = {{{\'C}iprijanovi{\'c}}, Aleksandra and {Lewis}, Ashia and {Pedro}, Kevin and {Madireddy}, Sandeep and {Nord}, Brian and {Perdue}, Gabriel N. and {Wild}, Stefan M.},
        title = "{Semi-Supervised Domain Adaptation for Cross-Survey Galaxy Morphology Classification and Anomaly Detection}",
      journal = {arXiv e-prints},
         year = 2022,
        month = nov,
          eid = {arXiv:2211.00677},
        pages = {arXiv:2211.00677},
          doi = {10.48550/arXiv.2211.00677},
archivePrefix = {arXiv},
       eprint = {2211.00677},
 primaryClass = {astro-ph.GA},
       adsurl = {https://ui.adsabs.harvard.edu/abs/2022arXiv221100677C}
}

@ARTICLE{Chongchitnan2014,
       author = {{Chongchitnan}, Sirichai and {Meiksin}, Avery},
        title = "{The effect of cosmic magnetic fields on the metagalactic ionization background inferred from the Lyman {\ensuremath{\alpha}} forest}",
      journal = {\mnras},
         year = 2014,
        month = feb,
       volume = {437},
       number = {4},
        pages = {3639-3646},
          doi = {10.1093/mnras/stt2169},
archivePrefix = {arXiv},
       eprint = {1311.1504},
 primaryClass = {astro-ph.CO},
       adsurl = {https://ui.adsabs.harvard.edu/abs/2014MNRAS.437.3639C}
}

@ARTICLE{Dave2016,
       author = {{Dav{\'e}}, Romeel and {Thompson}, Robert and {Hopkins}, Philip F.},
        title = "{MUFASA: galaxy formation simulations with meshless hydrodynamics}",
      journal = {\mnras},
         year = 2016,
        month = nov,
       volume = {462},
       number = {3},
        pages = {3265-3284},
          doi = {10.1093/mnras/stw1862},
archivePrefix = {arXiv},
       eprint = {1604.01418},
 primaryClass = {astro-ph.GA},
       adsurl = {https://ui.adsabs.harvard.edu/abs/2016MNRAS.462.3265D}
}

@ARTICLE{Dave2019,
       author = {{Dav{\'e}}, Romeel and {Angl{\'e}s-Alc{\'a}zar}, Daniel and {Narayanan}, Desika and {Li}, Qi and {Rafieferantsoa}, Mika H. and {Appleby}, Sarah},
        title = "{SIMBA: Cosmological simulations with black hole growth and feedback}",
      journal = {\mnras},
         year = 2019,
        month = jun,
       volume = {486},
       number = {2},
        pages = {2827-2849},
          doi = {10.1093/mnras/stz937},
archivePrefix = {arXiv},
       eprint = {1901.10203},
 primaryClass = {astro-ph.GA},
       adsurl = {https://ui.adsabs.harvard.edu/abs/2019MNRAS.486.2827D}
}

@article{
  Deistler2022,
  title={Truncated proposals for scalable and hassle-free simulation-based inference},
  author={Deistler, Michael and Goncalves, Pedro J and Macke, Jakob H},
  journal={Advances in neural information processing systems},
  volume={35},
  pages={23135--23149},
  year={2022}
}

@ARTICLE{DESI_DR2,
       author = {{Abdul Karim}, M. and {Aguilar}, J. and {Ahlen}, S. and {Alam}, S. and {Allen}, L. and {Allende Prieto}, C. and {Alves}, O. and {Anand}, A. and {Andrade}, U. and {Armengaud}, E. and {Aviles}, A. and {Bailey}, S. and {Baltay}, C. and {Bansal}, P. and {Bault}, A. and {Behera}, J. and {BenZvi}, S. and {Bianchi}, D. and {Blake}, C. and {Brieden}, S. and {Brodzeller}, A. and {Brooks}, D. and {Buckley-Geer}, E. and {Burtin}, E. and {Calderon}, R. and {Canning}, R. and {Rosell}, A. Carnero and {Carrilho}, P. and {Casas}, L. and {Castander}, F.~J. and {Charles}, M. and {Chaussidon}, E. and {Chaves-Montero}, J. and {Chebat}, D. and {Chen}, X. and {Claybaugh}, T. and {Cole}, S. and {Cooper}, A.~P. and {Cuceu}, A. and {Dawson}, K.~S. and {de la Macorra}, A. and {de Mattia}, A. and {Deiosso}, N. and {Della Costa}, J. and {Demina}, R. and {Dey}, A. and {Dey}, B. and {Ding}, Z. and {Doel}, P. and {Edelstein}, J. and {Eisenstein}, D.~J. and {Elbers}, W. and {Fagrelius}, P. and {Fanning}, K. and {Fern{\'a}ndez-Garc{\'\i}a}, E. and {Ferraro}, S. and {Font-Ribera}, A. and {Forero-Romero}, J.~E. and {Frenk}, C.~S. and {Garcia-Quintero}, C. and {Garrison}, L.~H. and {Gazta{\~n}aga}, E. and {Gil-Mar{\'\i}n}, H. and {Gontcho A Gontcho}, S. and {Gonzalez}, D. and {Gonzalez-Morales}, A.~X. and {Gordon}, C. and {Green}, D. and {Gutierrez}, G. and {Guy}, J. and {Hadzhiyska}, B. and {Hahn}, C. and {He}, S. and {Herbold}, M. and {Herrera-Alcantar}, H.~K. and {Ho}, M.-F. and {Honscheid}, K. and {Howlett}, C. and {Huterer}, D. and {Ishak}, M. and {Juneau}, S. and {Kamble}, N.~V. and {Kara{\c{c}}ayl{\i}}, N.~G. and {Kehoe}, R. and {Kent}, S. and {Kim}, A.~G. and {Kirkby}, D. and {Kisner}, T. and {Koposov}, S.~E. and {Kremin}, A. and {Krolewski}, A. and {Lahav}, O. and {Lamman}, C. and {Landriau}, M. and {Lang}, D. and {Lasker}, J. and {Le Goff}, J.~M. and {Le Guillou}, L. and {Leauthaud}, A. and {Levi}, M.~E. and {Li}, Q. and {Li}, T.~S. and {Lodha}, K. and {Lokken}, M. and {Lozano-Rodr{\'\i}guez}, F. and {Magneville}, C. and {Manera}, M. and {Martini}, P. and {Matthewson}, W.~L. and {Meisner}, A. and {Mena-Fern{\'a}ndez}, J. and {Menegas}, A. and {Mergulh{\~a}o}, T. and {Miquel}, R. and {Moustakas}, J. and {Mu{\~n}oz-Guti{\'e}rrez}, A. and {Mu{\~n}oz-Santos}, D. and {Myers}, A.~D. and {Nadathur}, S. and {Naidoo}, K. and {Napolitano}, L. and {Newman}, J.~A. and {Niz}, G. and {Noriega}, H.~E. and {Paillas}, E. and {Palanque-Delabrouille}, N. and {Pan}, J. and {Peacock}, J.~A. and {Pellejero Ibanez}, M. and {Percival}, W.~J. and {P{\'e}rez-Fern{\'a}ndez}, A. and {P{\'e}rez-R{\`a}fols}, I. and {Pieri}, M.~M. and {Poppett}, C. and {Prada}, F. and {Rabinowitz}, D. and {Raichoor}, A. and {Ram{\'\i}rez-P{\'e}rez}, C. and {Rashkovetskyi}, M. and {Ravoux}, C. and {Rich}, J. and {Rocher}, A. and {Rockosi}, C. and {Rohlf}, J. and {Rom{\'a}n-Herrera}, J.~O. and {Ross}, A.~J. and {Rossi}, G. and {Ruggeri}, R. and {Ruhlmann-Kleider}, V. and {Samushia}, L. and {Sanchez}, E. and {Sanders}, N. and {Schlegel}, D. and {Schubnell}, M. and {Seo}, H. and {Shafieloo}, A. and {Sharples}, R. and {Silber}, J. and {Sinigaglia}, F. and {Sprayberry}, D. and {Tan}, T. and {Tarl{\'e}}, G. and {Taylor}, P. and {Turner}, W. and {Ure{\~n}a-L{\'o}pez}, L.~A. and {Vaisakh}, R. and {Valdes}, F. and {Valogiannis}, G. and {Vargas-Maga{\~n}a}, M. and {Verde}, L. and {Walther}, M. and {Weaver}, B.~A. and {Weinberg}, D.~H. and {White}, M. and {Wolfson}, M. and {Y{\`e}che}, C. and {Yu}, J. and {Zaborowski}, E.~A. and {Zarrouk}, P. and {Zhai}, Z. and {Zhang}, H. and {Zhao}, C. and {Zhao}, G.~B. and {Zhou}, R. and {Zou}, H. and {DESI Collaboration}},
        title = "{DESI DR2 results. II. Measurements of baryon acoustic oscillations and cosmological constraints}",
      journal = {\prd},
         year = 2025,
        month = oct,
       volume = {112},
       number = {8},
          eid = {083515},
        pages = {083515},
          doi = {10.1103/tr6y-kpc6},
archivePrefix = {arXiv},
       eprint = {2503.14738},
 primaryClass = {astro-ph.CO},
       adsurl = {https://ui.adsabs.harvard.edu/abs/2025PhRvD.112h3515A}
}

@ARTICLE{Dong2024,
       author = {{Dong}, Chenze and {Lee}, Khee-Gan and {Cui}, Weiguang and {Dav{\'e}}, Romeel and {Sorini}, Daniele},
        title = "{The effect of AGN feedback on the Lyman-{\ensuremath{\alpha}} forest signature of galaxy protoclusters at z  2.3}",
      journal = {\mnras},
         year = 2024,
        month = aug,
       volume = {532},
       number = {4},
        pages = {4876-4888},
          doi = {10.1093/mnras/stae1830},
archivePrefix = {arXiv},
       eprint = {2402.13568},
 primaryClass = {astro-ph.GA},
       adsurl = {https://ui.adsabs.harvard.edu/abs/2024MNRAS.532.4876D}
}

@ARTICLE{FaucherGiguere2009,
       author = {{Faucher-Gigu{\`e}re}, Claude-Andr{\'e} and {Lidz}, Adam and {Zaldarriaga}, Matias and {Hernquist}, Lars},
        title = "{A New Calculation of the Ionizing Background Spectrum and the Effects of He II Reionization}",
      journal = {\apj},
         year = 2009,
        month = oct,
       volume = {703},
       number = {2},
        pages = {1416-1443},
          doi = {10.1088/0004-637X/703/2/1416},
archivePrefix = {arXiv},
       eprint = {0901.4554},
 primaryClass = {astro-ph.CO},
       adsurl = {https://ui.adsabs.harvard.edu/abs/2009ApJ...703.1416F}
}

@ARTICLE{ForemanMackey2013,
       author = {{Foreman-Mackey}, Daniel and {Hogg}, David W. and {Lang}, Dustin and {Goodman}, Jonathan},
        title = "{emcee: The MCMC Hammer}",
      journal = {\pasp},
         year = 2013,
        month = mar,
       volume = {125},
       number = {925},
        pages = {306},
          doi = {10.1086/670067},
archivePrefix = {arXiv},
       eprint = {1202.3665},
 primaryClass = {astro-ph.IM},
       adsurl = {https://ui.adsabs.harvard.edu/abs/2013PASP..125..306F}
}

@ARTICLE{Gaikwad2021,
       author = {{Gaikwad}, Prakash and {Srianand}, Raghunathan and {Haehnelt}, Martin G. and {Choudhury}, Tirthankar Roy},
        title = "{A consistent and robust measurement of the thermal state of the IGM at 2 {\ensuremath{\leq}} z {\ensuremath{\leq}} 4 from a large sample of Ly {\ensuremath{\alpha}} forest spectra: evidence for late and rapid He II reionization}",
      journal = {\mnras},
         year = 2021,
        month = sep,
       volume = {506},
       number = {3},
        pages = {4389-4412},
          doi = {10.1093/mnras/stab2017},
archivePrefix = {arXiv},
       eprint = {2009.00016},
 primaryClass = {astro-ph.CO},
       adsurl = {https://ui.adsabs.harvard.edu/abs/2021MNRAS.506.4389G}
}

@ARTICLE{GarciaGallego2025,
       author = {{Garcia-Gallego}, Olga and {Ir{\v{s}}i{\v{c}}}, Vid and {Haehnelt}, Martin G. and {Viel}, Matteo and {Bolton}, James S.},
        title = "{Constraining mixed dark matter models with high-redshift Lyman-alpha forest data}",
      journal = {\prd},
         year = 2025,
        month = aug,
       volume = {112},
       number = {4},
          eid = {043502},
        pages = {043502},
          doi = {10.1103/4k29-h99l},
archivePrefix = {arXiv},
       eprint = {2504.06367},
 primaryClass = {astro-ph.CO},
       adsurl = {https://ui.adsabs.harvard.edu/abs/2025PhRvD.112d3502G}
}

@ARTICLE{Garzilli2012,
       author = {{Garzilli}, A. and {Bolton}, J.~S. and {Kim}, T.-S. and {Leach}, S. and {Viel}, M.},
        title = "{The intergalactic medium thermal history at redshift z = 1.7-3.2 from the Ly{\ensuremath{\alpha}} forest: a comparison of measurements using wavelets and the flux distribution}",
      journal = {\mnras},
         year = 2012,
        month = aug,
       volume = {424},
       number = {3},
        pages = {1723-1736},
          doi = {10.1111/j.1365-2966.2012.21223.x},
archivePrefix = {arXiv},
       eprint = {1202.3577},
 primaryClass = {astro-ph.CO},
       adsurl = {https://ui.adsabs.harvard.edu/abs/2012MNRAS.424.1723G}
}

@ARTICLE{Garzilli2020,
       author = {{Garzilli}, Antonella and {Theuns}, Tom and {Schaye}, Joop},
        title = "{Measuring the temperature and profiles of Ly {\ensuremath{\alpha}} absorbers}",
      journal = {\mnras},
         year = 2020,
        month = feb,
       volume = {492},
       number = {2},
        pages = {2193-2207},
          doi = {10.1093/mnras/stz3585},
archivePrefix = {arXiv},
       eprint = {1808.06646},
 primaryClass = {astro-ph.CO},
       adsurl = {https://ui.adsabs.harvard.edu/abs/2020MNRAS.492.2193G}
}

@ARTICLE{Gilda2024,
       author = {{Gilda}, Sankalp and {de Mathelin}, Antoine and {Bellstedt}, Sabine and {Richard}, Guillaume},
        title = "{Unsupervised Domain Adaptation for Constraining Star Formation Histories}",
      journal = {Astronomy},
         year = 2024,
        month = jul,
       volume = {3},
       number = {3},
        pages = {189-207},
          doi = {10.3390/astronomy3030012},
archivePrefix = {arXiv},
       eprint = {2112.14072},
 primaryClass = {astro-ph.GA},
       adsurl = {https://ui.adsabs.harvard.edu/abs/2024Astro...3..189G}
}

@ARTICLE{Gluck2024,
       author = {{Gluck}, Naomi and {Oppenheimer}, Benjamin D. and {Nagai}, Daisuke and {Villaescusa-Navarro}, Francisco and {Angl{\'e}s-Alc{\'a}zar}, Daniel},
        title = "{An observationally driven multifield approach for probing the circum-galactic medium with convolutional neural networks}",
      journal = {\mnras},
         year = 2024,
        month = feb,
       volume = {527},
       number = {4},
        pages = {10038-10058},
          doi = {10.1093/mnras/stad3784},
archivePrefix = {arXiv},
       eprint = {2309.07912},
 primaryClass = {astro-ph.GA},
       adsurl = {https://ui.adsabs.harvard.edu/abs/2024MNRAS.52710038G}
}

@ARTICLE{HaardtMadau2012,
       author = {{Haardt}, Francesco and {Madau}, Piero},
        title = "{Radiative Transfer in a Clumpy Universe. IV. New Synthesis Models of the Cosmic UV/X-Ray Background}",
      journal = {\apj},
         year = 2012,
        month = feb,
       volume = {746},
       number = {2},
          eid = {125},
        pages = {125},
          doi = {10.1088/0004-637X/746/2/125},
       adsurl = {https://ui.adsabs.harvard.edu/abs/2012ApJ...746..125H}
}

@ARTICLE{Hahn2019,
       author = {{Hahn}, ChangHoon and {Beutler}, Florian and {Sinha}, Manodeep and {Berlind}, Andreas and {Ho}, Shirley and {Hogg}, David W.},
        title = "{Likelihood non-Gaussianity in large-scale structure analyses}",
      journal = {\mnras},
         year = 2019,
        month = may,
       volume = {485},
       number = {2},
        pages = {2956-2969},
          doi = {10.1093/mnras/stz558},
archivePrefix = {arXiv},
       eprint = {1803.06348},
 primaryClass = {astro-ph.CO},
       adsurl = {https://ui.adsabs.harvard.edu/abs/2019MNRAS.485.2956H}
}

@ARTICLE{Hahn2022,
       author = {{Hahn}, ChangHoon and {Melchior}, Peter},
        title = "{Accelerated Bayesian SED Modeling Using Amortized Neural Posterior Estimation}",
      journal = {\apj},
         year = 2022,
        month = oct,
       volume = {938},
       number = {1},
          eid = {11},
        pages = {11},
          doi = {10.3847/1538-4357/ac7b84},
archivePrefix = {arXiv},
       eprint = {2203.07391},
 primaryClass = {astro-ph.GA},
       adsurl = {https://ui.adsabs.harvard.edu/abs/2022ApJ...938...11H}
}

@ARTICLE{Heckman2014,
       author = {{Heckman}, Timothy M. and {Best}, Philip N.},
        title = "{The Coevolution of Galaxies and Supermassive Black Holes: Insights from Surveys of the Contemporary Universe}",
      journal = {\araa},
         year = 2014,
        month = aug,
       volume = {52},
        pages = {589-660},
          doi = {10.1146/annurev-astro-081913-035722},
archivePrefix = {arXiv},
       eprint = {1403.4620},
 primaryClass = {astro-ph.GA},
       adsurl = {https://ui.adsabs.harvard.edu/abs/2014ARA&A..52..589H}
}

@article{  
  Hermans2022,
  title={A crisis in simulation-based inference? beware, your posterior approximations can be unfaithful},
  author={Hermans, Joeri and Delaunoy, Arnaud and Rozet, Fran{\c{c}}ois and Wehenkel, Antoine and Louppe, Gilles},
  journal={Transactions on Machine Learning Research},
  year={2022},
  publisher={OpenReview, Amherst, United States-Massachusetts}
}

@ARTICLE{Hopkins2011,
       author = {{Hopkins}, Philip F. and {Quataert}, Eliot and {Murray}, Norman},
        title = "{Self-regulated star formation in galaxies via momentum input from massive stars}",
      journal = {\mnras},
         year = 2011,
        month = oct,
       volume = {417},
       number = {2},
        pages = {950-973},
          doi = {10.1111/j.1365-2966.2011.19306.x},
archivePrefix = {arXiv},
       eprint = {1101.4940},
 primaryClass = {astro-ph.CO},
       adsurl = {https://ui.adsabs.harvard.edu/abs/2011MNRAS.417..950H}
}

@ARTICLE{Hopkins2015,
       author = {{Hopkins}, Philip F.},
        title = "{A new class of accurate, mesh-free hydrodynamic simulation methods}",
      journal = {\mnras},
         year = 2015,
        month = jun,
       volume = {450},
       number = {1},
        pages = {53-110},
          doi = {10.1093/mnras/stv195},
archivePrefix = {arXiv},
       eprint = {1409.7395},
 primaryClass = {astro-ph.CO},
       adsurl = {https://ui.adsabs.harvard.edu/abs/2015MNRAS.450...53H}
}

@ARTICLE{Ho2024,
       author = {{Ho}, Matthew and {Bartlett}, Deaglan J. and {Chartier}, Nicolas and {Cuesta-Lazaro}, Carolina and {Ding}, Simon and {Lapel}, Axel and {Lemos}, Pablo and {Lovell}, Christopher C. and {Makinen}, T. Lucas and {Modi}, Chirag and {Pandya}, Viraj and {Pandey}, Shivam and {Perez}, Lucia A. and {Wandelt}, Benjamin and {Bryan}, Greg L.},
        title = "{LtU-ILI: An All-in-One Framework for Implicit Inference in Astrophysics and Cosmology}",
      journal = {The Open Journal of Astrophysics},
         year = 2024,
        month = jul,
       volume = {7},
          eid = {54},
        pages = {54},
          doi = {10.33232/001c.120559},
archivePrefix = {arXiv},
       eprint = {2402.05137},
 primaryClass = {astro-ph.IM},
       adsurl = {https://ui.adsabs.harvard.edu/abs/2024OJAp....7E..54H}
}

@ARTICLE{IglesiasNavarro2025,
       author = {{Iglesias-Navarro}, Patricia and {Huertas-Company}, Marc and {P{\'e}rez-Gonz{\'a}lez}, Pablo and {Knapen}, Johan H. and {Hahn}, ChangHoon and {Koekemoer}, Anton M. and {Finkelstein}, Steven L. and {Villanueva}, Natalia and {Asensio Ramos}, Andr{\'e}s},
        title = "{Simulation-based inference of galaxy properties from JWST pixels}",
      journal = {\aap},
         year = 2025,
        month = nov,
       volume = {703},
          eid = {A229},
        pages = {A229},
          doi = {10.1051/0004-6361/202555810},
archivePrefix = {arXiv},
       eprint = {2506.04336},
 primaryClass = {astro-ph.GA},
       adsurl = {https://ui.adsabs.harvard.edu/abs/2025A&A...703A.229I}
}

@ARTICLE{Irsic2017,
       author = {{Ir{\v{s}}i{\v{c}}}, Vid and {Viel}, Matteo and {Haehnelt}, Martin G. and {Bolton}, James S. and {Becker}, George D.},
        title = "{First Constraints on Fuzzy Dark Matter from Lyman-{\ensuremath{\alpha}} Forest Data and Hydrodynamical Simulations}",
      journal = {\prl},
         year = 2017,
        month = jul,
       volume = {119},
       number = {3},
          eid = {031302},
        pages = {031302},
          doi = {10.1103/PhysRevLett.119.031302},
archivePrefix = {arXiv},
       eprint = {1703.04683},
 primaryClass = {astro-ph.CO},
       adsurl = {https://ui.adsabs.harvard.edu/abs/2017PhRvL.119c1302I}
}

@ARTICLE{Inman2019,
       author = {{Inman}, Derek and {Ali-Ha{\"\i}moud}, Yacine},
        title = "{Early structure formation in primordial black hole cosmologies}",
      journal = {\prd},
         year = 2019,
        month = oct,
       volume = {100},
       number = {8},
          eid = {083528},
        pages = {083528},
          doi = {10.1103/PhysRevD.100.083528},
archivePrefix = {arXiv},
       eprint = {1907.08129},
 primaryClass = {astro-ph.CO},
       adsurl = {https://ui.adsabs.harvard.edu/abs/2019PhRvD.100h3528I}
}

@ARTICLE{Ivanov2025,
       author = {{Ivanov}, Mikhail M. and {Toomey}, Michael W. and {Kara{\c{c}}ayl{\i}}, Naim G{\"o}ksel},
        title = "{Fundamental Physics with the Lyman-Alpha Forest: Constraints on the Growth of Structure and Neutrino Masses from SDSS with Effective Field Theory}",
      journal = {\prl},
         year = 2025,
        month = mar,
       volume = {134},
       number = {9},
          eid = {091001},
        pages = {091001},
          doi = {10.1103/PhysRevLett.134.091001},
archivePrefix = {arXiv},
       eprint = {2405.13208},
 primaryClass = {astro-ph.CO},
       adsurl = {https://ui.adsabs.harvard.edu/abs/2025PhRvL.134i1001I}
}

@ARTICLE{JimenesRezende2015,
       author = {{Jimenez Rezende}, Danilo and {Mohamed}, Shakir},
        title = "{Variational Inference with Normalizing Flows}",
      journal = {arXiv e-prints},
         year = 2015,
        month = may,
          eid = {arXiv:1505.05770},
        pages = {arXiv:1505.05770},
          doi = {10.48550/arXiv.1505.05770},
archivePrefix = {arXiv},
       eprint = {1505.05770},
 primaryClass = {stat.ML},
       adsurl = {https://ui.adsabs.harvard.edu/abs/2015arXiv150505770J}
}

@ARTICLE{Karacayli2022,
       author = {{Kara{\c{c}}ayl{\i}}, Naim G{\"o}ksel and {Padmanabhan}, Nikhil and {Font-Ribera}, Andreu and {Ir{\v{s}}i{\v{c}}}, Vid and {Walther}, Michael and {Brooks}, David and {Gazta{\~n}aga}, Enrique and {Kehoe}, Robert and {Levi}, Michael and {Ntelis}, Pierros and {Palanque-Delabrouille}, Nathalie and {Tarl{\'e}}, Gregory},
        title = "{Optimal 1D Ly {\ensuremath{\alpha}} forest power spectrum estimation - II. KODIAQ, SQUAD, and XQ-100}",
      journal = {\mnras},
         year = 2022,
        month = jan,
       volume = {509},
       number = {2},
        pages = {2842-2855},
          doi = {10.1093/mnras/stab3201},
archivePrefix = {arXiv},
       eprint = {2108.10870},
 primaryClass = {astro-ph.CO},
       adsurl = {https://ui.adsabs.harvard.edu/abs/2022MNRAS.509.2842K}
}

@ARTICLE{Karacayli2025,
       author = {{Kara{\c{c}}ayl{\i}}, Naim G{\"o}ksel and {Martini}, Paul and {Aguilar}, J. and {Ahlen}, S. and {Armengaud}, E. and {Bailey}, S. and {Bault}, A. and {Bianchi}, D. and {Brodzeller}, A. and {Brooks}, D. and {Chaves-Montero}, J. and {Claybaugh}, T. and {Cuceu}, A. and {de la Macorra}, A. and {Dey}, A. and {Dey}, B. and {Doel}, P. and {Ferraro}, S. and {Font-Ribera}, A. and {Forero-Romero}, J.~E. and {Gazta{\~n}aga}, E. and {Gontcho}, S. Gontcho A. and {Gutierrez}, G. and {Guy}, J. and {Hahn}, C. and {Herrera-Alcantar}, H.~K. and {Honscheid}, K. and {Ishak}, M. and {Kehoe}, R. and {Kirkby}, D. and {Kremin}, A. and {Landriau}, M. and {Le Goff}, J.~M. and {Le Guillou}, L. and {Levi}, M.~E. and {Manera}, M. and {Meisner}, A. and {Miquel}, R. and {Montero-Camacho}, P. and {Nadathur}, S. and {Niz}, G. and {Palanque-Delabrouille}, N. and {Pan}, Z. and {Percival}, W.~J. and {Pieri}, Matthew M. and {Prada}, F. and {P{\'e}rez-R{\`a}fols}, I. and {Ravoux}, C. and {Rossi}, G. and {Sanchez}, E. and {Saulder}, C. and {Schlegel}, D. and {Schubnell}, M. and {Seo}, H. and {Siudek}, M. and {Sprayberry}, D. and {Tan}, T. and {Tang}, Ji-Jia and {Tarl{\'e}}, G. and {Walther}, M. and {Weaver}, B.~A. and {Yu}, J. and {Zhou}, R. and {Zou}, H.},
        title = "{DESI DR1 Ly{\ensuremath{\alpha}} 1D power spectrum: the optimal estimator measurement}",
      journal = {\jcap},
         year = 2025,
        month = oct,
       volume = {2025},
       number = {10},
          eid = {004},
        pages = {004},
          doi = {10.1088/1475-7516/2025/10/004},
archivePrefix = {arXiv},
       eprint = {2505.07974},
 primaryClass = {astro-ph.CO},
       adsurl = {https://ui.adsabs.harvard.edu/abs/2025JCAP...10..004K}
}

@ARTICLE{Katz1996,
       author = {{Katz}, Neal and {Weinberg}, David H. and {Hernquist}, Lars},
        title = "{Cosmological Simulations with TreeSPH}",
      journal = {\apjs},
         year = 1996,
        month = jul,
       volume = {105},
        pages = {19},
          doi = {10.1086/192305},
archivePrefix = {arXiv},
       eprint = {astro-ph/9509107},
 primaryClass = {astro-ph},
       adsurl = {https://ui.adsabs.harvard.edu/abs/1996ApJS..105...19K}
}

@inproceedings{Kingma2015,
  author = {Kingma, Diederik P. and Ba, Jimmy},
  booktitle = {ICLR (Poster)},
  editor = {Bengio, Yoshua and LeCun, Yann},
  ee = {http://arxiv.org/abs/1412.6980},
  title = {Adam: A Method for Stochastic Optimization.},
  year = 2015
}

@ARTICLE{Khaire2024,
       author = {{Khaire}, Vikram and {Hu}, Teng and {Hennawi}, Joseph F. and {Burchett}, Joseph N. and {Walther}, Michael and {Davies}, Frederick},
        title = "{Searching for the imprints of AGN feedback on the Lyman alpha forest around luminous red galaxies}",
      journal = {\mnras},
         year = 2024,
        month = oct,
       volume = {534},
       number = {1},
        pages = {465-484},
          doi = {10.1093/mnras/stae1981},
archivePrefix = {arXiv},
       eprint = {2311.08470},
 primaryClass = {astro-ph.GA},
       adsurl = {https://ui.adsabs.harvard.edu/abs/2024MNRAS.534..465K}
}

@ARTICLE{Krumholz2011,
       author = {{Krumholz}, Mark R. and {Gnedin}, Nickolay Y.},
        title = "{A Comparison of Methods for Determining the Molecular Content of Model Galaxies}",
      journal = {\apj},
         year = 2011,
        month = mar,
       volume = {729},
       number = {1},
          eid = {36},
        pages = {36},
          doi = {10.1088/0004-637X/729/1/36},
archivePrefix = {arXiv},
       eprint = {1011.4065},
 primaryClass = {astro-ph.CO},
       adsurl = {https://ui.adsabs.harvard.edu/abs/2011ApJ...729...36K}
}

@ARTICLE{Yeche2017,
       author = {{Y{\`e}che}, Christophe and {Palanque-Delabrouille}, Nathalie and {Baur}, Julien and {du Mas des Bourboux}, H{\'e}lion},
        title = "{Constraints on neutrino masses from Lyman-alpha forest power spectrum with BOSS and XQ-100}",
      journal = {\jcap},
         year = 2017,
        month = jun,
       volume = {2017},
       number = {6},
          eid = {047},
        pages = {047},
          doi = {10.1088/1475-7516/2017/06/047},
archivePrefix = {arXiv},
       eprint = {1702.03314},
 primaryClass = {astro-ph.CO},
       adsurl = {https://ui.adsabs.harvard.edu/abs/2017JCAP...06..047Y}
}

@article{LedoitWolf2004,
  title={A well-conditioned estimator for large-dimensional covariance matrices},
  author={Ledoit, Olivier and Wolf, Michael},
  journal={Journal of Multivariate Analysis},
  volume={88},
  number={2},
  pages={365--411},
  year={2004},
  publisher={Elsevier}
}

@ARTICLE{Lemos2023,
       author = {{Lemos}, Pablo and {Coogan}, Adam and {Hezaveh}, Yashar and {Perreault-Levasseur}, Laurence},
        title = "{Sampling-Based Accuracy Testing of Posterior Estimators for General Inference}",
      journal = {40th International Conference on Machine Learning},
         year = 2023,
        month = jan,
       volume = {202},
        pages = {19256-19273},
          doi = {10.48550/arXiv.2302.03026},
archivePrefix = {arXiv},
       eprint = {2302.03026},
 primaryClass = {stat.ML},
       adsurl = {https://ui.adsabs.harvard.edu/abs/2023PMLR..20219256L}
}

@ARTICLE{Lidz2010,
       author = {{Lidz}, Adam and {Faucher-Gigu{\`e}re}, Claude-Andr{\'e} and {Dall'Aglio}, Aldo and {McQuinn}, Matthew and {Fechner}, Cora and {Zaldarriaga}, Matias and {Hernquist}, Lars and {Dutta}, Suvendra},
        title = "{A Measurement of Small-scale Structure in the 2.2 <= z <= 4.2 Ly{\ensuremath{\alpha}} Forest}",
      journal = {\apj},
         year = 2010,
        month = jul,
       volume = {718},
       number = {1},
        pages = {199-230},
          doi = {10.1088/0004-637X/718/1/199},
archivePrefix = {arXiv},
       eprint = {0909.5210},
 primaryClass = {astro-ph.CO},
       adsurl = {https://ui.adsabs.harvard.edu/abs/2010ApJ...718..199L}
}

@ARTICLE{Lopez2016,
       author = {{L{\'o}pez}, S. and {D'Odorico}, V. and {Ellison}, S.~L. and {Becker}, G.~D. and {Christensen}, L. and {Cupani}, G. and {Denney}, K.~D. and {P{\^a}ris}, I. and {Worseck}, G. and {Berg}, T.~A.~M. and {Cristiani}, S. and {Dessauges-Zavadsky}, M. and {Haehnelt}, M. and {Hamann}, F. and {Hennawi}, J. and {Ir{\v{s}}i{\v{c}}}, V. and {Kim}, T.-S. and {L{\'o}pez}, P. and {Lund Saust}, R. and {M{\'e}nard}, B. and {Perrotta}, S. and {Prochaska}, J.~X. and {S{\'a}nchez-Ram{\'\i}rez}, R. and {Vestergaard}, M. and {Viel}, M. and {Wisotzki}, L.},
        title = "{XQ-100: A legacy survey of one hundred 3.5 {\ensuremath{\lesssim}} z {\ensuremath{\lesssim}} 4.5 quasars observed with VLT/X-shooter}",
      journal = {\aap},
         year = 2016,
        month = oct,
       volume = {594},
          eid = {A91},
        pages = {A91},
          doi = {10.1051/0004-6361/201628161},
archivePrefix = {arXiv},
       eprint = {1607.08776},
 primaryClass = {astro-ph.GA},
       adsurl = {https://ui.adsabs.harvard.edu/abs/2016A&A...594A..91L}
}

@ARTICLE{Lovell2025,
       author = {{Lovell}, Christopher C. and {Starkenburg}, Tjitske and {Ho}, Matthew and {Angl{\'e}s-Alc{\'a}zar}, Daniel and {Dav{\'e}}, Romeel and {Gabrielpillai}, Austen and {Iyer}, Kartheik G. and {Matthews}, Alice E. and {Roper}, William J. and {Somerville}, Rachel S. and {Sommovigo}, Laura and {Villaescusa-Navarro}, Francisco},
        title = "{Learning the Universe: cosmological and astrophysical parameter inference with galaxy luminosity functions and colours}",
      journal = {\mnras},
         year = 2025,
        month = dec,
       volume = {544},
       number = {4},
        pages = {3949-3979},
          doi = {10.1093/mnras/staf1888},
archivePrefix = {arXiv},
       eprint = {2411.13960},
 primaryClass = {astro-ph.GA},
       adsurl = {https://ui.adsabs.harvard.edu/abs/2025MNRAS.544.3949L}
}

@ARTICLE{McDonald2006,
       author = {{McDonald}, Patrick and {Seljak}, Uro{\v{s}} and {Burles}, Scott and {Schlegel}, David J. and {Weinberg}, David H. and {Cen}, Renyue and {Shih}, David and {Schaye}, Joop and {Schneider}, Donald P. and {Bahcall}, Neta A. and {Briggs}, John W. and {Brinkmann}, J. and {Brunner}, Robert J. and {Fukugita}, Masataka and {Gunn}, James E. and {Ivezi{\'c}}, {\v{Z}}eljko and {Kent}, Stephen and {Lupton}, Robert H. and {Vanden Berk}, Daniel E.},
        title = "{The Ly{\ensuremath{\alpha}} Forest Power Spectrum from the Sloan Digital Sky Survey}",
      journal = {\apjs},
         year = 2006,
        month = mar,
       volume = {163},
       number = {1},
        pages = {80-109},
          doi = {10.1086/444361},
archivePrefix = {arXiv},
       eprint = {astro-ph/0405013},
 primaryClass = {astro-ph},
       adsurl = {https://ui.adsabs.harvard.edu/abs/2006ApJS..163...80M}
}

@ARTICLE{McQuinn2016,
       author = {{McQuinn}, Matthew},
        title = "{The Evolution of the Intergalactic Medium}",
      journal = {\araa},
         year = 2016,
        month = sep,
       volume = {54},
        pages = {313-362},
          doi = {10.1146/annurev-astro-082214-122355},
archivePrefix = {arXiv},
       eprint = {1512.00086},
 primaryClass = {astro-ph.CO},
       adsurl = {https://ui.adsabs.harvard.edu/abs/2016ARA&A..54..313M}
}

@article{
 Miller2021,
  title={Truncated marginal neural ratio estimation},
  author={Miller, Benjamin K and Cole, Alex and Forr{\'e}, Patrick and Louppe, Gilles and Weniger, Christoph},
  journal={Advances in Neural Information Processing Systems},
  volume={34},
  pages={129--143},
  year={2021}
}

@ARTICLE{Murgia2018,
       author = {{Murgia}, Riccardo and {Ir{\v{s}}i{\v{c}}}, Vid and {Viel}, Matteo},
        title = "{Novel constraints on noncold, nonthermal dark matter from Lyman-{\ensuremath{\alpha}} forest data}",
      journal = {\prd},
         year = 2018,
        month = oct,
       volume = {98},
       number = {8},
          eid = {083540},
        pages = {083540},
          doi = {10.1103/PhysRevD.98.083540},
archivePrefix = {arXiv},
       eprint = {1806.08371},
 primaryClass = {astro-ph.CO},
       adsurl = {https://ui.adsabs.harvard.edu/abs/2018PhRvD..98h3540M}
}

@ARTICLE{Murphy2019,
       author = {{Murphy}, Michael T. and {Kacprzak}, Glenn G. and {Savorgnan}, Giulia A.~D. and {Carswell}, Robert F.},
        title = "{The UVES Spectral Quasar Absorption Database (SQUAD) data release 1: the first 10 million seconds}",
      journal = {\mnras},
         year = 2019,
        month = jan,
       volume = {482},
       number = {3},
        pages = {3458-3479},
          doi = {10.1093/mnras/sty2834},
archivePrefix = {arXiv},
       eprint = {1810.06136},
 primaryClass = {astro-ph.GA},
       adsurl = {https://ui.adsabs.harvard.edu/abs/2019MNRAS.482.3458M}
}

@ARTICLE{Nasir2016,
       author = {{Nasir}, Fahad and {Bolton}, James S. and {Becker}, George D.},
        title = "{Inferring the IGM thermal history during reionization with the Lyman {\ensuremath{\alpha}} forest power spectrum at redshift z = 5}",
      journal = {\mnras},
         year = 2016,
        month = dec,
       volume = {463},
       number = {3},
        pages = {2335-2347},
          doi = {10.1093/mnras/stw2147},
archivePrefix = {arXiv},
       eprint = {1605.04155},
 primaryClass = {astro-ph.CO},
       adsurl = {https://ui.adsabs.harvard.edu/abs/2016MNRAS.463.2335N}
}

@ARTICLE{Nasir2024,
       author = {{Nasir}, Fahad and {Gaikwad}, Prakash and {Davies}, Frederick B. and {Bolton}, James S. and {Puchwein}, Ewald and {Bosman}, Sarah E.~I.},
        title = "{Deep learning the intergalactic medium using Lyman-alpha forest at 4 {\ensuremath{\leq}} z {\ensuremath{\leq}} 5}",
      journal = {\mnras},
         year = 2024,
        month = oct,
       volume = {534},
       number = {2},
        pages = {1299-1316},
          doi = {10.1093/mnras/stae2153},
archivePrefix = {arXiv},
       eprint = {2404.05794},
 primaryClass = {astro-ph.CO},
       adsurl = {https://ui.adsabs.harvard.edu/abs/2024MNRAS.534.1299N}
}

@ARTICLE{Nelson2019,
       author = {{Nelson}, Dylan and {Springel}, Volker and {Pillepich}, Annalisa and {Rodriguez-Gomez}, Vicente and {Torrey}, Paul and {Genel}, Shy and {Vogelsberger}, Mark and {Pakmor}, Ruediger and {Marinacci}, Federico and {Weinberger}, Rainer and {Kelley}, Luke and {Lovell}, Mark and {Diemer}, Benedikt and {Hernquist}, Lars},
        title = "{The IllustrisTNG simulations: public data release}",
      journal = {Computational Astrophysics and Cosmology},
         year = 2019,
        month = may,
       volume = {6},
       number = {1},
          eid = {2},
        pages = {2},
          doi = {10.1186/s40668-019-0028-x},
archivePrefix = {arXiv},
       eprint = {1812.05609},
 primaryClass = {astro-ph.GA},
       adsurl = {https://ui.adsabs.harvard.edu/abs/2019ComAC...6....2N}
}

@ARTICLE{Ntampaka2025,
       author = {{Ntampaka}, Michelle and {Ciprijanovic}, A. and {Delgado}, Ana Maria and {Soltis}, John and {Wu}, John F. and {Yunus}, Mikaeel and {ZuHone}, John},
        title = "{The Importance of Being Adaptable: An Exploration of the Power and Limitations of Domain Adaptation for Simulation-Based Inference with Galaxy Clusters}",
      journal = {arXiv e-prints},
         year = 2025,
        month = oct,
          eid = {arXiv:2510.09748},
        pages = {arXiv:2510.09748},
          doi = {10.48550/arXiv.2510.09748},
archivePrefix = {arXiv},
       eprint = {2510.09748},
 primaryClass = {astro-ph.IM},
       adsurl = {https://ui.adsabs.harvard.edu/abs/2025arXiv251009748N}
}

@ARTICLE{Omeara2017,
       author = {{O'Meara}, J.~M. and {Lehner}, N. and {Howk}, J.~C. and {Prochaska}, J.~X. and {Fox}, A.~J. and {Peeples}, M.~S. and {Tumlinson}, J. and {O'Shea}, B.~W.},
        title = "{The Second Data Release of the KODIAQ Survey}",
      journal = {\aj},
         year = 2017,
        month = sep,
       volume = {154},
       number = {3},
          eid = {114},
        pages = {114},
          doi = {10.3847/1538-3881/aa82b8},
archivePrefix = {arXiv},
       eprint = {1707.07905},
 primaryClass = {astro-ph.GA},
       adsurl = {https://ui.adsabs.harvard.edu/abs/2017AJ....154..114O}
}

@ARTICLE{Palanque2013,
       author = {{Palanque-Delabrouille}, Nathalie and {Y{\`e}che}, Christophe and {Borde}, Arnaud and {Le Goff}, Jean-Marc and {Rossi}, Graziano and {Viel}, Matteo and {Aubourg}, {\'E}ric and {Bailey}, Stephen and {Bautista}, Julian and {Blomqvist}, Michael and {Bolton}, Adam and {Bolton}, James S. and {Busca}, Nicol{\'a}s G. and {Carithers}, Bill and {Croft}, Rupert A.~C. and {Dawson}, Kyle S. and {Delubac}, Timoth{\'e}e and {Font-Ribera}, Andreu and {Ho}, Shirley and {Kirkby}, David and {Lee}, Khee-Gan and {Margala}, Daniel and {Miralda-Escud{\'e}}, Jordi and {Muna}, Demitri and {Myers}, Adam D. and {Noterdaeme}, Pasquier and {P{\^a}ris}, Isabelle and {Petitjean}, Patrick and {Pieri}, Matthew M. and {Rich}, James and {Rollinde}, Emmanuel and {Ross}, Nicholas P. and {Schlegel}, David J. and {Schneider}, Donald P. and {Slosar}, An{\v{z}}e and {Weinberg}, David H.},
        title = "{The one-dimensional Ly{\ensuremath{\alpha}} forest power spectrum from BOSS}",
      journal = {\aap},
         year = 2013,
        month = nov,
       volume = {559},
          eid = {A85},
        pages = {A85},
          doi = {10.1051/0004-6361/201322130},
archivePrefix = {arXiv},
       eprint = {1306.5896},
 primaryClass = {astro-ph.CO},
       adsurl = {https://ui.adsabs.harvard.edu/abs/2013A&A...559A..85P}
}

@ARTICLE{Palanque2015,
       author = {{Palanque-Delabrouille}, Nathalie and {Y{\`e}che}, Christophe and {Lesgourgues}, Julien and {Rossi}, Graziano and {Borde}, Arnaud and {Viel}, Matteo and {Aubourg}, Eric and {Kirkby}, David and {LeGoff}, Jean-Marc and {Rich}, James and {Roe}, Natalie and {Ross}, Nicholas P. and {Schneider}, Donald P. and {Weinberg}, David},
        title = "{Constraint on neutrino masses from SDSS-III/BOSS Ly{\ensuremath{\alpha}} forest and other cosmological probes}",
      journal = {\jcap},
         year = 2015,
        month = feb,
       volume = {2015},
       number = {2},
        pages = {045-045},
          doi = {10.1088/1475-7516/2015/02/045},
archivePrefix = {arXiv},
       eprint = {1410.7244},
 primaryClass = {astro-ph.CO},
       adsurl = {https://ui.adsabs.harvard.edu/abs/2015JCAP...02..045P}
}

@ARTICLE{Palanque2015b,
       author = {{Palanque-Delabrouille}, Nathalie and {Y{\`e}che}, Christophe and {Baur}, Julien and {Magneville}, Christophe and {Rossi}, Graziano and {Lesgourgues}, Julien and {Borde}, Arnaud and {Burtin}, Etienne and {LeGoff}, Jean-Marc and {Rich}, James and {Viel}, Matteo and {Weinberg}, David},
        title = "{Neutrino masses and cosmology with Lyman-alpha forest power spectrum}",
      journal = {\jcap},
         year = 2015,
        month = nov,
       volume = {2015},
       number = {11},
        pages = {011-011},
          doi = {10.1088/1475-7516/2015/11/011},
archivePrefix = {arXiv},
       eprint = {1506.05976},
 primaryClass = {astro-ph.CO},
       adsurl = {https://ui.adsabs.harvard.edu/abs/2015JCAP...11..011P}
}

@ARTICLE{Palanque2020,
       author = {{Palanque-Delabrouille}, Nathalie and {Y{\`e}che}, Christophe and {Sch{\"o}neberg}, Nils and {Lesgourgues}, Julien and {Walther}, Michael and {Chabanier}, Sol{\`e}ne and {Armengaud}, Eric},
        title = "{Hints, neutrino bounds, and WDM constraints from SDSS DR14 Lyman-{\ensuremath{\alpha}} and Planck full-survey data}",
      journal = {\jcap},
         year = 2020,
        month = apr,
       volume = {2020},
       number = {4},
          eid = {038},
        pages = {038},
          doi = {10.1088/1475-7516/2020/04/038},
archivePrefix = {arXiv},
       eprint = {1911.09073},
 primaryClass = {astro-ph.CO},
       adsurl = {https://ui.adsabs.harvard.edu/abs/2020JCAP...04..038P}
}

@ARTICLE{Pederson2021,
       author = {{Pedersen}, Christian and {Font-Ribera}, Andreu and {Rogers}, Keir K. and {McDonald}, Patrick and {Peiris}, Hiranya V. and {Pontzen}, Andrew and {Slosar}, An{\v{z}}e},
        title = "{An emulator for the Lyman-{\ensuremath{\alpha}} forest in beyond-{\ensuremath{\Lambda}}CDM cosmologies}",
      journal = {\jcap},
         year = 2021,
        month = may,
       volume = {2021},
       number = {5},
          eid = {033},
        pages = {033},
          doi = {10.1088/1475-7516/2021/05/033},
archivePrefix = {arXiv},
       eprint = {2011.15127},
 primaryClass = {astro-ph.CO},
       adsurl = {https://ui.adsabs.harvard.edu/abs/2021JCAP...05..033P}
}

@ARTICLE{Pedersen2023,
       author = {{Pedersen}, Christian and {Font-Ribera}, Andreu and {Gnedin}, Nickolay Y.},
        title = "{Compressing the Cosmological Information in One-dimensional Correlations of the Lyman-{\ensuremath{\alpha}} Forest}",
      journal = {\apj},
         year = 2023,
        month = feb,
       volume = {944},
       number = {2},
          eid = {223},
        pages = {223},
          doi = {10.3847/1538-4357/acb433},
archivePrefix = {arXiv},
       eprint = {2209.09895},
 primaryClass = {astro-ph.CO},
       adsurl = {https://ui.adsabs.harvard.edu/abs/2023ApJ...944..223P}
}

@ARTICLE{Papamakarios2016,
       author = {{Papamakarios}, George and {Murray}, Iain},
        title = "{Fast $\epsilon$-free Inference of Simulation Models with Bayesian Conditional Density Estimation}",
      journal = {arXiv e-prints},
         year = 2016,
        month = may,
          eid = {arXiv:1605.06376},
        pages = {arXiv:1605.06376},
          doi = {10.48550/arXiv.1605.06376},
archivePrefix = {arXiv},
       eprint = {1605.06376},
 primaryClass = {stat.ML},
       adsurl = {https://ui.adsabs.harvard.edu/abs/2016arXiv160506376P}
}

@ARTICLE{Papamakarios2017,
       author = {{Papamakarios}, George and {Pavlakou}, Theo and {Murray}, Iain},
        title = "{Masked Autoregressive Flow for Density Estimation}",
      journal = {arXiv e-prints},
         year = 2017,
        month = may,
          eid = {arXiv:1705.07057},
        pages = {arXiv:1705.07057},
          doi = {10.48550/arXiv.1705.07057},
archivePrefix = {arXiv},
       eprint = {1705.07057},
 primaryClass = {stat.ML},
       adsurl = {https://ui.adsabs.harvard.edu/abs/2017arXiv170507057P}
}

@ARTICLE{Paramakarios2019,
       author = {{Papamakarios}, George},
        title = "{Neural Density Estimation and Likelihood-free Inference}",
      journal = {arXiv e-prints},
         year = 2019,
        month = oct,
          eid = {arXiv:1910.13233},
        pages = {arXiv:1910.13233},
          doi = {10.48550/arXiv.1910.13233},
archivePrefix = {arXiv},
       eprint = {1910.13233},
 primaryClass = {stat.ML},
       adsurl = {https://ui.adsabs.harvard.edu/abs/2019arXiv191013233P}
}

@ARTICLE{Pavicevic2025,
       author = {{Pavi{\v{c}}evi{\'c}}, Mak and {Ir{\v{s}}i{\v{c}}}, Vid and {Viel}, Matteo and {Bolton}, James S. and {Haehnelt}, Martin G. and {Martin-Alvarez}, Sergio and {Puchwein}, Ewald and {Ralegankar}, Pranjal},
        title = "{Constraints on Primordial Magnetic Fields from the Lyman-{\ensuremath{\alpha}} Forest}",
      journal = {\prl},
         year = 2025,
        month = aug,
       volume = {135},
       number = {7},
          eid = {071001},
        pages = {071001},
          doi = {10.1103/77rd-vkpz},
archivePrefix = {arXiv},
       eprint = {2501.06299},
 primaryClass = {astro-ph.CO},
       adsurl = {https://ui.adsabs.harvard.edu/abs/2025PhRvL.135g1001P}
}

@ARTICLE{Pedersen2020,
       author = {{Pedersen}, Christian and {Font-Ribera}, Andreu and {Kitching}, Thomas D. and {McDonald}, Patrick and {Bird}, Simeon and {Slosar}, An{\v{z}}e and {Rogers}, Keir K. and {Pontzen}, Andrew},
        title = "{Massive neutrinos and degeneracies in Lyman-alpha forest simulations}",
      journal = {\jcap},
         year = 2020,
        month = apr,
       volume = {2020},
       number = {4},
          eid = {025},
        pages = {025},
          doi = {10.1088/1475-7516/2020/04/025},
archivePrefix = {arXiv},
       eprint = {1911.09596},
 primaryClass = {astro-ph.CO},
       adsurl = {https://ui.adsabs.harvard.edu/abs/2020JCAP...04..025P}
}

@ARTICLE{Pillepich2018,
       author = {{Pillepich}, Annalisa and {Springel}, Volker and {Nelson}, Dylan and {Genel}, Shy and {Naiman}, Jill and {Pakmor}, R{\"u}diger and {Hernquist}, Lars and {Torrey}, Paul and {Vogelsberger}, Mark and {Weinberger}, Rainer and {Marinacci}, Federico},
        title = "{Simulating galaxy formation with the IllustrisTNG model}",
      journal = {\mnras},
         year = 2018,
        month = jan,
       volume = {473},
       number = {3},
        pages = {4077-4106},
          doi = {10.1093/mnras/stx2656},
archivePrefix = {arXiv},
       eprint = {1703.02970},
 primaryClass = {astro-ph.GA},
       adsurl = {https://ui.adsabs.harvard.edu/abs/2018MNRAS.473.4077P}
}

@ARTICLE{Puchwein2023,
       author = {{Puchwein}, Ewald and {Bolton}, James S. and {Keating}, Laura C. and {Molaro}, Margherita and {Gaikwad}, Prakash and {Kulkarni}, Girish and {Haehnelt}, Martin G. and {Ir{\v{s}}i{\v{c}}}, Vid and {{\v{S}}oltinsk{\'y}}, Tom{\'a}{\v{s}} and {Viel}, Matteo and {Aubert}, Dominique and {Becker}, George D. and {Meiksin}, Avery},
        title = "{The Sherwood-Relics simulations: overview and impact of patchy reionization and pressure smoothing on the intergalactic medium}",
      journal = {\mnras},
         year = 2023,
        month = mar,
       volume = {519},
       number = {4},
        pages = {6162-6183},
          doi = {10.1093/mnras/stac3761},
archivePrefix = {arXiv},
       eprint = {2207.13098},
 primaryClass = {astro-ph.CO},
       adsurl = {https://ui.adsabs.harvard.edu/abs/2023MNRAS.519.6162P}
}

@ARTICLE{Rahmati2013,
       author = {{Rahmati}, Alireza and {Pawlik}, Andreas H. and {Rai{\v{c}}evi{\'c}}, Milan and {Schaye}, Joop},
        title = "{On the evolution of the H I column density distribution in cosmological simulations}",
      journal = {\mnras},
         year = 2013,
        month = apr,
       volume = {430},
       number = {3},
        pages = {2427-2445},
          doi = {10.1093/mnras/stt066},
archivePrefix = {arXiv},
       eprint = {1210.7808},
 primaryClass = {astro-ph.CO},
       adsurl = {https://ui.adsabs.harvard.edu/abs/2013MNRAS.430.2427R}
}

@ARTICLE{Ravoux2025,
       author = {{Ravoux}, Corentin and {Abdul-Karim}, Marie-Lynn and {Le Goff}, Jean-Marc and {Armengaud}, Eric and {Aguilar}, Jessica N. and {Ahlen}, Steven and {Bailey}, Stephen and {Bianchi}, Davide and {Brodzeller}, Allyson and {Brooks}, David and {Chaves-Montero}, Jon{\'a}s and {Claybaugh}, Todd and {Cuceu}, Andrei and {de Belsunce}, Roger and {de la Macorra}, Axel and {Dey}, Arjun and {Ding}, Zhejie and {Doel}, Peter and {Ferraro}, Simone and {Font-Ribera}, Andreu and {Forero-Romero}, Jaime E. and {Gazta{\~n}aga}, Enrique and {Kara{\c{c}}ayl{\i}}, Naim G{\"o}ksel and {Gontcho}, Satya Gontcho A. and {Gutierrez}, Gaston and {Guy}, Julien and {Herrera-Alcantar}, Hiram K. and {Ishak}, Mustapha and {Kehoe}, Robert and {Kirkby}, David and {Kisner}, Theodore and {Kremin}, Anthony and {Landriau}, Martin and {Le Guillou}, Laurent and {Levi}, Michael E. and {Manera}, Marc and {Martini}, Paul and {Meisner}, Aaron and {Miquel}, Ramon and {Montero-Camacho}, Paulo and {Mu{\~n}oz-Guti{\'e}rrez}, Andrea and {Nadathur}, Seshadri and {Niz}, Gustavo and {Palanque-Delabrouille}, Nathalie and {Pan}, Zhiwei and {Percival}, Will J. and {P{\'e}rez-R{\`a}fols}, Ignasi and {Pieri}, Matthew M. and {Prada}, Francisco and {Rossi}, Graziano and {Sanchez}, Eusebio and {Saulder}, Christoph and {Schlegel}, David and {Schubnell}, Michael and {Seo}, Hee-Jong and {Silber}, Joseph H. and {Siudek}, Ma{\l}gorzata and {Sprayberry}, David and {Tan}, Ting and {Tang}, Ji-Jia and {Tarl{\'e}}, Gregory and {Walther}, Michael and {Weaver}, Benjamin A. and {Y{\`e}che}, Christophe and {Yu}, Jiaxi and {Zhou}, Rongpu and {Zou}, Hu},
        title = "{DESI DR1 Ly{\ensuremath{\alpha}} 1D power spectrum: the Fast Fourier Transform estimator measurement}",
      journal = {\jcap},
         year = 2025,
        month = nov,
       volume = {2025},
       number = {11},
          eid = {079},
        pages = {079},
          doi = {10.1088/1475-7516/2025/11/079},
archivePrefix = {arXiv},
       eprint = {2505.09493},
 primaryClass = {astro-ph.CO},
       adsurl = {https://ui.adsabs.harvard.edu/abs/2025JCAP...11..079R}
}

@ARTICLE{Roncoli2023,
       author = {{Roncoli}, Andrea and {{\'C}iprijanovi{\'c}}, Aleksandra and {Voetberg}, Maggie and {Villaescusa-Navarro}, Francisco and {Nord}, Brian},
        title = "{Domain Adaptive Graph Neural Networks for Constraining Cosmological Parameters Across Multiple Data Sets}",
      journal = {arXiv e-prints},
         year = 2023,
        month = nov,
          eid = {arXiv:2311.01588},
        pages = {arXiv:2311.01588},
          doi = {10.48550/arXiv.2311.01588},
archivePrefix = {arXiv},
       eprint = {2311.01588},
 primaryClass = {astro-ph.CO},
       adsurl = {https://ui.adsabs.harvard.edu/abs/2023arXiv231101588R}
}

@ARTICLE{Rose2025,
       author = {{Rose}, Jonah C. and {Torrey}, Paul and {Villaescusa-Navarro}, Francisco and {Lisanti}, Mariangela and {Nguyen}, Tri and {Roy}, Sandip and {Kollmann}, Kassidy E. and {Vogelsberger}, Mark and {Cyr-Racine}, Francis-Yan and {Medvedev}, Mikhail V. and {Genel}, Shy and {Angl{\'e}s-Alc{\'a}zar}, Daniel and {Kallivayalil}, Nitya and {Wang}, Bonny Y. and {Costanza}, Bel{\'e}n and {O'Neil}, Stephanie and {Roche}, Cian and {Karmakar}, Soumyodipta and {Garcia}, Alex M. and {Low}, Ryan and {Lin}, Shurui and {Mostow}, Olivia and {Cruz}, Akaxia and {Caputo}, Andrea and {Farahi}, Arya and {Mu{\~n}oz}, Julian B. and {Necib}, Lina and {Teyssier}, Romain and {Dalcanton}, Julianne J. and {Spergel}, David},
        title = "{Introducing the DREAMS Project: DaRk mattEr and Astrophysics with Machine Learning and Simulations}",
      journal = {\apj},
         year = 2025,
        month = apr,
       volume = {982},
       number = {2},
          eid = {68},
        pages = {68},
          doi = {10.3847/1538-4357/adb8e5},
archivePrefix = {arXiv},
       eprint = {2405.00766},
 primaryClass = {astro-ph.GA},
       adsurl = {https://ui.adsabs.harvard.edu/abs/2025ApJ...982...68R}
}

@ARTICLE{Rossi2017,
       author = {{Rossi}, Graziano},
        title = "{Impact of Massive Neutrinos and Dark Radiation on the High-redshift Cosmic Web. I. Ly{\ensuremath{\alpha}} Forest Observables}",
      journal = {\apjs},
         year = 2017,
        month = nov,
       volume = {233},
       number = {1},
          eid = {12},
        pages = {12},
          doi = {10.3847/1538-4365/aa93d6},
archivePrefix = {arXiv},
       eprint = {1712.00230},
 primaryClass = {astro-ph.CO},
       adsurl = {https://ui.adsabs.harvard.edu/abs/2017ApJS..233...12R}
}

@ARTICLE{Smith2017,
       author = {{Smith}, Britton D. and {Bryan}, Greg L. and {Glover}, Simon C.~O. and {Goldbaum}, Nathan J. and {Turk}, Matthew J. and {Regan}, John and {Wise}, John H. and {Schive}, Hsi-Yu and {Abel}, Tom and {Emerick}, Andrew and {O'Shea}, Brian W. and {Anninos}, Peter and {Hummels}, Cameron B. and {Khochfar}, Sadegh},
        title = "{GRACKLE: a chemistry and cooling library for astrophysics}",
      journal = {\mnras},
         year = 2017,
        month = apr,
       volume = {466},
       number = {2},
        pages = {2217-2234},
          doi = {10.1093/mnras/stw3291},
archivePrefix = {arXiv},
       eprint = {1610.09591},
 primaryClass = {astro-ph.CO},
       adsurl = {https://ui.adsabs.harvard.edu/abs/2017MNRAS.466.2217S}
}

@ARTICLE{Springel2010,
       author = {{Springel}, Volker},
        title = "{E pur si muove: Galilean-invariant cosmological hydrodynamical simulations on a moving mesh}",
      journal = {\mnras},
         year = 2010,
        month = jan,
       volume = {401},
       number = {2},
        pages = {791-851},
          doi = {10.1111/j.1365-2966.2009.15715.x},
archivePrefix = {arXiv},
       eprint = {0901.4107},
 primaryClass = {astro-ph.CO},
       adsurl = {https://ui.adsabs.harvard.edu/abs/2010MNRAS.401..791S}
}

@ARTICLE{SpringelHernquist2003,
       author = {{Springel}, Volker and {Hernquist}, Lars},
        title = "{Cosmological smoothed particle hydrodynamics simulations: a hybrid multiphase model for star formation}",
      journal = {\mnras},
         year = 2003,
        month = feb,
       volume = {339},
       number = {2},
        pages = {289-311},
          doi = {10.1046/j.1365-8711.2003.06206.x},
archivePrefix = {arXiv},
       eprint = {astro-ph/0206393},
 primaryClass = {astro-ph},
       adsurl = {https://ui.adsabs.harvard.edu/abs/2003MNRAS.339..289S}
}

@ARTICLE{Springel2005,
       author = {{Springel}, Volker},
        title = "{The cosmological simulation code GADGET-2}",
      journal = {\mnras},
         year = 2005,
        month = dec,
       volume = {364},
       number = {4},
        pages = {1105-1134},
          doi = {10.1111/j.1365-2966.2005.09655.x},
archivePrefix = {arXiv},
       eprint = {astro-ph/0505010},
 primaryClass = {astro-ph},
       adsurl = {https://ui.adsabs.harvard.edu/abs/2005MNRAS.364.1105S}
}

@article{Tabak2010,
author = {Tabak, Esteban and Vanden-Eijnden, Eric},
year = {2010},
month = {03},
pages = {},
title = {Density estimation by dual ascent of the log-likelihood},
volume = {8},
journal = {Communications in Mathematical Sciences - COMMUN MATH SCI},
doi = {10.4310/CMS.2010.v8.n1.a11}
}

@article{Tabak2013,
author = {Tabak, E. G. and Turner, Cristina V.},
title = {A Family of Nonparametric Density Estimation Algorithms},
journal = {Communications on Pure and Applied Mathematics},
volume = {66},
number = {2},
pages = {145-164},
doi = {https://doi.org/10.1002/cpa.21423},
url = {https://onlinelibrary.wiley.com/doi/abs/10.1002/cpa.21423},
eprint = {https://onlinelibrary.wiley.com/doi/pdf/10.1002/cpa.21423},
year = {2013}
}

@article{TejeroCantero2020,
  doi = {10.21105/joss.02505},
  url = {https://doi.org/10.21105/joss.02505},
  year = {2020},
  publisher = {The Open Journal},
  volume = {5},
  number = {52},
  pages = {2505},
  author = {Alvaro Tejero-Cantero and Jan Boelts and Michael Deistler and Jan-Matthis Lueckmann and Conor Durkan and Pedro J. Gonçalves and David S. Greenberg and Jakob H. Macke},
  title = {sbi: A toolkit for simulation-based inference},
  journal = {Journal of Open Source Software}
}

@ARTICLE{Tillman2025,
       author = {{Tillman}, Megan Taylor and {Burkhart}, Blakesley and {Tonnesen}, Stephanie and {Bird}, Simeon and {Bryan}, Greg L.},
        title = "{The Effects of Active Galactic Nuclei Feedback on the Ly{\ensuremath{\alpha}} Forest Flux Power Spectrum}",
      journal = {\apj},
         year = 2025,
        month = feb,
       volume = {980},
       number = {1},
          eid = {72},
        pages = {72},
          doi = {10.3847/1538-4357/ada5f7},
archivePrefix = {arXiv},
       eprint = {2410.05383},
 primaryClass = {astro-ph.CO},
       adsurl = {https://ui.adsabs.harvard.edu/abs/2025ApJ...980...72T}
}

@ARTICLE{Torrey2014,
       author = {{Torrey}, Paul and {Vogelsberger}, Mark and {Genel}, Shy and {Sijacki}, Debora and {Springel}, Volker and {Hernquist}, Lars},
        title = "{A model for cosmological simulations of galaxy formation physics: multi-epoch validation}",
      journal = {\mnras},
         year = 2014,
        month = mar,
       volume = {438},
       number = {3},
        pages = {1985-2004},
          doi = {10.1093/mnras/stt2295},
archivePrefix = {arXiv},
       eprint = {1305.4931},
 primaryClass = {astro-ph.CO},
       adsurl = {https://ui.adsabs.harvard.edu/abs/2014MNRAS.438.1985T}
}

@ARTICLE{Uria2016,
       author = {{Uria}, Benigno and {C{\^o}t{\'e}}, Marc-Alexandre and {Gregor}, Karol and {Murray}, Iain and {Larochelle}, Hugo},
        title = "{Neural Autoregressive Distribution Estimation}",
      journal = {arXiv e-prints},
         year = 2016,
        month = may,
          eid = {arXiv:1605.02226},
        pages = {arXiv:1605.02226},
          doi = {10.48550/arXiv.1605.02226},
archivePrefix = {arXiv},
       eprint = {1605.02226},
 primaryClass = {cs.LG},
       adsurl = {https://ui.adsabs.harvard.edu/abs/2016arXiv160502226U}
}

@ARTICLE{Viel2004,
       author = {{Viel}, Matteo and {Haehnelt}, Martin G. and {Springel}, Volker},
        title = "{Inferring the dark matter power spectrum from the Lyman {\ensuremath{\alpha}} forest in high-resolution QSO absorption spectra}",
      journal = {\mnras},
         year = 2004,
        month = nov,
       volume = {354},
       number = {3},
        pages = {684-694},
          doi = {10.1111/j.1365-2966.2004.08224.x},
archivePrefix = {arXiv},
       eprint = {astro-ph/0404600},
 primaryClass = {astro-ph},
       adsurl = {https://ui.adsabs.harvard.edu/abs/2004MNRAS.354..684V}
}

@ARTICLE{Viel2005,
       author = {{Viel}, Matteo and {Lesgourgues}, Julien and {Haehnelt}, Martin G. and {Matarrese}, Sabino and {Riotto}, Antonio},
        title = "{Constraining warm dark matter candidates including sterile neutrinos and light gravitinos with WMAP and the Lyman-{\ensuremath{\alpha}} forest}",
      journal = {\prd},
         year = 2005,
        month = mar,
       volume = {71},
       number = {6},
          eid = {063534},
        pages = {063534},
          doi = {10.1103/PhysRevD.71.063534},
archivePrefix = {arXiv},
       eprint = {astro-ph/0501562},
 primaryClass = {astro-ph},
       adsurl = {https://ui.adsabs.harvard.edu/abs/2005PhRvD..71f3534V}
}

@ARTICLE{Viel2006,
       author = {{Viel}, Matteo and {Haehnelt}, Martin G.},
        title = "{Cosmological and astrophysical parameters from the Sloan Digital Sky Survey flux power spectrum and hydrodynamical simulations of the Lyman {\ensuremath{\alpha}} forest}",
      journal = {\mnras},
         year = 2006,
        month = jan,
       volume = {365},
       number = {1},
        pages = {231-244},
          doi = {10.1111/j.1365-2966.2005.09703.x},
archivePrefix = {arXiv},
       eprint = {astro-ph/0508177},
 primaryClass = {astro-ph},
       adsurl = {https://ui.adsabs.harvard.edu/abs/2006MNRAS.365..231V}
}

@ARTICLE{Viel2013,
       author = {{Viel}, Matteo and {Becker}, George D. and {Bolton}, James S. and {Haehnelt}, Martin G.},
        title = "{Warm dark matter as a solution to the small scale crisis: New constraints from high redshift Lyman-{\ensuremath{\alpha}} forest data}",
      journal = {\prd},
         year = 2013,
        month = aug,
       volume = {88},
       number = {4},
          eid = {043502},
        pages = {043502},
          doi = {10.1103/PhysRevD.88.043502},
archivePrefix = {arXiv},
       eprint = {1306.2314},
 primaryClass = {astro-ph.CO},
       adsurl = {https://ui.adsabs.harvard.edu/abs/2013PhRvD..88d3502V}
}

@ARTICLE{Viel2013b,
       author = {{Viel}, Matteo and {Schaye}, Joop and {Booth}, C.~M.},
        title = "{The impact of feedback from galaxy formation on the Lyman {\ensuremath{\alpha}} transmitted flux}",
      journal = {\mnras},
         year = 2013,
        month = feb,
       volume = {429},
       number = {2},
        pages = {1734-1746},
          doi = {10.1093/mnras/sts465},
archivePrefix = {arXiv},
       eprint = {1207.6567},
 primaryClass = {astro-ph.CO},
       adsurl = {https://ui.adsabs.harvard.edu/abs/2013MNRAS.429.1734V}
}

@ARTICLE{Villasenor2022,
       author = {{Villasenor}, Bruno and {Robertson}, Brant and {Madau}, Piero and {Schneider}, Evan},
        title = "{Inferring the Thermal History of the Intergalactic Medium from the Properties of the Hydrogen and Helium Ly{\ensuremath{\alpha}} Forest}",
      journal = {\apj},
         year = 2022,
        month = jul,
       volume = {933},
       number = {1},
          eid = {59},
        pages = {59},
          doi = {10.3847/1538-4357/ac704e},
archivePrefix = {arXiv},
       eprint = {2111.00019},
 primaryClass = {astro-ph.CO},
       adsurl = {https://ui.adsabs.harvard.edu/abs/2022ApJ...933...59V}
}

@ARTICLE{Villasenor2023,
       author = {{Villasenor}, Bruno and {Robertson}, Brant and {Madau}, Piero and {Schneider}, Evan},
        title = "{New constraints on warm dark matter from the Lyman-{\ensuremath{\alpha}} forest power spectrum}",
      journal = {\prd},
         year = 2023,
        month = jul,
       volume = {108},
       number = {2},
          eid = {023502},
        pages = {023502},
          doi = {10.1103/PhysRevD.108.023502},
archivePrefix = {arXiv},
       eprint = {2209.14220},
 primaryClass = {astro-ph.CO},
       adsurl = {https://ui.adsabs.harvard.edu/abs/2023PhRvD.108b3502V}
}

@ARTICLE{VillaescusaNavarro2021,
       author = {{Villaescusa-Navarro}, Francisco and {Angl{\'e}s-Alc{\'a}zar}, Daniel and {Genel}, Shy and {Spergel}, David N. and {Somerville}, Rachel S. and {Dave}, Romeel and {Pillepich}, Annalisa and {Hernquist}, Lars and {Nelson}, Dylan and {Torrey}, Paul and {Narayanan}, Desika and {Li}, Yin and {Philcox}, Oliver and {La Torre}, Valentina and {Maria Delgado}, Ana and {Ho}, Shirley and {Hassan}, Sultan and {Burkhart}, Blakesley and {Wadekar}, Digvijay and {Battaglia}, Nicholas and {Contardo}, Gabriella and {Bryan}, Greg L.},
        title = "{The CAMELS Project: Cosmology and Astrophysics with Machine-learning Simulations}",
      journal = {\apj},
         year = 2021,
        month = jul,
       volume = {915},
       number = {1},
          eid = {71},
        pages = {71},
          doi = {10.3847/1538-4357/abf7ba},
archivePrefix = {arXiv},
       eprint = {2010.00619},
 primaryClass = {astro-ph.CO},
       adsurl = {https://ui.adsabs.harvard.edu/abs/2021ApJ...915...71V}
}

@ARTICLE{CAMELS_DR1,
author = {{Villaescusa-Navarro}, Francisco and {Genel}, Shy and {Angl{\'e}s-Alc{\'a}zar}, Daniel and {Perez}, Lucia A. and {Villanueva-Domingo}, Pablo and {Wadekar}, Digvijay and {Shao}, Helen and {Mohammad}, Faizan G. and {Hassan}, Sultan and {Moser}, Emily and {Lau}, Erwin T. and {Machado Poletti Valle}, Luis Fernando and {Nicola}, Andrina and {Thiele}, Leander and {Jo}, Yongseok and {Philcox}, Oliver H.~E. and {Oppenheimer}, Benjamin D. and {Tillman}, Megan and {Hahn}, ChangHoon and {Kaushal}, Neerav and {Pisani}, Alice and {Gebhardt}, Matthew and {Delgado}, Ana Maria and {Caliendo}, Joyce and {Kreisch}, Christina and {Wong}, Kaze W.~K. and {Coulton}, William R. and {Eickenberg}, Michael and {Parimbelli}, Gabriele and {Ni}, Yueying and {Steinwandel}, Ulrich P. and {La Torre}, Valentina and {Dave}, Romeel and {Battaglia}, Nicholas and {Nagai}, Daisuke and {Spergel}, David N. and {Hernquist}, Lars and {Burkhart}, Blakesley and {Narayanan}, Desika and {Wandelt}, Benjamin and {Somerville}, Rachel S. and {Bryan}, Greg L. and {Viel}, Matteo and {Li}, Yin and {Irsic}, Vid and {Kraljic}, Katarina and {Marinacci}, Federico and {Vogelsberger}, Mark},
title = "{The CAMELS Project: Public Data Release}",
journal = {\apjs},
year = 2023,
month = apr,
volume = {265},
number = {2},
eid = {54},
pages = {54},
doi = {10.3847/1538-4365/acbf47},
archivePrefix = {arXiv},
eprint = {2201.01300},
primaryClass = {astro-ph.CO},
adsurl = {https://ui.adsabs.harvard.edu/abs/2023ApJS..265...54V}
}

@ARTICLE{VillaescusaNavaroo2022,
       author = {{Villaescusa-Navarro}, Francisco and {Ding}, Jupiter and {Genel}, Shy and {Tonnesen}, Stephanie and {La Torre}, Valentina and {Spergel}, David N. and {Teyssier}, Romain and {Li}, Yin and {Heneka}, Caroline and {Lemos}, Pablo and {Angl{\'e}s-Alc{\'a}zar}, Daniel and {Nagai}, Daisuke and {Vogelsberger}, Mark},
        title = "{Cosmology with One Galaxy?}",
      journal = {\apj},
         year = 2022,
        month = apr,
       volume = {929},
       number = {2},
          eid = {132},
        pages = {132},
          doi = {10.3847/1538-4357/ac5d3f},
archivePrefix = {arXiv},
       eprint = {2201.02202},
 primaryClass = {astro-ph.CO},
       adsurl = {https://ui.adsabs.harvard.edu/abs/2022ApJ...929..132V}
}

@ARTICLE{Vogelsberger2013,
       author = {{Vogelsberger}, Mark and {Genel}, Shy and {Sijacki}, Debora and {Torrey}, Paul and {Springel}, Volker and {Hernquist}, Lars},
        title = "{A model for cosmological simulations of galaxy formation physics}",
      journal = {\mnras},
         year = 2013,
        month = dec,
       volume = {436},
       number = {4},
        pages = {3031-3067},
          doi = {10.1093/mnras/stt1789},
archivePrefix = {arXiv},
       eprint = {1305.2913},
 primaryClass = {astro-ph.CO},
       adsurl = {https://ui.adsabs.harvard.edu/abs/2013MNRAS.436.3031V}
}

@ARTICLE{Walther2019,
       author = {{Walther}, Michael and {O{\~n}orbe}, Jose and {Hennawi}, Joseph F. and {Luki{\'c}}, Zarija},
        title = "{New Constraints on IGM Thermal Evolution from the Ly{\ensuremath{\alpha}} Forest Power Spectrum}",
      journal = {\apj},
         year = 2019,
        month = feb,
       volume = {872},
       number = {1},
          eid = {13},
        pages = {13},
          doi = {10.3847/1538-4357/aafad1},
archivePrefix = {arXiv},
       eprint = {1808.04367},
 primaryClass = {astro-ph.CO},
       adsurl = {https://ui.adsabs.harvard.edu/abs/2019ApJ...872...13W}
}

@ARTICLE{Walther2025,
       author = {{Walther}, Michael and {Sch{\"o}neberg}, Nils and {Chabanier}, Sol{\`e}ne and {Armengaud}, Eric and {Sexton}, Jean and {Y{\`e}che}, Christophe and {Lesgourgues}, Julien and {Mosbech}, Markus R. and {Ravoux}, Corentin and {Palanque-Delabrouille}, Nathalie and {Luki{\'c}}, Zarija},
        title = "{Emulating the Lyman-Alpha forest 1D power spectrum from cosmological simulations: new models and constraints from the eBOSS measurement}",
      journal = {\jcap},
         year = 2025,
        month = may,
       volume = {2025},
       number = {5},
          eid = {099},
        pages = {099},
          doi = {10.1088/1475-7516/2025/05/099},
archivePrefix = {arXiv},
       eprint = {2412.05372},
 primaryClass = {astro-ph.CO},
       adsurl = {https://ui.adsabs.harvard.edu/abs/2025JCAP...05..099W}
}

@ARTICLE{Wang2013,
       author = {{Wang}, Mei-Yu and {Croft}, Rupert A.~C. and {Peter}, Annika H.~G. and {Zentner}, Andrew R. and {Purcell}, Chris W.},
        title = "{Lyman-{\ensuremath{\alpha}} forest constraints on decaying dark matter}",
      journal = {\prd},
         year = 2013,
        month = dec,
       volume = {88},
       number = {12},
          eid = {123515},
        pages = {123515},
          doi = {10.1103/PhysRevD.88.123515},
archivePrefix = {arXiv},
       eprint = {1309.7354},
 primaryClass = {astro-ph.CO},
       adsurl = {https://ui.adsabs.harvard.edu/abs/2013PhRvD..88l3515W}
}

@ARTICLE{Weinberger2017,
       author = {{Weinberger}, Rainer and {Springel}, Volker and {Hernquist}, Lars and {Pillepich}, Annalisa and {Marinacci}, Federico and {Pakmor}, R{\"u}diger and {Nelson}, Dylan and {Genel}, Shy and {Vogelsberger}, Mark and {Naiman}, Jill and {Torrey}, Paul},
        title = "{Simulating galaxy formation with black hole driven thermal and kinetic feedback}",
      journal = {\mnras},
         year = 2017,
        month = mar,
       volume = {465},
       number = {3},
        pages = {3291-3308},
          doi = {10.1093/mnras/stw2944},
archivePrefix = {arXiv},
       eprint = {1607.03486},
 primaryClass = {astro-ph.GA},
       adsurl = {https://ui.adsabs.harvard.edu/abs/2017MNRAS.465.3291W}
}

@ARTICLE{Weinberger2020,
       author = {{Weinberger}, Rainer and {Springel}, Volker and {Pakmor}, R{\"u}diger},
        title = "{The AREPO Public Code Release}",
      journal = {\apjs},
         year = 2020,
        month = jun,
       volume = {248},
       number = {2},
          eid = {32},
        pages = {32},
          doi = {10.3847/1538-4365/ab908c},
archivePrefix = {arXiv},
       eprint = {1909.04667},
 primaryClass = {astro-ph.IM},
       adsurl = {https://ui.adsabs.harvard.edu/abs/2020ApJS..248...32W}
}

@ARTICLE{Wiersma2009,
       author = {{Wiersma}, Robert P.~C. and {Schaye}, Joop and {Theuns}, Tom and {Dalla Vecchia}, Claudio and {Tornatore}, Luca},
        title = "{Chemical enrichment in cosmological, smoothed particle hydrodynamics simulations}",
      journal = {\mnras},
         year = 2009,
        month = oct,
       volume = {399},
       number = {2},
        pages = {574-600},
          doi = {10.1111/j.1365-2966.2009.15331.x},
archivePrefix = {arXiv},
       eprint = {0902.1535},
 primaryClass = {astro-ph.CO},
       adsurl = {https://ui.adsabs.harvard.edu/abs/2009MNRAS.399..574W}
}

\end{document}